\documentstyle[11pt,aaspp4]{article}

\def\lae{\mathrel{<\kern-1.0em\lower0.9ex\hbox{$\sim$}}}
\def\gae{\mathrel{>\kern-1.0em\lower0.9ex\hbox{$\sim$}}}

\def\phan2{\phantom{2}}

\def\etal{$et~al.~$}
\def\kms{km~s$^{-1}$}
 
\slugcomment{Accepted for publication in the Astrophysical Journal}
 
\lefthead{C\^ot\'e \etal}
\righthead{Dynamics of M49 Globular Clusters}
 
\begin{document}
 
\title{Dynamics of the Globular Cluster System Associated with M49 (NGC~4472): 
Cluster Orbital Properties and the Distribution of Dark Matter\altaffilmark{1}}
 
\author{Patrick C\^ot\'e\altaffilmark{2}, Dean~E.~McLaughlin\altaffilmark{3},
Judith~G.~Cohen\altaffilmark{4} and John~P.~Blakeslee\altaffilmark{5}}
 
\altaffiltext{1}{Based on observations obtained at the W.M. Keck Observatory 
which is operated by the California Association for Research in Astronomy, a 
scientific partnership between the California Institute of Technology, the 
University of California, and the National Aeronautics and Space Administration.}
 
\altaffiltext{2}{Department of Physics and Astronomy, Rutgers University,
136 Frelinghuysen Road, Piscataway, NJ 08854-8019; pcote@physics.rutgers.edu}
 
\altaffiltext{3}{Space Telescope Science Institute, 3700 San Martin
Drive, Baltimore, MD 21218; deanm@stsci.edu}
 
\altaffiltext{4}{California Institute of Technology, Mail Stop 105-24,
Pasadena, CA 91125; jlc@astro.caltech.edu}
 
\altaffiltext{5}{Department of Physics and Astronomy, Center for
Astrophysical Sciences, Johns Hopkins University,
3400 North Charles Street, Baltimore, MD 21218; jpb@pha.jhu.edu}
 
\begin{abstract}
 
Using the Low Resolution
Imaging Spectrometer on the Keck I and II telescopes, we have measured radial
velocities for 196 globular clusters
(GCs) around M49 (NGC~4472), the brightest member of the Virgo Cluster. 
Combined with published data, they bring the total number of GCs with measured
radial velocities in this galaxy to 263. In terms of sample size, spatial coverage, 
velocity precision, and the availability of metallicity estimates from Washington
photometry, this radial velocity database resembles that presented
recently for M87 (NGC~4486), Virgo's cD galaxy and its second-ranked member.
 
We extract the projected kinematics of the full sample of GCs and of
separate subsamples of 158 metal-poor and 105 metal-rich GCs.
In agreement with previous results for the global GC kinematics 
based on smaller datasets, we find that the GC system as a whole
exhibits a slow overall rotation that is due almost entirely to a net
rotation of the metal-poor GC subsystem alone. In
a spatial average, the metal-rich GCs shows essentially no rotation. 
As a function of galactocentric position,
the metal-poor GCs rotate roughly about the photometric minor axis
of M49 and at an approximately constant level of $\Omega R \sim 100$--150~km~s$^{-1}$ 
out to $R\simeq 2 R_{\rm eff}$. The metal-rich GC system shows some
evidence (at roughly 1-$\sigma$ significance) for weak rotation 
($\Omega R \sim 50$ \kms) beyond $R\gae
0.5 R_{\rm eff}$, also about the galaxy's minor-axis, {\it but in the opposite
direction from the metal-poor GCs}. 
Outside of $R \sim R_{\rm eff}$,
the line-of-sight velocity dispersion of the metal-poor GCs exceeds that of
their metal-rich counterparts by $\sim$ 50\%.
We also note the presence of a well defined grouping of 10 metal-rich GCs
that are located at opposite poles 
along the galaxy's major axis and which appear to be rotating at nearly 
300 \kms\ about the minor axis. This grouping may be the relic of a past
merger or accretion event.
 
The dynamics of the GC system are modeled by using published catalogs and
number counts to define three-dimensional GC density distributions as input
to a Jeans-equation analysis. We show that the GC radial velocities alone
point unequivocally --- and independently of X-ray observations --- to the need for
a massive dark halo associated with M49 and the Virgo B subcluster around
it. We then use a mass model for M49/Virgo B, constructed without reference
to any GC data and described in detail in a forthcoming paper, to infer the 
orbital properties of the M49 globulars. {\it The GC system as a whole is 
shown to be consistent with an almost perfectly isotropic velocity ellipsoid.}
It is more difficult to draw any firm conclusions on the orbital (an)isotropy 
of the two metallicity subsamples, owing to the large uncertainties in their 
individual spatial density profiles and to the poorly observationally 
defined kinematics of the metal-rich GCs in particular.
 
After M87, M49 is the second elliptical galaxy for which we have been
able to demonstrate velocity isotropy in the globular cluster system
overall, when no division based on GC color or metallicity is attempted.
Thus, the data for these two galaxies lend support the general {\it assumption} of 
isotropy when using GC kinematics to study the dark-matter distribution 
in early-type galaxies.  We also compare the kinematic properties of the GC
system of M49 to those of M87, M31, and the Milky Way: the other galaxies
for which samples of one hundred or more GC velocities have been accumulated. 
We argue that, contrary to the traditional view of GCs
as non-rotating, or slowly rotating, systems, rotation may in fact be 
common byproduct of the formation of GC systems. However, the quantitative 
details of the rotation are still not clear, particularly
with regard to the question of possible differences between metal-poor
and metal-rich globulars. 
 
\end{abstract}

 
\section{Introduction}
 
Systems of globular clusters (GCs) can be found in galaxies of virtually all
types, from the brightest members of rich galaxy clusters to faint dwarfs in
loose groups. Recent studies suggest that, over a range of roughly ten 
thousand in total luminosity, about 0.25\% of the baryonic mass in galaxies 
takes the form of GCs (Blakeslee, Tonry \& Metzger 1997; McLaughlin 1999a;
Blakeslee 1999).  Given this ubiquity, and the relative ease with which they 
can be detected and studied in external galaxies, GCs have long been 
recognized as invaluable probes of galaxy formation and evolution (see, $e.g.,$ 
Harris 1991 and references therein). To date, observational studies of
extragalactic GC systems have most commonly relied on GC specific frequencies,
metallicities and, in a few cases, ages, to constrain formation models for GC
systems and their host galaxies ($e.g.$, Cohen, Blakeslee \& Rhyzov 1998;
Kundu \etal 1999; Puzia \etal 1999; Gebhardt \& Kissler-Patig 1999; 
Larsen \etal 2001; Jord\'an \etal 2002; Cohen, Blakeslee \& C\^ot\'e 2003). 
However, additional insights into the process of galaxy formation may, at 
least in principle, be gleaned from the dynamical properties of GC systems.
 
Not only do radial velocity surveys of extragalactic GCs provide a way of
tracing out the distribution of dark matter within the host galaxy but,
with sufficiently large samples, they offer the hope of measuring directly
the shape of the GC velocity ellipsoid. Not surprisingly, the first dynamical
studies of an extragalactic GC system focused on the rich GC population of
M87=NGC 4486 ($e.g.,$ Huchra \& Brodie 1987; Mould, Oke \& Nemec 1987; Mould \etal
1990). However, with
radial velocities available for just 44 clusters, only very weak constraints
could be placed on the mass distribution --- and then only if some assumption
({\it i.e.}, isotropy) were made {\it a priori} regarding the GC orbital properties
(see Merritt \& Tremblay 1993). As Merritt \& Tremblay show, dauntingly large
samples of order one thousand velocities would be required to constrain,
simultaneously and self-consistently in the absence of any other information,
both the velocity anisotropy of a GC system and the underlying mass
distribution of its parent galaxy.
 
Only recently have multi-object spectrographs on 4m- and 8m-class telescopes
begun to produce anywhere near the large numbers of radial velocities
needed to execute such a program in any galaxy. Hanes \etal (2001) compiled a radial
velocity database of 278 GCs associated with M87 after combining new CFHT
velocities with published data from Mould \etal (1990), Cohen \& Ryzhov (1997)
and Cohen (2000). By using an {\it independent} mass model for M87 and the
surrounding ``Virgo A'' cluster (McLaughlin 1999b), C\^ot\'e \etal (2001)
showed from these data that the M87 GC system as a whole has an almost
perfectly isotropic velocity dispersion tensor. Considered separately, the
metal-poor and metal-rich GC subsystems showed some evidence for weak
tangential and radial biases, respectively.
 
Since M87 contains the richest and most thoroughly studied GC system in the
Local Supercluster, it was the obvious first choice for such studies.  However, its
unique location --- at the dynamical center of the Virgo Cluster as traced by
both galaxies and intracluster gas --- complicates the interpretation of the
dynamical properties of its GC system. For instance, C\^ot\'e \etal (2001)
noted that the apparent rotation of the metal-poor GCs in M87 might be better
described as a velocity ``shear'' similar to that exhibited on larger scales
by the surrounding Virgo cluster {\it galaxies}. If correct, this would
suggest a connection between the M87 GCs and the surrounding population
of (mostly dwarf) galaxies, perhaps due to the slow infall of the material
onto M87 itself.
 
If we are to understand which properties of the M87 GC system are generic to
early-type galaxies, and which are unique to M87 as a result of its central
location in Virgo, then we require dynamical studies of additional GC systems.
An appealing second target is that of M49 (= NGC~4472). As the brightest
member of the Virgo cluster, M49 contains $\sim$ 6000 GCs projected within
100 kpc of its center (Lee \etal 1998; McLaughlin 1999a; Rhode \& Zepf
2001) --- just the number expected for an elliptical galaxy
with the luminosity of M49
and a ``normal'' specific frequency. M87, by contrast, has an integrated
luminosity about 80\% that of M49 but contains some 13,500 GCs projected
within 100 kpc of its center. This factor-of-three difference in GC specific
frequency between the two galaxies may indicate that some as yet
unidentified ``second parameter" has played a role in the formation and/or
evolution of their GC systems (which otherwise appear remarkably similar; see
C\^ot\'e 2003). However, the relevance of the observed specific frequency
difference recently has been called into question by McLaughlin (1999a), who
showed that the number of GCs associated with M87 and M49 are both consistent
with a ``universal'' ratio of total GC mass to total baryonic galaxy mass,
$\epsilon_{\rm GC} = 0.26\pm0.05\%$, once the contribution of X-ray gas mass
in each galaxy is included. The principal distinction between the two may
then be the fact that M87 sits at the gas-rich center of the main Virgo
cluster, while M49 occupies the core of the relatively gas-poor,
``Virgo B'' subcluster (Binggeli, Tammann \&
Sandage 1987; Binggeli, Popescu \& Tammann 1993; Schindler, Binggeli \&
B\"ohringer 1999).
 
The first radial velocity measurements for M49 GCs were presented by Mould
\etal (1990), who reported a global velocity dispersion of $340\pm50$ \kms\
based on a sample of 26 GCs. Building on this work, Sharples \etal (1998)
presented radial velocities for an additional 47 GCs, found a difference in
velocity dispersion between the metal-poor and metal-rich subsystems (320 and
240 \kms, respectively), and noted the presence of apparent rotation among the
metal-poor subsystem. These conclusions were reinforced by Zepf \etal
(2000), who measured velocities for another 87 GCs and found global velocity
dispersions of 356$\pm$25 \kms\ and 221$\pm$22 \kms\ for the
metal-poor and metal-rich subsystems, with projected rotation amplitudes
of 101 and 15 \kms\ for the two populations. These findings are in stark
contrast with those for M87, where both GC populations were found to be
rapidly rotating, with projected rotation amplitudes of roughly 170 \kms\
(C\^ot\'e \etal 2001). Zepf \etal also calculated the mass distribution around
M49 implied by the velocity-dispersion profile of the full GC system under the
{\it assumption} of orbital isotropy, and showed this to resemble
the gravitating-mass profile inferred from X-ray observations (Irwin \&
Sarazin 1996).
 
In this paper, we present 196 new radial velocities for GC candidates in M49
obtained with the Keck I and II telescopes. Combining our new radial
velocities with those from the literature, we isolate a sample of 263 distinct
objects whose colors and radial velocities are consistent with their being GCs
belonging to M49. This database is comparable to that presented recently for
M87 (Hanes \etal 2001) in terms of sample size, spatial coverage
(from about 0.1 to 3 effective radii),
velocity precision, and the availability of metallicities from Washington
photometry. With it, we largely confirm the earlier conclusions
summarized above on the global, spatially averaged kinematics of the GC
system as a whole and of its metal-poor and metal-rich components.
However, with our expanded dataset, we are better able to examine trends in
the GC kinematics as a function of galactocentric position. We see some
evidence that the bulk of the
metal-rich GC system may in fact rotate---albeit very slowly and with low
statistical significance---in a direction opposite to the metal-poor GCs.
 
We then go on to carry out a dynamical analysis of the three GC samples
(full, metal-poor, and metal-rich) based on the Jeans equation following
the methodology laid out in McLaughlin (1999b) and C\^ot\'e \etal (2001).
We adopt a mass model for M49/Virgo B that we have constructed completely
independently of any GC data (McLaughlin \& C\^ot\'e 2003), which allows
us to treat the velocity anisotropy of the GC system as a
free parameter to be constrained. Our main new result is the 
demonstration that velocity isotropy suffices fully to 
describe the velocity ellipsoid of the full globular cluster 
system ($i.e.$, with no attempt made to separate red GCs from
blue). M49 is the second galaxy (after M87) in which
we have attempted this sort of analysis, and it is the second in which
isotropy has been found to hold to a high level of precision. Quite apart
from any bearing that GC dynamics will have on issues of galaxy and globular
cluster formation, this suggests that it could very well be reasonable
to {\it assume} orbital isotropy in other GC systems in order to
use them as accurate probes of the total mass distribution in early-type
galaxies whose mass profiles have not been secured through independent means.

\section{Observations, Reductions and the Composite Database}
 
In this section, we describe the criteria used to target candidate GCs for
observation with the Low Resolution Imaging Spectrometer (LRIS; Oke \etal
1995), the procedures used to acquire and reduce the  LRIS spectra, and
the methods used to define the composite GC database for M49.
 
\subsection{Sample Selection}
 
Before embarking upon a radial velocity survey, we require a sample of objects
whose magnitudes and colors are consistent with those expected for
GCs associated with the program galaxy. We used the photometric catalog of
Geisler, Lee \& Kim (1996) to select candidate GCs for spectroscopic
observations. This catalog, which contains GC candidates spread over an area of
$\sim$ 256 arcmin$^2$ centered on M49, has the benefit that the imaging on which it is based
was carried out in Washington $C$ and $T_1$ filters, a combination that is known to
provide excellent metallicity sensitivity for old stellar populations.
A total of 1774 GC candidates brighter than $T_1 = 23$ make up the catalog.
Since accurate coordinates are needed to design the slit-masks, we
combined the GC pixel positions given in Geisler \etal (1996)
with the coordinates reported by Sharples \etal (1998) for 61 GCs to
derive astrometric positions for all objects in the
Geisler \etal (1996) catalog.
 
\subsection{Keck Spectroscopy}
 
Spectra for candidate GCs were obtained in the course of four observing runs
at the W.M. Keck Observatory during the 1998, 1999 and 2000 observing seasons.
The observing log given in Table~\ref{tab1} records the identification
number, observing dates, grating, central wavelength and
resolving power for each observing run. A total of 14 LRIS masks were
designed, targeting GC candidates in the range $1 \le (C-T_1) \le 2.25$
and 20 $\le T_1 \le$ 22.25.\footnotemark\footnotetext{For 
$\langle C-T_1 \rangle = 1.60$ --- the mean color of our 263 GCs --- this
magnitude range is roughly equivalent to $20.5 \lae V \lae 22.7$ (Geisler 1996).}
Individual masks contained between 19 and 29 slits,
measuring 1\farcs0 in width and having a minimum length of 8\farcs0. Two
different instrumental setups were employed. During runs 1 and
4, we used a 600~line~mm$^{-1}$ grating blazed at 5000 \AA\ and
centered at wavelengths of 5200 \AA\ and 4860 \AA , respectively. During runs
2 and 3, we used a 600 line mm$^{-1}$ grating blazed at 7500 \AA\ and
centered at 8550 \AA . For both configurations, the dispersion was 1.28 \AA\
pixel$^{-1}$, giving a spectral resolution of 6 \AA\  and a wavelength
coverage of $\sim$2600 \AA. For each mask, we obtained a pair of exposures
ranging between 2400s and 3000s, with comparison spectra for Hg-Ne-Ar-Kr-Xe
and quartz lamps taken before and/or after each exposure. On each night,
long-slit spectra for 2--4 IAU radial velocity standard stars were obtained
during twilight. The extracted, wavelength-calibrated spectra for the
candidate GCs were cross-correlated against a master template spectrum
created from these spectra. All reductions were performed within the IRAF
environment.\footnotemark\footnotetext{IRAF is distributed by the
National Optical Astronomy Observatories, which are operated by the
Association of Universities for Research in Astronomy, Inc., under
contract to the National Science Foundation.}
 
Uncertainties for radial velocities measured from spectra acquired during runs
1--3 were derived from the cross-correlation functions using the method of
Tonry \& Davis (1979). A total of 147 radial velocities were measured during
these observing runs; the mean and median uncertainties for these measurements
are 104 and 97 km s$^{-1}$. Spectra obtained during run 4 were typically of
the highest quality, as only the brightest GC candidates were selected during
this run ($i.e.,$ objects with 20.1 $\le T_1 \le$ 21.51).
Radial velocities were measured from these spectra following the methods
outlined in Cohen \& Ryzhov (1997), and representative uncertainties of 25 or
50 km~s$^{-1}$ were assigned based on the repeatability of measurements based on
multiple strong absorption lines. Including these measurements,
the total number of new radial velocities presented here is 196, with
mean and median uncertainties of 86 and 79 km s$^{-1}$, respectively.
 
\subsection{The Composite Database}
 
Before examining the GC kinematics or carrying out a dynamical analysis,
we must combine the new velocities with published measurements to isolate a
sample of bona fide GCs. In order to correct for possible systematic
differences between radial velocities measured during different runs, and with
different instruments or telescopes, we have
transformed the raw velocities from April 1998 and April 1999 (runs 1 and 3,
respectively) onto the system defined by
our April 2000 observations (run 4) via the relations,
\begin{equation}
\begin{array}{rrrrr}
v_{p,1} & = & (1.023)v^{\prime}_{p,1} & + & 114~{\rm km~s^{-1}} \\
v_{p,3} & = & (0.950)v^{\prime}_{p,3} & + &  46~{\rm km~s^{-1}} \ ,
\end{array}
\label{eq1}
\end{equation}
which were derived from GCs whose velocities were measured during run 4, and either of
runs 1 and 3. Since there are no objects in common between runs 2 and 4, no
correction was applied to the former dataset. Likewise, no corrections were
applied to the velocities of Sharples \etal (1998) and Zepf \etal (2000),
since both datasets show good agreement with our measurements. However, we found it
necessary to apply a transformation to the velocities of Mould \etal (1990):
\begin{equation}
\begin{array}{rrrrr}
v_{p,M} & = & (0.904)v^{\prime}_{p,M} & + & 215~{\rm km~s^{-1}} \ .
\end{array}
\label{eq2}
\end{equation}
Here $v_{p,M}$ and $v^{\prime}_{p,M}$ denote the corrected and
uncorrected
velocities from Mould \etal (1990). Both the scale and zero-point terms
in this relation agree with those found by Hanes \etal (2001) from a
comparison of Mould \etal (1990) radial velocities for M87 GCs with those
measured at Keck and CFHT. Figure~\ref{fig01} compares velocities
measured during run 4 with previously published velocities and those
obtained during other Keck runs,
after applying the above transformations. The dashed line shows the
one-to-one relation, while the dotted line indicates the line of best fit.
 
After transforming the radial velocities onto a common system, the
various datasets were merged and a weighted mean velocity,
$\langle v_p\rangle$, was calculated from all measurements for each object.
Unfortunately, given the galaxy's low systemic velocity
($v_{\rm gal} = 997\pm7$ \kms\ according to the NASA Extragalactic
Database) and the large line-of-sight velocity dispersion of the GC system
($\sigma_p\sim300$ \kms; see \S3 below), there is some ambiguity in
distinguishing low-velocity GCs from foreground Galactic stars. After some
experimentation, we decided to discard those objects with radial velocities
outside of the interval,
\begin{equation}
250 \le \langle v_p\rangle \le 1950~{\rm km~s}^{-1},
\label{eq3}
\end{equation}
where the upper limit was chosen to include a few unambiguous GCs with radial
velocities of $\langle v_p\rangle \gae 1900$ \kms. Requiring this
range to be symmetric about the systemic velocity would give a low-end cutoff
of $\sim$ 50 \kms. Such a cutoff would almost certainly result
in the inclusion of some halo stars in our sample. If the 
velocity dispersion of the M49 GC system is $\sim$ 300 \kms\ then, in a
sample of the present size, we expect at most two M49 globulars will fall below 
the low-end velocity limit in equation (\ref{eq3}).
 
In addition to the above selection on radial velocity, we imposed the
condition that true M49 GCs must have colors in the range,
\begin{equation}
1.0 \le (C-T_1) \le 2.25~{\rm mag},
\label{eq4}
\end{equation}
and adopted a foreground extinction of $E(B-V) = 0.022$ based on the
DIRBE maps of Schlegel, Finkbeiner \& Davies (1998). According to the
relation of Secker \etal (1995), this extinction is equivalent to
$E(C-T_1)$ = 0.045. With the color-metallicity calibration of
Geisler \& Forte (1990), this color range translates to a metallicity
interval of $-$2.15~$\lae$~[Fe/H]~$\lae$~+0.8~dex. From optical
spectroscopy,
Beasley \etal (2000) showed the M49 GC system to span a range of
$-$1.6 $\lae$ [Fe/H] $\lae$ 0~dex, so our selection limits should
include the vast majority of GCs in M49.
 
Figure \ref{fig02} presents an illustration of these selection criteria.
The upper left panel shows color histograms for the 1774 objects
with $T_1 \le 23$ in the the catalog of Geisler \etal (1996), as well as for
the 276 GC  candidates with measured velocities (upper and lower histograms,
respectively). The dashed vertical lines show the adopted cutoffs on $(C-T_1)$
color, while the dotted vertical lines indicates the value of $(C-T_1) =
1.625$ used to divide the sample into metal-poor and metal-rich components.
According to the color-metallicity relation of Geisler \& Forte (1990),
this dividing point corresponds to a metallicity of
[Fe/H]~=~$-$0.68~dex.\footnotemark\footnotetext{Globular cluster metallicities quoted
in this paper are $\simeq$ 0.1 dex lower
than those reported in Geisler \etal (1996) since we have adopted
the Schlegel \etal (1998) extinction of $E(B-V) = 0.022$; 
Geisler \etal (1996) assumed $E(B-V) = 0$.}
 
The upper right panel of Figure~\ref{fig02} shows a radial velocity histogram
for the same sample of 276 objects. Note that two objects (identification
numbers \#4497 and \#1982 in Geisler \etal 1996) fall outside the plotted
region. The vertical lines show the adopted cutoffs on radial velocity, while
the arrow indicates the velocity of M49 itself. The lower panel of
Figure~\ref{fig02} shows the joint constraints on color and radial velocity in
the form of a color-velocity diagram. Filled circles show all objects with
measured radial velocities, with the exception of \#4497 and \#1982 which,
once again, fall outside the plotted region. The dashed box indicates
the adopted selection criteria for color and radial velocity. Note that \#2256,
the lone circled object inside the boxed region, was not included in the
final sample of GCs since two independent velocity measurements for this
object differ by more than 700 km s$^{-1}$. Similarly, two independent
measurements for \#1982 disagree on whether this object is a true GC or
a background galaxy. Discarding it, and the 12 objects that fall outside of
the dashed region in Figure~\ref{fig02}, brings the final sample to 263 GCs.
 
The final GC database is given in Table~\ref{tab2}. The
first five columns of this table record the object identification number from
Geisler \etal (1996), right ascension, declination, distance from
the center of M49 in arcseconds, $R$, and position angle, $\Theta$,
in degrees East of North. In calculating $R$ and $\Theta$, we take 
the center of M49 to be $\alpha$(J2000) = 12:29:47.5 and
$\delta$(J2000) = +08:00:10.5  following Sharples \etal (1998).
The next three columns of Table~\ref{tab2}
give the $T_1$ magnitude, $(C-T_1)$ color, and metallicity based on
the color-metallicity relation of Geisler \& Forte (1990).
Individual radial velocity measurements and their source are presented
in the next two columns. The final column gives the weighted mean
velocity and uncertainty for each object. The 263 confirmed GCs are listed
first in the table; objects classified as foreground stars or
background galaxies, and objects with discrepant radial velocity
measurements, are given at the end.
 
The spatial distribution of the final sample of GCs is shown in
Figure~\ref{fig03}.
Metal-poor GCs are indicated by circles, and metal-rich GCs by squares.
Open and filled symbols indicate GCs with positive and negative velocity
residuals, ${\Delta}v_p \equiv \langle{v_p}\rangle - v_{\rm gal}$,
with respect to M49. Symbol sizes are proportional to the absolute value of
the residual velocity. The dashed lines show the photometric major and minor
axes of the galaxy, with respective position angles of
$\Theta = 155^{\circ}$ and $65^{\circ}$ (Kim \etal 2000).
The large circle shows our determination of galaxy's effective radius,
$R_{\rm eff} = 3\farcm1$, based on our modeling of the Kim \etal surface
photometry (see \S4.2 below). At our adopted distance of 15~Mpc,
1\arcmin = 4.363 kpc, so that $R_{\rm eff}=13.5$ kpc. The full GC sample
spans a range of projected galactocentric radius $0\farcm4 \lae R \lae
9\farcm5$, equivalent to $1.9 \lae R \lae 41$ kpc or $\simeq$ 0.1--3$R_{\rm eff}$.
A comparison of
Figure~\ref{fig03} with Figure~2 of C\^ot\'e \etal (2001) reveals the
azimuthal distribution of radial velocity measurements to be somewhat more
uniform in M49 than was the case for M87.
 
\section{Kinematics of the Globular Cluster System}
 
Before proceeding with a dynamical analysis of the GC system, we examine its
kinematics in a model-independent way. Our aim is to determine the basic
parameters that describe the GC system: the average line-of-sight velocity,
projected velocity dispersion and, if rotation is important, its
amplitude and the position angle of the rotation axis. In this section,
we consider both the global kinematic properties of the GC system
and their behavior as a function of galactocentric distance.
 
We fit the observed line-of-sight velocities of the GCs with the
function,
\begin{equation}
\langle v_p\rangle = v_{\rm sys} + (\Omega R) \sin(\Theta - \Theta_0)\ ,
\label{eq5}
\end{equation}
where $\Theta$ is the projected position angle from a reference axis (taken
here to be the North--South direction, such that $\Theta$ is measured in
degrees East of North) and $\Theta_0$ is the orientation of the
rotation axis of the GC system. A complete discussion of the implications of
fitting sine curves of this type to projected ($\langle{v_p}\rangle$,
$\Theta$) data is given in C\^ot\'e \etal (2001). Briefly, in so doing we
assume that the GC system is spherically symmetric with an intrinsic
angular velocity field stratified on spheres, and that the GC rotation
axis lies exactly in the plane of the sky (in other words, that 
the galaxy is being viewed
``edge-on''). In principle, the rotation amplitude, $\Omega R$, may then be
any function of galactocentric radius $R$; fitting equation (\ref{eq5})
to our data does not imply an assumption of solid-body rotation or of
cylindrical symmetry in the velocity field. Our assumption of spherical
symmetry for the GC system is reasonable given its modest projected ellipticity
(0.16 according to Lee \etal 1998).
 
\subsection{Global Kinematic Properties}
 
Kinematic properties of various subsets of the GC system are summarized
in Table~\ref{tab3}. The first five columns in this table record the radial
range spanned by the cluster sample in question, the median galactocentric
distance, $\langle R\rangle$, and number, $N$, of GCs in each sample, the
average radial velocity (the biweight ``location'' of Beers,
Flynn, \& Gebhardt 1990), $\overline{v_p}$, and the rms dispersion about this
average (the biweight ``scale'' of Beers \etal 1990), $\sigma_p$.
Quoted uncertainties represent 68\% (1-$\sigma$) confidence intervals,
determined from a numerical bootstrap procedure in which 1000 artificial
datasets are individually analyzed after choosing $N$ clusters at random 
from the actual subsample under consideration.
 
Rather than let $v_{\rm sys}$ be a free parameter in the fits of equation
(\ref{eq5}) to the $\langle v_p\rangle$--$\Theta$ data, we make use of the
expectation (confirmed by the results in Column 4 of Table \ref{tab3}) that the
average velocity of the GC system should equal that of M49 itself:
$v_{\rm sys}\equiv v_{\rm gal}=997$ \kms. In the remaining columns of Table
\ref{tab3}, we therefore give the position angle of the rotation axis, 
$\Theta_0$, the rotation amplitude, $({\Omega}R)$, and the dispersion 
about the best-fit sine curve, $\sigma_{p,r}$, obtained with this 
constraint placed on $v_{\rm sys}$.
 
The first line of each the three subsections of Table \ref{tab3} presents
the results of immediate interest: the global kinematics ($i.e.$, those determined
using all appropriate GC data with no discrimination on the basis of
galactocentric position) for our full GC sample and for the separate
metal-poor and metal-rich subsystems. The remainder of the table applies to
further division of the GC data into four wide radial bins; these results
are discussed in the next subsection.
 
The global velocity dispersion of the metal-poor GC system is
significantly larger than that of the metal-rich GC system: 
$\sigma_{p,r} = 342$ \kms\ versus 265 km~s$^{-1}$.  The global velocity dispersion
of the entire GC system, $\sigma_{p,r} = 312$ \kms, is naturally
intermediate to these values; it falls somewhat closer to the
metal-poor value because that sample is $\sim$ 50\% larger than the metal-rich
one. These results are consistent with the previous measurements of Sharples
\etal (1998) and Zepf \etal (2000) based on smaller samples.
 
Figure \ref{fig04} plots the individual
velocities $\langle v_p \rangle$ against projected position angle
$\Theta$ for GCs at all galactocentric radii in each of our three metallicity
samples. A horizontal, broken line in each panel of this figure marks the
velocity of M49, $v_{\rm gal}=997$ \kms. It is immediately apparent that the
lower dispersion of the metal-rich GC velocities relative to the
metal-poor sample stems in part from the absence of even a single GC
with $(C-T_1)\geq 1.625$ and $\langle v_p\rangle > 1500$
\kms --- a fact that is also reflected in
the empty, upper right-hand corner of the box used to identify globular
clusters in the bottom panel of Figure~\ref{fig02}. It would appear that this
effect is real: if the metal-rich GC velocity dispersion were the same
($\simeq$ 340 \kms) as the metal-poor one, then given a sample of 105
metal-rich clusters with $\langle v_p\rangle<1500$ \kms, we should have also
found 8--9 with $\langle v_p\rangle>1500$ km~s$^{-1}$. There is no obvious
deficiency in the spatial coverage of the full radial velocity survey that
might account for the discrepancy (see Figure~\ref{fig03}).
 
The bold sine curves in the two upper panels of Figure~\ref{fig04} trace the
best fits of equation (\ref{eq5}) to the entire GC dataset and the metal-poor
subset when $v_{\rm sys}$ is held fixed and equal to $v_{\rm gal}=997$ \kms\
({\it i.e.}, with $\Omega R$ and $\Theta_0$ given in Columns 6 and 7 of Table
\ref{tab3}). For the full sample, the best-fit sine
curve has $\Omega R = 53$ \kms\ and $\Theta_0 = 105^{\circ}$.
More importantly, we find that $\Omega R = 93$ \kms\ and
$\Theta_0 = 100^{\circ}$ for the metal-poor sample, while
$\Omega R$ is easily consistent with zero for the metal-rich GC system
(hence, the fit to it is not drawn on Figure~\ref{fig04}). Thus, averaged over
$R\lae3 R_{\rm eff}$ in M49, the metal-rich GC system shows essentially no
net rotation, while the metal-poor GC system does. Our result for the
metal-poor rotation amplitude is consistent with that of Zepf \etal (2000),
who found $\Omega R = 101$ \kms\ for a sample of 93 blue GCs. Our
finding for the metal-rich GC is similarly in keeping with previous
analyses (see also Sharples \etal 1998).
 
The position angle of the galaxy's photometric minor axis is
$\Theta_{\rm min} = 65^{\circ}$ (Kim \etal 2000) and, to within its
(large) 1-$\sigma$ errorbars, our fitted rotation axis for the metal-poor GC
system is roughly consistent with this. However, it is incompatible at the 
1.5-$\sigma$ level with alignment along the photometric {\it major} axis, which 
is the solution favored by Zepf \etal (2000) for their smaller sample of
metal-poor GCs (and their full GC system). Though also similar to the
photometric minor axis of M49, the fitted $\Theta_0$ for the metal-rich
GCs is effectively meaningless because of the formally null rotation
amplitude.
 
To summarize these {\it global, spatially averaged} impressions, the
average velocity of each GC sample is in very good agreement with that of M49
itself. The global velocity dispersions of the metal-poor and metal-rich
samples differ formally by 76 \kms\ (although the 2-$\sigma$ errorbars
overlap) with the metal-poor GCs being dynamically hotter than their
metal-rich counterparts. The GC system appears as a whole to be slowly
rotating, but this is purely the result of a net signal from the
metal-poor GC subsystem alone; the metal-rich subsystem has none.
The velocity dispersions of the full and the metal-poor GC samples are
not changed by the correction for rotation, as is apparent from a 
comparison of the values of $\sigma_{p,r}$ and $\sigma_p$ in 
Table~\ref{tab3}. Thus, regardless of its statistical significance, 
rotation is not dynamically important in the M49 GC system.
And although it is only loosely determined, the average kinematic axis of the
metal-poor GC system does not differ significantly from the photometric
minor axis of M49.
 
\subsection{Kinematic Properties as a Function of Projected Radius}
 
Table \ref{tab3} also presents the GC kinematics that result from further
dividing each of the three metallicity samples that we have defined into
four broad radial bins:
(1) $25\arcsec \leq R < 150\arcsec$;
(2) $150\arcsec \leq R < 250\arcsec$;
(3) $250\arcsec \leq R < 350\arcsec$; and
(4) $350\arcsec \leq R < 570\arcsec$.
Although the parameters in some bins with small GC numbers must be
viewed with caution, the results suggest that there are no drastic
changes in the  average and rms velocities for each sample.
 
It is also clear that the lower velocity dispersion of the metal-rich GCs
relative to the metal-poor GCs, noted above on the basis of globally
averaged kinematics, is only apparent outside of the first
radial bin; the metal-poor and metal-rich clusters within
$R\leq 150\arcsec\simeq0.8 R_{\rm eff}$ have the same $\sigma_p$ and
$\sigma_{p,r}$.
Figures \ref{fig05} and \ref{fig06} demonstrate these points 
visually. Figure \ref{fig05} is a plot of individual GC velocities
versus projected galactocentric radii, for the entire sample regardless of
color and for the metal-poor and metal-rich subsamples separately.
Overplotted are large, open squares at the median radius and average
velocity of each radial bin, as reported in Columns 2 and 4 of Table
\ref{tab3}. The vertical errorbars on these large squares represent the
dispersion about the average velocity (Column 5 of Table \ref{tab3}), and
the horizontal errorbars delimit the bins themselves. The broken horizontal
line through each panel shows the systemic velocity of the galaxy.
 
The scatter of points about the line $\langle v_p\rangle=997$ \kms\ in the
top panel of Figure~\ref{fig05} implies that the average and dispersion of
$\langle v_p\rangle$ in the full GC system are indeed essentially constant
as functions of galactocentric position --- until the region beyond $R\gae
7\farcm5\simeq33$ kpc, where there is an obvious dearth of high-velocity
GCs in our sample. We suspect that this is an artifact of the small sample
size, but we cannot rule out with certainty that is a real, physical effect.
Until the
question can be addressed directly (with still more GC velocity measurements
at large radii), any kinematics referring to $R\gae 30\ {\rm kpc}$ in the M49
GC system should perhaps be viewed as provisional. Meanwhile, the
middle and bottom panels of Figure~\ref{fig05} suggest that the average
velocities of the metal-poor and metal-rich subsamples agree separately, at
all $R\lae 30$ kpc, with the galaxy's systemic velocity. The bottom panel
also shows again the absence of any red GCs with $\langle v_p\rangle > 1500$
\kms, and it further reveals a deficit of clusters with $v\lae 700$ \kms\ at
radii $R\gae 3\arcmin\sim R_{\rm eff}$.

We have also constructed ``smoothed'', binning-independent radial
profiles of average and rms GC velocity, following the procedure
of C\^ot\'e \etal (2001): a radial bin of fixed width (chosen here to be
$\Delta R = 120\arcsec \simeq 8.7$~kpc) is slid through the GC dataset,
centering on each globular in turn, and if the number of objects falling
within this spatial bin exceeds a minimum of 20 then the biweight average
and rms velocity at that position are calculated. (Equation [\ref{eq5}] is
also fit to the $\langle v_p\rangle$--$\Theta$ data in the bin around every
point; the rotation profiles that result are discussed below.) Confidence
intervals about the best-fit kinematics are defined by the same bootstrap
method that was used to calculate the errorbars in Table~\ref{tab3}.
 
The profile of average GC velocity as a function of galactocentric position
that results from this procedure only confirms that $\overline{v_p}$
is constant (inside $R\lae 30$ kpc) and equal to $v_{\rm gal}$ for each
metallicity sample. The velocity dispersion profiles that we obtain are
shown in Figure~\ref{fig06} for the full GC system and the metal-poor and
metal-rich subsets (filled squares represent the dispersion $\sigma_p$ of
velocities about the average; open squares, the dispersion $\sigma_{p, r}$
about the best-fit sine curve at each point). The horizontal errorbar in the
top panel represents the 2\arcmin\ width of the sliding radial bin that we
have used; only points separated by this distance in these plots are
statistically independent. The dotted and solid lines around the points in
each panel denote the 68\% and 95\% confidence intervals
for each measurement. All in all, the impressions gleaned from Table
\ref{tab3} and Figure~\ref{fig05} are confirmed in Figure~\ref{fig06}.

The velocity dispersion of the full GC system is remarkably constant as a
function of radius inside $R\lae
30$ kpc, never straying by more than $1 \sigma$ from the global average of
312 \kms\ (dashed horizontal line in the top panel). We do not see
the gradual decline in $\sigma_p(R)$ with radius that was inferred by Zepf
\etal (2000) on the basis of a heavily spatially smoothed representation of
a smaller dataset. The dispersion profile of the metal-poor
GCs is also relatively flat; if anything, it tends to increase towards larger
radii (although this trend is not highly significant). This behavior is
opposite to that found by Zepf \etal (2000).

The metal-rich GC system is
roughly as dynamically hot as the metal-poor GC sample at small
galactocentric radii, but it becomes colder beyond $R\sim 3\arcmin \simeq
13\ {\rm kpc}\simeq 1 R_{\rm eff}$ before possibly increasing slightly again
towards $R\sim 6\arcmin\sim2 R_{\rm eff}$. None of this variation, however, is
significant at more than the 68\% confidence level. Also, note
that we are unable to trace the metal-rich GC kinematics beyond $R=6\arcmin$
in the present approach, because there are fewer than 20 red clusters between
that radius and $R\simeq9\farcm5$, whereas every point in Figure~\ref{fig06}
relies on data for at least 20 GCs within a 2\arcmin\ interval in $R$.
Thus, we have included in the bottom panel of Figure~\ref{fig06} (as a large
open square with vertical errorbars corresponding to 68\% and 95\% confidence
intervals) the estimate for $\sigma_p$ in the outermost radial bin of Table
\ref{tab3}. Zepf \etal (2000) presented a smoothed velocity dispersion
profile for the metal-rich GC system that appears somewhat flatter than ours
inside $R\lae 6\arcmin$.

The ``rotation-corrected'' GC velocity dispersions, $\sigma_{p, r}$,
are nowhere, in any of the three metallicity samples, dramatically different from
the dispersions, $\sigma_p$, about the average velocity. This is yet another
demonstration that rotation is
dynamically unimportant in the globular cluster system of this galaxy.
Nevertheless, it is of some interest to consider the rotation properties of
the GCs as a function of galactocentric radius, as they appear to be
somewhat more subtle than suggested by the global view of \S3.1, in which the
metal-poor GCs show a statistically significant, if small, net rotation while
the metal-rich ones do not.
 
We show in Figure~\ref{fig07} plots of GC velocity versus projected
position angle, for each of the four
radial bins defined in Table \ref{tab3}. A broken, horizontal line in
each panel again indicates the overall velocity of M49, $v_{\rm gal}=997$
\kms, and the bold sine curves trace the best fits of equation (\ref{eq5}) to
the metal-poor (filled points and solid curves) and metal-rich (open points
and broken curves) GCs in each bin. (The fit parameters are again those
in Columns 6 and 7 of Table \ref{tab3}.) This gives some indication of the
quality of the sine fits we are able to achieve with the data so finely
divided, and the extent to which such fits are even feasible given the data.
Note that we do not present any rotation fit for the
metal-rich GCs in the bin $250\arcsec\leq R < 350\arcsec$, for reasons that we
shall discuss shortly.

For more detail, Figures~\ref{fig08}--\ref{fig10} plot the runs of $\Omega R$
and $\Theta_0$ with galactocentric radius in the full GC system and in the
metal-poor and metal-rich subsystems, obtained during the same ``smoothing''
process that resulted in the velocity dispersion profiles in Figure~\ref{fig06}.
 
Figure \ref{fig08}, which presents the results for our full sample of 263
GCs, is included primarily for completeness, as it is somewhat more profitable
to consider the metal-poor and metal-rich subsamples separately. Worthy of note
here, however, are the large uncertainties at every radius in the $\Theta_0$
and $\Omega R$ determinations (as in Figure~\ref{fig06}, 68\% and 95\%
confidence intervals are indicated here), even when our full dataset is
analyzed in bins as wide as 2\arcmin. Much larger numbers of GC velocities
will be required if the rotation properties of the system are to be characterized
point-by-point in this way with any real confidence.
 
Figure \ref{fig09} shows the rotation axis and amplitude as functions of $R$
in the metal-poor GC system only. Inside $R\lae25$~kpc,
$\Theta_0$ is essentially constant and generally within $1 \sigma$ of the
photometric minor axis of M49 (drawn as the bold, solid lines in
the top panel of Figure~\ref{fig09}). Indeed, it is made clear in this plot that
the rotation axis of the metal-poor GC system by and large falls closer to
the galaxy's minor axis than its major axis. The $\Omega R$ profile is
similarly rather flat, holding roughly steady (again, within the 1-$\sigma$
uncertainties) at the level of 100--150 \kms. It is nonzero almost everywhere
at better than the 68\% confidence level, but less than the 2-$\sigma$ level.
The obvious exception is at very small radii, $R<2\arcmin$. The
strong negative rotation suggested in the innermost radial bin of Table
\ref{tab3}, and illustrated in the top panel of Figure~\ref{fig07}, is seen
here simply not to be significant; it is an artifact of the small sample
size.

It could be of considerable interest that the metal-poor rotation axis
appears to switch suddenly to align with the photometric {\it major} axis of
M49 (indicated by the bold, dashed lines in the top panel of Figure~\ref{fig09}) 
at the largest radii probed by our sample, $R\gae 6\arcmin$. This is clearly 
responsible for a similar, though slightly more muted, trend in the full GC system
(Figure~\ref{fig08}). However, since every point plotted in these graphs
incorporates all GC data from within 1\arcmin\ on either side of it, the
measurements beyond 6\arcmin\ all rely on some velocities from
$R>7\arcmin\simeq30$ kpc, which is the regime that was mentioned as somewhat
suspect on the basis of Figure~\ref{fig05}. A larger dataset is required to
confirm this potential major-axis rotation at large radii in the metal-poor
GC system.
 
Figure \ref{fig10} presents the analysis as applied to the metal-rich GC
system. Since the ``smoothed'' profiles cannot be
calculated beyond $R\gae 6\arcmin$ in this case, a single large data
point is used to represent the sine-fit parameters for all metal-rich GCs
with $350\arcsec\leq R<570\arcsec$ ($18\ {\rm kpc}\lae R\lae 41\ {\rm kpc}$)
together. At small radii $R\lae 4\arcmin$, the axis of GC rotation is
poorly determined but is consistent, within the 95\% confidence bands,
with alignment everywhere along the photometric minor axis (or coincidence
with the formal best fit, $\Theta_0=100^{\circ}$, to the metal-poor
GC rotation). Notably, and despite the vanishing net rotation of the globally
averaged metal-rich GC sample, $\Omega R$ at these small radii differs from
zero at the $\sim$ 1-$\sigma$ level and is of {\it opposite sign} to that found for
the metal-poor GCs. Thus, there is marginal evidence for some counter-rotation
in the metal-rich GC subsystem, relative to the metal-poor one, inside
$R\lae 4\farcm5\sim1.5 R_{\rm eff}$ in M49.

The situation appears in Figure~\ref{fig10} to change suddenly around $R\sim
4\farcm5$, with $\Omega R$ becoming positive, large,
and significant at the $>95\%$ confidence level. The rotation axis
also appears to come sharply into line with the minor axis of the galaxy.
However, the reversion to a negative rotation of much lower significance in
the outermost bin of our sample suggests that this effect may be related to
some curiosity in the data between $4\arcmin\lae R\lae 6\arcmin$.
Indeed, inspection of the metal-rich GC distribution in the third panel of
Figure~\ref{fig07} shows immediately that the azimuthal sampling of metal-rich
GC velocities in this radial range is highly incomplete: all but a few data
points lie within a couple tens of degrees of the photometric major axis on
opposite sides of the galaxy ($\Theta=155^{\circ}$ and $335^{\circ}$). 

The strong ``prograde'' rotation that suddenly
appears in Figure~\ref{fig10} comes entirely from a small group of 10
metal-rich clusters, all contained in the thin annulus $302\arcsec\leq R
\leq 337\arcsec$ ($22.0\ {\rm kpc}\leq R\leq 24.5\ {\rm kpc}$) and located
at opposite poles of the photometric major axis. In ($\langle v_p\rangle$,
$\Theta$) space, these GCs trace almost perfectly (with a dispersion of only
$\sim$100 \kms) a sine curve with peak and trough at the position angles
of the major axis and an amplitude $\Omega R=+300$ \kms. Because all 10
objects are so closely spaced around their median radius of $\langle R\rangle=
319\arcsec$, all 10 are included in every point with
$4\farcm5 \leq R\leq 6\farcm0$ in Figure~\ref{fig10}, and they are thus
entirely responsible for the behavior in those portions of the smoothed
$\Theta_0$ and $\Omega R$ profiles (recall that every point in these profiles
is computed using all GC velocities within $\pm1\arcmin$ of the indicated
position). Similarly, the same 10 objects single-handedly erase a
(weak) net negative-rotation signal from the rest of the metal-rich
GC system.

In the top panel of Figure \ref{fig11}, we show the velocity-position
curve of the
GCs in question. The sine fit indicated has $v_{\rm sys}=997$ \kms\ and
$\Theta_0=65^{\circ}$ fixed, and a fitted amplitude of $\Omega R=
296$ \kms\ (2-$\sigma$ lower limit: 145 \kms). The ID's of the ten objects
from Table \ref{tab2} are: 677, 830, 929, 1508, 6905, 7043, 7894, 8164,
8740, and 8890. Aside from their galactocentric radius, there is no obvious,
common characteristic to distinguish them from any other data, although we
do note that five of the 10 clusters have super-solar metallicities, and 
that the one [\#6905] falling farthest from the fitted sine curve is also 
the most metal-poor.

The bottom panel of Figure~\ref{fig11} is an attempt to delineate the rotation
curve of the metal-rich GC system with the clusters at $302\arcsec\leq R\leq
337\arcsec$ isolated from the rest of the sample. On the basis of the smoothed
profiles at $R<4\arcmin$ and the single-bin datapoints at $R>6\arcmin$ in
Figure~\ref{fig10}, we fix the position angle of the rotation axis to coincide
with the galaxy minor axis for definiteness here ($\Theta_0\equiv65^{\circ}$).
The same smoothing procedure is applied as before, to fit for
$\Omega R$ in equation (\ref{eq5}), but only using GC data interior to
$R\leq300\arcsec$. The result, shown as the small, solid squares
between $2\arcmin\lae R\leq 4\arcmin$ in Figure~\ref{fig11} (but incorporating
data from $1\arcmin\lae R\leq 5\arcmin$), can hardly be called a profile any
longer, as now only the two endpoints are independent. It does, however, give
an impression of the (small) impact that a particular choice of bin boundaries
has on the derived GC rotation amplitude. The open square at $R=5\farcm3=
23.2$ kpc represents the clusters in the top panel of Figure~\ref{fig11}, while
the open square at $R=7\farcm25$ is at the rotation amplitude fit to the GCs
with $350\arcsec\leq R<570\arcsec$ with $v_{\rm sys}=997$ \kms\ and
$\Theta_0=65^{\circ}$ both held fixed.

As we suggested just above, the removal of the small group of apparently
rapidly rotating clusters around $R=22$--24.5 kpc leaves a metal-rich GC
system that, beyond $R\sim2\arcmin\simeq0.6 R_{\rm eff}$ appears to be
rotating --- albeit slowly, and with less than impressive statistical
significance --- roughly around the minor axis in a direction counter to the
metal-poor GC system. Whether this feature is real or not --- and,
indeed, whether the highly contaminating influence of the 10 isolated
GCs is an unlucky byproduct of sampling statistics or an indication of
a real, physical grouping of newly recognized objects --- can only be decided
by the acquisition of further velocity data. Additional observations of these
clusters would certainly be useful, as they may be the relic
of a past merger or accretion event.
In any case, it is clear that globally averaged kinematics are bound to miss
potentially significant and complex radial variations within a GC system.
Any arguments about GC and galaxy formation and evolution that draw only
on globally averaged rotation should be regarded in this light.

We have checked that excluding the aberrant metal-rich GCs does not lead
to significant changes (no more than a few \kms\ for velocities, or degrees
in $\Theta_0$) in the derived kinematics of the full GC velocity sample,
either in Table \ref{tab3} or Figure~\ref{fig08}. It does, of course, alter
the global rotation of the metal-rich GC sample from that quoted in Table
\ref{fig03}, and the revised numbers are given in Table \ref{tab4} below
(\S5). Their exclusion also leaves too few GCs in the bin $250\arcsec\leq R<
350\arcsec$ to derive a reliable rotation solution there, which is why none
is given in Table \ref{tab3}. However, the spatial sampling of points
around an annulus at a given radius should not affect grossly the estimation
of a velocity dispersion there, and thus the metal-rich $\sigma_p(R)$ profile
in the bottom panel of Figure~\ref{fig06} (which excludes no data) is likely
still reasonable enough---although the points between $4\arcmin\lae R\lae
6\arcmin$ should obviously be regarded with due caution.

Finally, we plot in Figure~\ref{fig12} the absolute value of the ratio of
rotation amplitude to velocity dispersion as a function of radius in the full
GC system, the metal-poor subsystem, and the metal-rich sample.  Note that the
latter profile corresponds to the rotation solutions as obtained in the 
bottom panel of Figure~\ref{fig11}, rather than Figure~\ref{fig10}. For the
metal-poor subsystem, a constant $|\Omega R|/\sigma_p \simeq0.3$--0.4 could 
be taken as an adequate description, within
the 1-$\sigma$ uncertainties, of the metal-poor GC data at all radii (except
perhaps around $7\arcmin\simeq 30$ kpc). Thus, as the low
statistical significance of all the rotation amplitudes has already 
suggested, rotation is not an important source of dynamical support in 
the M49 GC system. 

The weighted average for the full GC system, from either of
Tables \ref{tab3} or \ref{tab4}, is $|\Omega R|/\sigma_p \sim$ 0.15--0.2. 
A similar, though perhaps slightly higher, ratio also applies in the metal-poor 
GC system. And, with only the exception of the 10 extreme clusters at 
$R\simeq23.2$ kpc, the metal-rich GC system is consistent everywhere with 
$|\Omega R|/\sigma_p \sim0.1$--0.2.
In the next section, we therefore explore the dynamics of the M49 GC
system through a Jeans-equation analysis that ignores altogether the presence
of rotation. Nowhere should this expose us to uncertainty or error at any more
than a $\sim$10\% level.
 
\section{Dynamical Models}
 
We make the simplifying assumption of spherical symmetry to proceed with a
dynamical analysis based on the Jeans equation in the absence of rotation:
\begin{equation}
{d\over{dr}}\, n_{\rm cl}(r) \sigma_r^2(r) +
{{2\,\beta_{\rm cl}(r)}\over{r}}\, n_{\rm cl}(r) \sigma_r^2(r) =
- n_{\rm cl}(r)\,{{G M_{\rm tot}(r)}\over{r^2}}\ .
\label{eq6}
\end{equation}
Here $n_{\rm cl}(r)$ is the three dimensional density profile of the GC
system; $\sigma_r(r)$ is its intrinsic velocity dispersion in the radial
direction; $\beta_{\rm cl}(r)\equiv 1-\sigma_\theta^2(r)/\sigma_r^2(r)$
is a measure of its velocity anisotropy; and $M_{\rm tot}(r)$ is the
total gravitating mass enclosed within a sphere of radius $r$.

Our analysis of the M49 GC system parallels that undertaken for
M87 by C\^ot\'e \etal (2001), and it differs from the approach normally
taken in dynamical studies of extragalactic GC systems ($e.g.,$ Huchra \& 
Brodie 1987; Mould \etal 1990; Cohen \& Ryzhov 1997; Sharples \etal 1998;
Kissler-Patig \& Gebhardt 1998; Zepf \etal 2000), in which the measured
line-of-sight velocity dispersions are used to infer the deprojected
profile $\sigma_r(r)$ so as to solve equation (\ref{eq6}), under the
assumption of orbital isotropy [$\beta_{\rm cl}(r)\equiv 0$], for the
gravitating mass distribution $M_{\rm tot}(r)$. We opt instead
to determine a mass model for M49 and the Virgo B subcluster
{\it a priori} (and independently of any GC data) and then to use this model,
along with three-dimensional GC density profiles that fit published
number counts, to solve equation (\ref{eq6}) for the velocity
dispersion profile $\sigma_r(r)$ under a variety of assumptions on the
velocity anisotropy $\beta_{\rm cl}(r)$. These model profiles are then
numerically projected for comparison with the observed $\sigma_p(R)$ profiles
obtained in \S3.2, in order to examine more directly what range of
$\beta_{\rm cl}(r)$ is compatible with the data. (The observed dispersion
profiles are still too noisy to be deprojected and used to solve directly
for $\beta_{\rm cl}(r)$ with any confidence.) Romanowsky \& Kochanek
(2001) present an analysis of the M87 GC system that employs different
methodology for the same basic goal.

For later reference, the relevant projection integrals are
\begin{equation}
\sigma_p^2(R) =
{2\over{N_{\rm cl}(R)}}\,
\int_R^{\infty} n_{\rm cl} \sigma_r^2(r)
\left(1 - \beta_{\rm cl}\,{{R^2}\over{r^2}} \right)\,
{{r\,dr}\over{\sqrt{r^2 - R^2}}},
\label{eq7}
\end{equation}
where the surface density profile $N_{\rm cl}(R)$ is related to 
the three-dimensional $n_{\rm cl}(r)$ by
\begin{equation}
N_{\rm cl}(R) = 2\int_{R}^{\infty} n_{\rm cl}(r)
{{r\,dr}\over{\sqrt{r^2-R^2}}}\ .
\label{eq8}
\end{equation}

Throughout this Section, we discuss all distances, densities, masses, and
luminosities in physical units that assume a distance of 15 Mpc to M49.
 
\subsection{Density Profiles for the GC System}

Figure \ref{fig13} shows the projected number-density profiles, $N_{\rm cl}(R)$, for
GCs of all colors (metallicities), taken from the comprehensive
and independent studies of McLaughlin (1999a) and Rhode \& Zepf (2001).
These authors have already corrected the GC densities to remove contamination
by foreground stars and background galaxies, and to account for GCs not
directly counted because they are fainter than the limiting magnitudes of
their surveys (this latter correction assumes the GC luminosity function, or
number of clusters per unit magnitude, to have a Gaussian shape with a peak
at $V = 23.75$ mag and a dispersion $\sigma=1.3$ mag). Thus, the data points
shown here genuinely reflect the total number of GCs per unit area on the
sky as a function of projected radius in M49.

For use in equation (\ref{eq6}), we treat both sets of data together as a
single number-density profile, to which we fit the projections of two
simple three-dimensional density functions.
Shown as the bold, solid curve in Figure~\ref{fig13} is the best-fit
projection of the density profile suggested by Navarro, Frenk, \& White (1997)
for dark-matter halos, {\it i.e.}, $n_{\rm cl}=n_0(r/b)^{-1}(1+r/b)^{-2}$:
\begin{equation}
n_{\rm cl}(r)=0.19\,{\rm kpc}^{-3}\ (r/11.7\,{\rm kpc})^{-1}
(1+r/11.7\,{\rm kpc})^{-2} \ ,
\label{eq9}
\end{equation}
with (correlated) uncertainties of $\pm0.07\,{\rm kpc}^{-3}$ in the
normalization and $\pm1.5$ kpc in the scale radius

The bold, dashed line in Figure~\ref{fig13} shows the
best-fit projection of one of the simple family of galaxy models developed by
Dehnen (1993): $n_{\rm cl}(r)=n_0{(r/b)^{-\gamma}}(1+r/b)^{\gamma-4}$. We assume 
$\gamma = 0$ and find
\begin{equation}
n_{\rm cl}(r)=0.40\,{\rm kpc}^{-3}\ (1+r/16.7\,{\rm kpc})^{-4}\ ,
\label{eq10}
\end{equation}
with correlated uncertainties of $\pm0.16\,{\rm kpc}^{-3}$ in the
normalization and $\pm 2.3$ kpc in the scale radius.

The reduced $\chi^2$ values of these two fits are comparable. We prefer the
Navarro \etal (NFW) form of equation (\ref{eq9}), because the total
baryonic mass density (stars and X-ray gas) at large radii in M49
appears to fall off roughly as $r^{-3}$ (McLaughlin \& C\^ot\'e 2003), and we
expect the GC system to follow this behavior (McLaughlin 1999a). However, the
particular form of the density profile in equation (\ref{eq10}) was chosen
specifically for its contrasting asymptotic behavior at both small and large
galactocentric radii; with it, we can check directly the extent to which the
functional form of $n_{\rm cl}(r)$ used in the Jeans equation
(\ref{eq6}) might influence our conclusions on the GC dynamics.

We wish also to model the dynamics of the metal-poor and metal-rich GC
subsystems separately. To define individual density profiles for them, we
make use of the catalog of Lee \etal (1998), which includes positions and
Washington $(C-T_1)$ colors for $\sim$ 2000 GC candidates around M49.
Although Lee \etal (1998) themselves presented separate density profiles for
the metal-poor and metal-rich GCs, the criteria they used to divide their
sample by color differ slightly from those adopted here, in \S2.3. We
therefore use their catalog directly to re-derive the surface density
profiles of those objects with $1\leq(C-T_1)<1.625$ and $1.625\leq(C-T_1)<
2.25$. We count only the points that are also brighter than $T_1 \le 23$,
to which magnitude the Lee \etal catalog is better than 97\% complete.
Because the catalog includes GCs only out to $R \simeq 8\arcmin$ from
the center of M49, whereas the full GC system is known to extend to
$R \gae 20\arcmin$ (Harris \& Petrie 1978; Rhode \& Zepf 2001), we are unable to
measure directly the level of its contamination by stars and background
galaxies. The exact choice of background is, however, constrained by the
fact that the corrected number density profile for all colors
$1\leq(C-T_1)<2.25$ should be consistent with the shape of 
$N_{\rm cl}(R)$ in Figure~\ref{fig13}.
The total background
density, to the limiting magnitude of $T_1=23$, estimated in this way is
$N_b\simeq2.0\pm0.2$ arcmin$^{-2}$, or $0.11\pm0.01$ kpc$^{-2}$. We then
{\it assume} that the contaminating backgrounds to be subtracted from the
separate metal-poor and metal-rich GC density profiles are 60\% and 40\% of
the total $N_b$. This 3:2 ratio is somewhat arbitrary, but it is consistent
with that predicted by the IAS Galaxy model (Bahcall \& Soneira 1980; Bahcall
1986) for foreground star counts to $T_1=23$, in our color ranges, in the
direction towards M49.

Having thus obtained corrected, metal-poor and metal-rich GC surface
density profiles, we fit each with projections of the same two three-dimensional
functions that we fit to the full GC system. The results are shown
in Figure~\ref{fig14}. The top panel presents fits of the NFW function,
with an adopted background of $N_b = 2.2$ arcmin$^{-2}$.
For the metal-poor and metal-rich GC samples in turn,
\begin{eqnarray}
n_{\rm cl}^{\rm MP}(r) & \propto &
  (r/20.7\,{\rm kpc})^{-1}(1+r/20.7\,{\rm kpc})^{-2}    \nonumber \\
n_{\rm cl}^{\rm MR}(r) & \propto &
  (r/5.54\,{\rm kpc})^{-1}(1+r/5.54\,{\rm kpc})^{-2}\ . \label{eq11}
\end{eqnarray}
The dashed curves running through the open datapoints in the top panel of
Figure~\ref{fig14} are the projections of these functions. The bolder dashed
curve is their sum; it agrees reasonably well both with the observed
density profile of metal-poor and metal-rich GCs {\it combined}
in the catalog of Lee \etal (1998) (filled circles), and with the NFW fit
(eq.~[\ref{eq9}]; the bold, solid line) to the combined data of McLaughlin
(1999a) and Rhode \& Zepf (2001) in Figure~\ref{fig13}.

The bottom panel of Figure~\ref{fig14} shows our projected fits of the
Dehnen (1993) density model to the metal-poor and metal-rich GC systems,
\begin{eqnarray}
n_{\rm cl}^{\rm MP}(r) & \propto &
  (1+r/20.1\,{\rm kpc})^{-4}        \nonumber\\
n_{\rm cl}^{\rm MR}(r) & \propto &
  (1+r/9.51\,{\rm kpc})^{-4}\ ,     \label{eq12}
\end{eqnarray}
where the adopted background in this case is $N_b = 1.9$ arcmin$^{-2}$. 
Again, the sum of the two fits is shown as the bold, dashed curve. 
It is fairly consistent with the observed total GC surface densities and
with the Dehnen-model fit of equation (\ref{eq10}).

Unlike the case for the full GC system in Figure~\ref{fig13}, we do not
consider the density profiles for the GC metallicity subsample to be
particularly well constrained. The limited spatial coverage of the
Lee \etal catalog leaves much room for erroneous extrapolations to larger
radii, and the lack of direct background density estimates makes the
$N_{\rm cl}(R)$ estimates for the metal-poor and metal-rich subsystems
too uncertain for our taste even at directly
observed radii. It is nevertheless necessary to
specify separate density profiles for the subsamples if their dynamics are
to be modeled individually. Note that estimates of the absolute
normalizations of $n_{\rm cl}^{\rm MP}(r)$ and $n_{\rm cl}^{\rm MR}(r)$
are not required in the following analysis, since these cancel out of 
equation (\ref{eq6}). 

\subsection{The Need for an Extended Dark-Matter Halo in M49/Virgo B}

Before going on to constrain the orbital parameters of the GCs in M49, we
make a brief aside regarding the evidence for dark matter in M49 and the
Virgo B subcluster around it, about which various claims have been made over
the years. Mould \etal (1990) argued on the basis
of radial velocities for 26 GCs that the galaxy does indeed contain a dark
halo, but they were unable to rule out the possibility of a constant
mass-to-light ratio due to their small velocity sample and the unknown
orbital properties of the GCs. Working with a sample of 144 GC velocities,
and assuming that GCs have isotropic orbits, Zepf \etal (2000) argued that
the mass-to-light ratio at $\sim$ 30 kpc (2--2.5 $R_{\rm eff}$) is at least five
times that at a distance of a few kpc, implying the presence of substantial
dark matter on large spatial scales.

Irwin \& Sarazin (1996) used ROSAT X-ray observations of the hot, gaseous
corona around M49 to derive a mass profile for M49 extending to
$R \sim 100\ {\rm kpc}\simeq 8 R_{\rm eff}$, leading them to conclude
that dark matter must be present on spatial scales of $R>R_{\rm eff}$.
Inside $R\sim 10$--15 kpc ($\sim R_{\rm eff}$), however, Irwin \& Sarazin
found mass-to-light ratios that are perfectly compatible with the canonical stellar
value in elliptical galaxies. Indeed, Brighenti \& Mathews (1997) later
decomposed Irwin \& Sarazin's (1996) mass distribution into separate stellar
and dark-matter components (as we also do, in \S4.3), and argued
that any dark-matter halo must contribute negligibly to the total mass inside
$R\lae 7.5\ {\rm kpc}\simeq0.5 R_{\rm eff}$.

Saglia \etal (1993) observed a flattening of the stellar velocity dispersion
profile in M49 at a much smaller projected radius, $R\simeq2$--$3\ {\rm kpc}
\sim 0.2 R_{\rm eff}$, and argued that this alone implied the presence of an
dynamically dominant, extended dark-matter halo. However, the large scatter
in independent measurements of the stellar kinematics beyond $R\gae1.5$ kpc
(see, {\it e.g.}, Caon, Macchetto \& Pastoriza 2000) rather weakens the conclusion;
and it is certainly possible to fit self-consistent galaxy models,
containing no dark matter, to the stellar kinematics if no other data
are considered ({\it e.g.}, Kronawitter \etal 2000).

Our improved GC velocity dispersion profile from \S3 can be used to look at
this issue again. We consider the hypothesis that light directly
traces mass in M49, so that $M_{\rm tot}(r)$ in equation (\ref{eq6}) comes
from integrating the stellar luminosity density profile and multiplying by a
constant, stellar mass-to-light ratio. We fit the surface brightness profile
of the galaxy taken from Kim \etal (2000) --- converted from their Washington
$T_1$ photometry to Cousins $R$-band according to the relation of Geisler
(1996) --- with the projection of the three-dimensional luminosity density
profile,
\begin{equation}
j(r) = {{(3-\gamma)(7-2\gamma)}\over{4}}\,
{L_{\rm tot}\over{\pi a^3}}\,
\left({r\over{a}}\right)^{-\gamma}\,
\left[1+\Bigl({r\over{a}}\Bigr)^{1/2} \right]^{2(\gamma-4)}\ .
\label{eq13}
\end{equation}
This slight modification of the models of Dehnen (1993) 
allows for a more gradual transition from the inner power-law behavior,
$j\sim r^{-\gamma}$, to the asymptotic $j\sim r^{-4}$ at large radii.

The projected best fit of equation (\ref{eq13}) to the surface-brightness
data of Kim \etal (1998) is shown in the
left panel of Figure~\ref{fig15}. For comparison, we also plot the
surface brightness profile measured by Caon, Capaccioli, \& D'Onofrio (1994),
converted from $B$-band to Cousins $R$. The fit has parameters
$\gamma = 0.7$, $a = 2.82$ kpc, and $L_{\rm tot} = 1.40\times10^{11}
L_{R,{\odot}}$. It is from this model that we derive a (projected) effective
radius of $R_{\rm eff}=3\farcm1\simeq13.5$ kpc for M49. Our measurement
of the galaxy's effective radius is intermediate to the values of
$R_{\rm eff}=3\farcm8$ found by Caon \etal (1994) and that of 
Kim \etal (2000), who obtained $R_{\rm eff}=2\farcm0$ by fitting
a de Vaucouleurs law to the same data analyzed here.
While the precise value of $R_{\rm eff}$ itself
has no bearing on our conclusions, we note that the surface brightness
profile of M49 is not well described by a de Vaucouleurs law, which
is why Kim \etal (2000) restricted their fit to the region 
$7^{\prime\prime} < R < 260^{\prime\prime}$. As their Figure~8 shows,
the fitted model falls well below the measured brightness profile
for $R \sim 260^{\prime\prime}$, leading to likely underestimates 
of the galaxy's total luminosity and effective radius.

The right panel of Figure~\ref{fig15} illustrates the three-dimensional
stellar mass density profile, $\rho_{\rm s}(r)=\Upsilon_0 j(r)$, with
$\Upsilon_0$ the $R$-band mass-to-light ratio. We take its value from
McLaughlin \&
C\^ot\'e (2003), who show that the stellar kinematics at projected radii
$0.1\leq R\leq 1$ kpc in M49 are best fit by a mass model with no dark matter
if $\Upsilon_0=5.9\ M_\odot\,L_{R,\odot}^{-1}$ {\it and} the velocity
ellipsoid of the stellar distribution is slightly radially biased:
$\beta_{\rm s}=0.3$. We note that this mass-to-light ratio corresponds
to a $B$-band value of $8.1\ M_\odot\,L_{B,\odot}^{-1}$, typical of the
cores of elliptical galaxies, and that this constant-$M/L$ mass model is
very similar to the self-consistent galaxy model of Kronawitter \etal (2000).

We are examining here the hypothesis that light traces mass, and thus the
total mass profile to be used in the Jeans equation is just the stellar mass
\begin{equation}
M_{\rm s}(r)  = 
    \Upsilon_0\int_{0}^{r}{4{\pi}x^2\,j(x)}dx
  = {\Upsilon_{0}}L_{\rm tot}\left[{(r/a)^{1/2}
    \over 1+(r/a)^{1/2}}\right]^{2(3-\gamma)}\left[{(7-2\gamma)+(r/a)^{1/2}
    \over 1+(r/a)^{1/2}}\right] \ ,
\label{eq14}
\end{equation}
where, again, $\Upsilon_0$ is independent of radius in the galaxy and has a
value of $5.9\ M_\odot\,L_{R,\odot}^{-1}$ when no additional dark-matter
component is invoked.
Taking $M_{\rm tot}(r)=M_{\rm s}(r)$, then, we solve equation (\ref{eq6}) for
the intrinsic radial velocity dispersion profile of the stars in M49 by
replacing $n_{\rm cl}(r)$ with $\rho_{\rm s}(r)\propto j(r)$ from equation
(\ref{eq13}), and by substituting $\beta_{\rm s}(r)\equiv+0.3$ (McLaughlin
\& C\^ot\'e 2003) for $\beta_{\rm cl}(r)$. The projection integrals of
equations (\ref{eq7}) and (\ref{eq8}) are then applied --- again with $j(r)$
in place of $n_{\rm cl}(r)$ --- to predict the stellar line-of-sight
velocity dispersion in the absence of an extended dark-matter halo around
M49.

This model is shown as a bold, dotted line in Figure~\ref{fig16}, where it is
compared with the stellar kinematic data of Davies \& Birkinshaw (1988),
Saglia \etal (1993), and Caon \etal (2000). It does a respectable job of
reproducing the observed velocity dispersions between $0.1\lae R\lae8$
kpc.\footnotemark\footnotetext{The failure of the model stellar
$\sigma_p(R)$ to describe the data at $R \lae 0.1$ kpc reflects a poor fit of
equation (\ref{eq13}) to the observed surface brightness profile at
$R \lae 2\arcsec$ in M49; a likely departure of the stellar anisotropy
from the average radial bias $\beta_{\rm s}(r)=0.3$; and the influence of
the nuclear black hole ($M_{\bullet}\simeq5\times10^8 M_\odot$; Merritt \&
Ferrarese 2001) at radii $R \lae 30$ pc. No attempt was made to
model these effects in detail, as the complexity in the inner 100 pc of
the galaxy has no bearing on the larger-scale dynamics of interest to us
here.}
Although the success relies in part on the considerable scatter among the
different datasets at $R\gae1$ kpc, the main point is that the stellar
kinematics in M49, by themselves, provide no conclusive 
evidence for the existence of any extended and massive dark-matter halo.

The bold, solid line in Figure~\ref{fig16} traces the line-of-sight
velocity dispersion profile $\sigma_p(R)$ derived from equations
(\ref{eq6})--(\ref{eq8}) for the full GC system, when the constant-$M/L$
mass model of equation (\ref{eq14}) is adopted, the GC density profile
is modeled by the NFW form for $n_{\rm cl}(r)$ (eq.~[\ref{eq9}]), and orbital
isotropy [$\beta_{\rm cl}(r)\equiv 0$] is assumed. It fails by a wide margin
to describe the observed dispersion profile of the GCs, which is taken from
the top panel of Figure~\ref{fig06}. The bold, short-dashed line is the expected
$\sigma_p(R)$ profile under the same main assumptions but taking the GC
density profile to be the Dehnen-model fit of equation (\ref{eq10}). The
lighter, long-dashed lines revert to the NFW GC density profile but assume
strongly radially and tangentially biased GC velocity ellipsoids (upper and
lower curves). None of these alternate models fares any better in accounting
for the observed GC velocity dispersion at
$R\gae6\ {\rm kpc}\sim 0.4$--$0.5 R_{\rm eff}$. Evidently, the mass-to-light
ratio must increase with galactocentric radius if the high velocity
dispersion of the GCs is to be understood.

Regardless of any assumptions on its velocity anisotropy or uncertainties in
its spatial distribution, the globular cluster system alone therefore
provides incontrovertible evidence, independent of any X-ray data, for the
presence of an extended dark-matter halo around M49. 
It is worth emphasizing that the GC data {\it alone} are
unambiguous on this point for spatial scales $R \lae R_{\rm eff}$.
 
\subsection{Total Mass Profile of M49/Virgo B}

Having established the presence of a dark-matter halo around M49,
we briefly describe a model for it that is developed and discussed in
detail by McLaughlin \& C\^ot\'e (2003). The model is constructed to match
the available stellar kinematics and X-ray data in M49/Virgo B. It makes
no reference to any aspect of the GC system, and it can thus be used as one
of two components contributing to $M_{\rm tot}(r)$ in the Jeans equation
(the other being the luminous component of \S4.2) as a
constraint on the velocity anisotropy of the GC system.
 
McLaughlin \& C\^ot\'e (2003) consider a variety of possible analytic
descriptions for the dark-matter density profile in M49/Virgo B. Among
these, the ``universal'' halo profile of Navarro \etal (1997),
$\rho_{\rm dm} = K\,(r/r_s)^{-1}(1+r/r_s)^{-2}$
(cf.~eqs.~[\ref{eq9}] and [\ref{eq11}]), is one that provides a good
match to all available data. We therefore adopt this form here, to specify
\begin{equation}
M_{\rm dm}(r) =
     \int_{0}^{r}{4{\pi}x^2\,\rho_{\rm dm}(x)}dx
   = 4{\pi}Kr_s^3
     \left[\ln{\left(1+{r \over r_s}\right)} - {r/r_s \over 1+r/r_s}\right]\ .
\label{eq15}
\end{equation}
The mass $M_{\rm tot}(r)$ interior to any (three-dimensional) radius
$r$ is then the sum of $M_{\rm dm}(r)$ and $M_{\rm s}(r)$ from equation
(\ref{eq14}). The unknown parameters in the total mass profile are the
dark-matter density normalization $K$ and scale length $r_s$, and the
stellar mass-to-light ratio $\Upsilon_0$. McLaughlin \& C\^ot\'e fix these
by first constraining the total $R$-band mass-to-light ratio at a radius
$r=1$ kpc: $M_{\rm tot}(r\leq 1\,{\rm kpc})/L(r\leq 1\,{\rm kpc}) =
5.9 M_\odot\,L_{R,\odot}^{-1}$, thereby guaranteeing that the final model will
adequately reproduce the observed stellar kinematics in the core of M49
(cf.~\S4.2). This leaves only two independent parameters in the mass
model, and these are fit by comparing to the X-ray mass measurements of
Irwin \& Sarazin (1996) and
Schindler \etal (1999).\footnotemark\footnotetext{The total mass in X-ray
gas within a sphere of radius $r$ = 100 kpc centered on M49 is only 
$\sim$ 1/15 the mass in stars and $\sim$ 1/200 that in dark matter, so
the gas mass itself is negligible. As it turns out, though,
the X-ray gas traces the dark-matter distribution
throughout M49/Virgo B (McLaughlin \& C\^ot\'e 2003), so its contribution
to the overall mass budget is actually included in the normalization of
$M_{\rm dm}(r)$.}

The best-fit mass model with an NFW halo is drawn as a bold,
solid line in Figure~\ref{fig17}. The lighter, dashed line is the mass profile
of the dark-matter halo alone, given by equation (\ref{eq15}) with
$K=1.12\times10^{-3} M_\odot\,{\rm pc}$ and $r_s = 196$ kpc. The light,
dotted line is the stellar mass distribution of equation (\ref{eq14}) with
a fitted $R$-band stellar mass-to-light ratio
$\Upsilon_0=5.7 M_\odot\,L_{R,\odot}^{-1}$ (and the galaxy parameters
$\gamma$, $a$, and $L_{\rm tot}$ already fixed by the fit to the surface
brightness profile in Figure~\ref{fig15}).
Explicitly, then, our final model for M49/Virgo B is:
\begin{eqnarray}
M_{\rm tot}(r) & = & M_{\rm s}(r) + M_{\rm dm}(r)    \nonumber \\
  M_{\rm s}(r) & = & 7.98\times10^{11}\,M_\odot
                      \left[{(r/2.82\,{\rm kpc})^{1/2}
                             \over 1+(r/2.82\,{\rm kpc})^{1/2}}\right]^{4.6}
                      \left[{5.6+(r/2.82\,{\rm kpc})^{1/2}
                             \over 1+(r/2.82\,{\rm kpc})^{1/2}}\right]
                      \label{eq16}   \\
 M_{\rm dm}(r) & = & 1.06\times10^{14}\,M_\odot
                       \left[\ln(1+r/196\,{\rm kpc}) -
                             {(r/196\,{\rm kpc}) \over (1+r/196\,{\rm kpc})}
                        \right] \ \ .   \nonumber
\end{eqnarray}

Note that the X-ray masses interior to $r\lae 10$--15 kpc in M49 are fully
consistent with the mass of the stellar
component alone there --- although they clearly do not require the complete
absence of any dark matter at the center of the galaxy, this component does
not begin to dominate the total mass budget until $r\gae 20\ {\rm kpc}\simeq
1.5 R_{\rm eff}$. Again, it is the GC velocity data that show the clearest
evidence on smaller spatial scales for the presence of a massive dark halo.

\subsection{Velocity Isotropy in the GC System}
 
Figure \ref{fig18} plots the stellar and GC velocity dispersion data as
Figure~\ref{fig16} did, but now with model curves computed using
$M_{\rm tot}(r)$
given by equation (\ref{eq16}) rather than $M_{\rm s}(r)$ alone. The mass
model was constructed by requiring, in part, that it agree with the stellar
kinematics, and the bold, dotted curve in Figure~\ref{fig18} verifies that
this is the case. It is worth noting that our model --- which now includes the
dark-matter halo that the GC data show must be present --- favors the stellar
data of Davies \& Birkinshaw (1988) and Saglia \etal (1993) over those
of Caon \etal (2000) (the small, open squares that scatter
well below the dotted curve in the Figure). The bold, solid line in this plot
is the velocity dispersion profile predicted for the full GC system if its
three-dimensional density profile $n_{\rm cl}(r)$ in equations
(\ref{eq6})--(\ref{eq8}) is given by the NFW functional fit of equation
(\ref{eq9}), and if velocity isotropy, $\beta_{\rm cl}(r)\equiv0$, is
assumed. The excellent agreement between this model curve and the GC data
over the full range $6\lae R\lae 30$ kpc implies that {\it the assumption of
isotropy is, in fact, essentially correct}. 

The bold, dashed line in Figure~\ref{fig18} is the model $\sigma_p(R)$
for the GC system if its density profile is taken from the alternate,
Dehnen-model fit of equation (\ref{eq10}). Isotropy is still assumed. This
model also compares favorably with the data, falling within the 68\%
confidence interval around the $\sigma_p$ estimate at almost every point. We
note again that we prefer on other grounds the shallower extrapolation of
the NFW density profile for the GC system, over the steeper fall-off of the
Dehnen profile (see \S4.1). However, the present comparison of the models
shows that our inference of velocity isotropy in the M49 GC system does
not rely on this particular choice.

Figure \ref{fig19} gives a close-up view of the comparison between
the velocity dispersion data and a number of models for the full GC system. 
The top panel shows the same data as Figures~\ref{fig18}, \ref{fig16}, and
\ref{fig06},
along with the usual 68\% and 95\% confidence bands. The bottom panel shows
the projected {\it aperture} dispersion profile, {\it i.e.}, the velocity
dispersion of all objects interior to a given projected radius $R$, rather
than just those in a narrower annulus. For a model, this cumulative
spatial average of $\sigma_p(R)$ is defined by
\begin{equation}
\sigma_{\rm ap}^2(\le R) =
  \left[ \int_{R_{\rm min}}^R N_{\rm cl}(R^\prime) \sigma_p^2(R^\prime)\,
         R^\prime\,dR^\prime \right]\,
  \left[\int_{R_{\rm min}}^R N_{\rm cl}(R^\prime)\,R^\prime\,dR^\prime
         \right]^{-1}\ ,
\label{eq17}
\end{equation}
with $R_{\rm min}$ the projected galactocentric radius of the innermost
datapoint in a velocity sample ($R_{\rm min}=1.9$ kpc for us). Although the
interpretation of the coarsely averaged $\sigma_{\rm ap}$ profile can be
subtler than that of the differential profile $\sigma_p(R)$, it has the
significant advantage of reduced noise at large $R$.

The bold, solid line in the top panel of Figure~\ref{fig19} is the same as that
in Figure~\ref{fig18}: the predicted $\sigma_p(R)$ profile for the full GC
system in the mass distribution of equation (\ref{eq16}), if $n_{\rm cl}(r)$
is given by the NFW profile of equation (\ref{eq9}) and
$\beta_{\rm cl}(r)\equiv0$. The same curve in the bottom panel is just
the result of averaging this model profile with equation (\ref{eq17}).
The bold, short-dashed curves in the two panels of the Figure
are analogous to the solid lines but assume the Dehnen-model density profile
of equation (\ref{eq10}) for the full GC system. The lighter, long-dashed
curves assume the NFW profile for $n_{\rm cl}(r)$ and show predictions for
velocity anisotropies ranging from extreme radial bias
to extreme tangential bias: from top to bottom, $\beta_{\rm cl}=
1-\sigma_\theta^2/\sigma_r^2=0.99$, 0.5, $-$1, and $-$99. These anisotropies
are included in the models as spatially constant, so that the curves
corresponding to them should be viewed as representative of more realistic,
spatially varying functions $\beta_{\rm cl}(r)$ which average out to
the constants that we have specified.

The close adherence of the isotropic models to the GC data is 
striking, particularly in the case of the aperture dispersion profile. The
spatial averaging involved there effectively blurs local, apparently
noise-driven features in the differential $\sigma_p(R)$ data ({\it e.g.},
the bump around $R\sim22$ kpc), such that the
$\sigma_{\rm ap}$ data never stray substantially from the predictions of
isotropy for either assumed density profile $n_{\rm cl}(r)$. The possibility
of even modestly radial or tangential average anisotropy ({\it i.e.},
$\beta_{\rm cl}=0.5$ or $-$1) can be ruled out at the 95\% confidence level
or better in the full GC system.
 
Aside from its intrinsic importance for models of GC formation and evolution, 
the finding of such a high degree of consistency with velocity isotropy in 
the full GC system could be of practical use for measurements of the 
gravitating mass profiles in other early-type galaxies, where data may not be 
available in the variety and quality required for the type of modeling that 
we described in \S4.3.  M49 is now the second galaxy (the other being M87) 
in which we have been able to show by direct means that an isotropic velocity 
ellipsoid is a highly accurate description of the GC kinematics. The 
{\it assumption} of isotropy in other GC systems may therefore be much
better justified than before, making it possible to use the Jeans equation
(\ref{eq6}) with $\beta_{\rm cl}(r)\equiv0$ --- provided that $n_{\rm cl}(r)$
is also well known --- to derive $M_{\rm tot}(r)$ from a set of
velocity-dispersion measurements.

Finally, we construct dynamical models for the metal-poor and metal-rich GC
subsystems separately. This is a somewhat riskier proposition, given the
noisier dispersion profiles of these smaller datasets and the much poorer
characterization of the density profiles $n_{\rm cl}^{\rm MP}(r)$ and
$n_{\rm cl}^{\rm MR}(r)$ (Figure~\ref{fig14} in \S4.1). With these caveats in
mind, we compare the observed and
predicted velocity dispersion profiles for the metal-poor and metal-rich
GCs in Figures \ref{fig20} and \ref{fig21}. The various
curves in these figures have
meanings analogous to those in Figure~\ref{fig19}. The bold, solid curves assume
orbital isotropy and adopt the NFW fits for the density profiles of the
subsystems (eq.~[\ref{eq11}]), while the bold, short-dashed curves assume isotropy
and the Dehnen-model density profiles of equation (\ref{eq12}). The light,
long-dashed curves assume NFW density profiles and spatially constant
anisotropies $\beta_{\rm cl}=0.99$, 0.5, $-$1, and $-$99.

The isotropic NFW density-model curve in Figure~\ref{fig20} provides quite an
acceptable description of the observed $\sigma_p(R)$ of the metal-poor
globulars until radii $R\gae20$--25 kpc, where the subsequent rise in the
data might suggest a move to a slight radial bias in the GC orbits. However,
the model remains within the 68\% confidence bands on the measurements nearly
until $R\sim30$ kpc, the limit of reliability. The observed
$\sigma_{\rm ap}$ data in the bottom panel adhere more closely to the
isotropic curve at still larger radii. The model assuming a Dehnen-type
density profile for the metal-poor GC system gives a stronger suggestion of
radial velocity anisotropy,\footnotemark\footnotetext{Given a fixed, observed velocity dispersion
for a tracer population in a fixed gravitational potential, a steeper density
profile always implies a more radial anisotropy through the Jeans equation.}
although the isotropic case is still barely consistent (within the 95\% confidence
bands) with the observed profile. Because of the increased noise in the measured dispersions
and, more importantly, the poorly constrained density profile,
it is difficult to draw any firm conclusions on the orbital properties
of the metal-poor GCs, beyond the obvious: (1) the hypothesis of isotropy is not
far wrong, particularly at small galactocentric distances $R\lae20\ {\rm kpc}
\simeq 1.5 R_{\rm eff}$; and (2) any possible radial bias present at larger
projected radii is likely to be modest.

An assessment of the orbital properties of the metal-rich GCs is necessarily more 
difficult.
Not only is the extrapolation of their observed density profile very
uncertain, but the difficulties posed by the 10 rapidly rotating objects
concentrated around $R\simeq23.2$ kpc, as discussed in \S3.2, leave
little in the way of data that can be compared reliably with any model. The
top panel shows our smoothed velocity-dispersion profile from
Figure~\ref{fig06}, with those parts relying on any GC velocities from within
$302\arcsec\leq R\leq 337\arcsec$ excised. The most that can be said is that,
at $R\sim R_{\rm eff}=13.5$ kpc, the data are once again
consistent with velocity isotropy, regardless of which GC density
profile $n_{\rm cl}^{\rm MR}(r)$ is assumed. Beyond this, however, the steep
drop in $\sigma_p(R)$ appears unsupportable in {\it any} dynamical model
that we have constructed. It is true that the outer-radius data point
suggests some return to understandable behavior in the metal-rich GC system
($i.e.$, 1-$\sigma$ consistency with isotropy for
either $n_{\rm cl}^{\rm MR}(r)$ profile), but the trend of the cumulative
$\sigma_{\rm ap}(\leq R)$ profile (towards model curves suggesting a
strong tangential bias in the orbital distribution) also suggests that the
low velocity dispersion may be problematic.

If the full GC system is perfectly isotropic, and the metal-poor
component is fairly closely so, then it stands to reason that the metal-rich GC system 
should be close to isotropic as well. Thus, its behavior beyond 
$\sim 1 R_{\rm eff}$ needs to be much better
understood empirically before any dynamical sense can be made of it. We noted
in \S3 (see, {\it e.g.}, Figure~\ref{fig05}) that the low velocity dispersion
of the metal-rich GCs is the manifestation of an absence of
red objects with $\langle v_p\rangle \gae 1500$ \kms\ from our velocity survey,
and a corresponding lack of metal-rich GCs with
$\langle v_p\rangle \lae 700$ \kms\ outside about one effective radius in the
galaxy. It would be useful to know whether these are artifacts of our
particular sample or indicative of something more profound. For instance, 
a velocity dispersion $\sigma_p \lae 250$ \kms\ at galactocentric radii
$R>R_{\rm eff}$ is even lower than we would expect for the {\it stellar light} of M49
(cf.~Figure~\ref{fig18}). Very closely related to this issue is the density
profile of the metal-rich GC system, which will have to be defined reliably
out to much larger radii than we have been able to do with existing databases.
Even the Dehnen-model density profile that we explored here, with its
behavior $n_{\rm cl}\rightarrow r^{-4}$ at large $r$, predicts GC
kinematics that seem considerably too ``hot'' compared with the metal-rich data. 
Either current radial velocity surveys have missed a number of high-velocity, 
metal-rich GCs at large galactocentric radii --- perhaps due to some observational bias
or simple bad luck --- or the density of metal-rich GCs  
system falls very rapidly indeed to zero beyond several tens of kpc.
Deep, wide-field imaging in a metallicity-sensitive filter combination can
directly address the latter possibility, and it will be essential to
developing any deeper understanding of GC dynamics in M49.

\section{Comparing M49, M87, M31 and the Milky Way}
 
A direct comparison between the kinematic and dynamic properties of GCs
in M49 and M87 seems in order, given the similarity in the respective datasets
and the methods of analysis.  In fact, the GC radial velocities
now amassed for these galaxies constitute the two largest radial
velocity samples yet assembled for pure Population II tracers in
early-type galaxies.
 
It is worth bearing in mind that that the 
GC systems of these two galaxies are similar in many respects.
For instance, both have bimodal metallicity distributions, with peaks at
[Fe/H] $\sim -1.5$ and $-$0.3 dex. The ratio of the number of metal poor 
to metal rich GCs is very nearly the same in both galaxies. The GC
formation efficiencies for the composite GC systems and the two subsystems
are quite similar in both cases ($e.g.$, McLaughlin 1999a; C\^ot\'e 2003). Likewise, 
the measured ages of the metal-poor and metal-rich GCs suggest that they are --- on the 
whole --- old and {\it roughly} coeval (Cohen \etal 1998; Puzia \etal 1999; Beasley \etal 
2000; Jord\'an \etal 2002; cf. Kundu \etal 1999 and Lee \& Kim 2000).

To some extent, this similarity extends to the dynamical properties of the 
two GC systems. Most notably, in both M49 and in M87, the full GC system has an
almost perfectly isotropic velocity dispersion tensor ($e.g.$, compare
Figure~\ref{fig19} above with Figure 13 of C\^ot\'e \etal 2001).
There are, nevertheless, some clear differences in the overall GC kinematics between
the two galaxies. For instance, the metal-poor and metal-rich subsystems
in M87 share an essentially identical projected velocity dispersion, while in M49,
the metal-rich GC sample appears to be significantly 
colder than the metal-poor component.

Table~\ref{tab4} summarizes the global kinematic properties of the two
GC systems. Note that the results for M87 differ slightly from those presented
in C\^ot\'e \etal (2001) since we have re-computed the various
kinematic properties after fixing the systemic velocity of each GC
subsample at the velocity of M87 ($v_{\rm gal} = 1277\pm2$ \kms;
van der Marel 1994), identical to the approach taken in \S3 for M49.
In addition, we have re-calculated the global quantities for the full GC
system and the metal-rich component in M49, after removing the 10 rapidly
rotating, metal-rich GCs at 22 $\lae R \lae$ 24.5 kpc.

As Table~\ref{tab4} shows, it is in their rotation 
properties that obvious differences between the two
systems appear. In M87, both the metal-poor and metal-rich GC subsystems show
relatively rapid rotation, with ${\Omega}R/\sigma_{p,r} \sim 0.45$. In M49,
however, rotation is somewhat more modest among the metal-poor GCs ($\Omega
R/\sigma_{p,r}=0.27$), and it is very weak indeed in the metal-rich GC system
($\Omega R/\sigma_{p,r}$ of order 0.1 and consistent with 0). The
orientation of the rotation axes in the two systems is also noteworthy.
In M87, the metal-rich GCs are found to rotate, at all radii, about the
galaxy's minor axis. This is also true of the metal-poor GCs beyond $R \sim
2R_{\rm eff}$ in that galaxy, but {\it inside} this radius, the metal-poor GCs
in M87 show {\it major-axis} rotation (see Table~\ref{tab4}, and C\^ot\'e
\etal 2001). In M49, the metal-poor GCs are consistent with minor-axis
rotation inside $R\lae2 R_{\rm eff}$ and suggest a possible transition to
major-axis rotation {\it outside} that radius. Interestingly, the position
angles for the rotation axes of the stellar light in M87 and M49 are found
to be $\Theta_0 = -23\pm10^\circ$ and $59\pm3^\circ$, respectively (Davies \&
Birkinshaw 1988). Thus, to within the measurement errors, and over the range
where the GC and stellar data overlap, the metal-poor GCs in both cases seem
to trace the rotation of the underlying galaxy. It would be interesting to
extend the stellar kinematic measurements to larger radii, to see if this
trend continues, {\it i.e.}, if the stellar rotation axis also ``flips''
by 90$^{\circ}$ in M87 especially.\footnotemark\footnotetext{At present, 
the best stellar kinematic data for M87 (Sembach \& Tonry 1996) extend 
only to about $1.5 R_{\rm eff}$. At $2 R_{\rm eff}$, the surface brightness 
of the galaxy is $\mu_V \sim 23.6$~mag~arcsec$^{-2}$, well within the range of 
modern long-slit spectrographs on large telescopes.} Finally, at the radii in M87
where the metal-poor and metal-rich GCs both rotate about the minor axis,
they do so in the same direction. By contrast, there is some evidence that
the metal-rich GC system of M49 may rotate {\it counter} to the metal-poor one.
We caution that this result is of low statistical significance, as we have a
sample of only $\sim$ 100 red GCs spread over an order-of-magnitude range in
galactocentric position, but it clearly merits further attention.

These differences between the GC systems of M49 and M87 become
all the more puzzling when one considers the results for the GC systems
of late-type galaxies. To date, in only two spiral galaxies have samples of
a hundred or more GC velocities been accumulated: the Milky Way and M31.
In the former case, ${\Omega}R/\sigma = 0.56\pm0.15$ for the GC system
as a whole, and ${\Omega}R/\sigma = 0.32\pm0.20$ and $1.05\pm0.28$ for
the metal-poor and metal-rich subsystems (C\^ot\'e 1999). In M31, the
global value for all GCs is ${\Omega}R/\sigma = 0.88\pm0.09$, with
the metal-poor and metal-rich subsystems having ${\Omega}R/\sigma =$
$0.85\pm0.09$ and $1.10\pm0.16$ (Perrett \etal 2002). Based on this very
small sample, it seems that rotation may be more important for the 
GC systems of disk galaxies than for those of ellipticals. Clearly, however, 
the four GC systems show considerable diversity in their
rotational properties. It seems that the lone common feature among the GC
systems of these four galaxies is the tendency
for the metal-poor GCs to form a rotating population, albeit with widely
varying levels of dynamical importance ($i.e.$, the percentage of
total kinetic energy stored in rotation varies by more than an order of magnitude,
from $\sim 3\%$ in M49 to $\sim 40\%$ for M31). Furthermore, the metal-rich
GCs of M87, M31, and the Milky Way {\it also} form rotating populations with
$\Omega R/\sigma$ ratios that are equal to or greater than those of their
metal-poor counterparts. In this regard, the metal-rich GC subsystem
in M49 may be something of an anomaly --- although such a conclusion 
may be premature given the current data. 
 
The kinematic properties of the GC systems studied to date
undoubtedly contain important clues to the formation histories of their
parent galaxies, although the full implications at this point remain unclear.
Nevertheless, it seems that the traditional picture of GC systems as
non-rotating, or slowly rotating, entities --- a view that can be traced to
the slow rotation exhibited by halo GCs outside of the central few
kpc in the Milky Way --- may be in need of revision. The net GC rotation detected 
in an ever-increasing number of galaxies, which we speculate may be a common
property of GC systems, probably reflects the outcome of a complex interplay between
any number of processes that could have shaped the observed GC velocity
fields: tidal torques from neighboring galaxies, spin-up during gravitational
collapse, angular momentum transfer and energy dissipation during gaseous
mergers, and the conversion of orbital angular momentum to rotation during
gas-poor mergers and accretions. 
The extraction of quantitative predictions
for GC dynamics from models of these processes will be a challenging
task --- as will be the complete and accurate empirical characterization of
kinematics to test those predictions.
 
\section{Summary}
 
We have reported radial velocities for 196 GCs associated with M49, the
brightest member of the Virgo cluster. Combining these velocities with previously
published measurements brings the total number of GCs with measured radial
velocities to 263. This sample is comparable to that assembled recently for M87
(Hanes \etal 2001) in terms of size, spatial coverage, velocity precision, and
the availability of metallicities from Washington photometry. Using
these data, we have carried out a kinematical and dynamical analysis of the M49 GC
system which mirrors that presented recently for M87 by C\^ot\'e \etal (2001).
 
We confirm previous findings that, when considered in its entirety, the GC system 
of M49 shows a modest net rotation that is due almost entirely to the rotation 
of the metal-poor subsystem. Likewise, we verify 
that, in a global average, the metal-rich subsystem shows essentially
no rotation. However, when examined in greater detail, there is some weak evidence
that the metal-rich GCs in this galaxy may, in fact, be slowly rotating, but in
the opposite direction to the metal-poor GCs. The overall negligible global rotation
of the metal-rich component is traced to a small sample of 10 clusters which appear to
show very rapid minor-axis rotation, in the same direction as the metal-poor GCs.
These objects merit further attention, as they may be the relic of a past
merger or accretion event.
Given the rather limited sample of metal-rich GCs with measured velocities, the 
significance of these findings must await additional
radial velocity measurements, but it is abundantly clear that conclusions
formed on the basis of globally averaged kinematics are likely to miss 
potentially interesting and complex variations within a given GC system.

A comparison with the results for M87, M31 and the Milky Way
suggests that rotation may, in fact, be generic to GC systems, a turnabout from
the traditional view of GC populations as slowly rotating systems. Any trends in
GC rotational properties with metallicity seem to be quite complex, and will
likely require GC radial velocity data for an expanded sample of galaxies before
they can be better characterized and understood. We note, however, that
in M49 and M87, there is a hint that the orientation
of the rotation axes for the metal-poor GCs trace those of the underlying galaxy.
 
Previous studies of M49, its GC system, and its X-ray halo have produced
some conflicting claims on the need for dark matter. We have shown that
the GC radial velocities and density profiles alone now provide unmistakable evidence
for a massive dark halo in M49/Virgo B. This result, which is corroborated by
analyses of its X-ray halo but is independent of the
X-ray observations, holds for any choice of GC orbital properties.
 
We have presented a mass model for M49/Virgo B that satisfies all existing
observational constraints on the mass distribution from optical and X-ray surface 
brightness profiles and stellar kinematical data.  A Jeans-equation analysis of 
the GC radial velocities and density profiles using this mass model reveals
the GC system as a whole to be perfectly well described by
an isotropic velocity dispersion tensor. This is also likely to be
true of the separate metal-poor and metal-rich subsystems, but definite conclusions 
must await the measurement of additional radial velocities and
improved surface density profiles for the subpopulations. Indeed, the poorly constrained
density distributions of metal-poor and metal-rich GCs on large scales may
now be the principal obstacle limiting our understanding their orbital properties.
In any case, the demonstration that the composite GC system of M49, like that of M87, has
an almost perfectly isotropic velocity dispersion tensor lends support to the 
general assumption of isotropy when using the GC systems of early-type galaxies 
to derive gravitational masses.
 
\acknowledgments
 
PC and JPB acknowledge support provided by the Sherman M. Fairchild Foundation during the early
stages of this work. PC also acknowledges support provided by NASA LTSA grant NAG5-11714.  This 
research has made use of the NASA/IPAC Extragalactic Database (NED) which is operated by the Jet
Propulsion Laboratory, California Institute of Technology, under contract with the National 
Aeronautics and Space Administration. The entire Keck/LRIS user communities owes a debt of
gratitude to Jerry Nelson, Gerry Smith, Bev Oke, and the many other people who worked 
to make the Keck telescopes a reality. 
We are grateful to the W. M. Keck Foundation for the vision to fund the construction of 
the W. M. Keck Observatory. The authors wish to extend special thanks to those of Hawaiian 
ancestry on whose sacred mountain we are privileged to be guests. 
Without their generous hospitality, none of the observations presented
herein would have been possible.

 
\begin{deluxetable}{crccr}
\tablecolumns{6}
\tablenum{1}
\tablewidth{0pc}
\tablecaption{Observing Log\label{tab1}}
\tablehead{
\colhead{Run} &
\colhead{Date} &
\colhead{Grating} &
\colhead{${\lambda}_{c}$} &
\colhead{${\lambda} / {\Delta}{\lambda}$} \nl
\colhead{} &
\colhead{(d/m/y)} &
\colhead{} &
\colhead{(\AA)} &
\colhead{}
}
\startdata
1 & 16-17/04/1998 & 600/5000 & 5200 &  870 \cr
2 & ~~~24/03/1999 & 600/7500 & 8550 & 1425 \cr
3 & ~~~~6/04/1999 & 600/7500 & 8550 & 1425 \cr
4 & ~~~30/04/2000 & 600/5000 & 4860 &  870 \cr
\enddata
\end{deluxetable}
 
\clearpage
 
\begin{deluxetable}{rrrrrrcrrcr}
\scriptsize
\tablecolumns{11}
\tablenum{2}
\tablewidth{0pc}
\tablecaption{Radial Velocities for Confirmed and Candidate M49 Globular
Clusters\label{tab2}}
\tablehead{
\colhead{ID} &
\colhead{$\alpha$(J2000)} &
\colhead{$\delta$(J2000)} &
\colhead{$R$} &
\colhead{$\Theta$} &
\colhead{$T_1$} &
\colhead{$(C-T_1)$} &
\colhead{[Fe/H]} &
\colhead{$v_{p,i}$}  &
\colhead{Source\tablenotemark{1}} &
\colhead{$\langle{v_{p}}\rangle$} \nl
\colhead{} &
\colhead{} &
\colhead{} &
\colhead{(arcsec)} &
\colhead{(deg)} &
\colhead{(mag)} &
\colhead{(mag)} &
\colhead{(dex)} &
\colhead{(km s$^{-1}$)} &
\colhead{}  &
\colhead{(km s$^{-1}$)}
}
\startdata
\multicolumn{11}{c}{} \nl
\multicolumn{11}{c}{\underbar{Globular Clusters}} \nl
\multicolumn{11}{c}{} \nl
  85 & 12:29:37.87 &  7:52:53.9 & 459.6 & 198.2 & 20.52 & 1.30 &
-1.43$\pm$0.05 &    826$\pm$152 & 1&   826$\pm$ 62\nl
     &             &            &       &       &       &     
&                &    826$\pm$ 68 & 3&             \nl
 117 & 12:29:33.24 &  7:53:02.6 & 477.7 & 206.4 & 21.78 & 1.99 & 
0.17$\pm$0.14 &    851$\pm$ 66 & 2&   851$\pm$ 66\nl
 150 & 12:29:58.39 &  7:53:12.8 & 447.7 & 158.9 & 21.75 & 1.21 &
-1.56$\pm$0.12 &    849$\pm$122 & 1&   849$\pm$122\nl
 170 & 12:29:37.75 &  7:53:18.0 & 437.4 & 199.5 & 21.19 & 1.73 &
-0.44$\pm$0.07 &   1243$\pm$ 69 & Z&  1243$\pm$ 69\nl
 179 & 12:29:26.74 &  7:53:21.0 & 513.0 & 217.1 & 20.91 & 1.35 &
-1.31$\pm$0.11 &    580$\pm$ 83 & 2&   580$\pm$ 83\nl
 258 & 12:29:57.99 &  7:53:44.1 & 416.4 & 158.1 & 22.01 & 2.12 & 
0.48$\pm$0.16 &   1281$\pm$155 & 1&  1281$\pm$155\nl
 282 & 12:30:05.42 &  7:53:49.8 & 464.3 & 145.1 & 20.38 & 1.65 &
-0.61$\pm$0.13 &    751$\pm$ 39 & Z&   751$\pm$ 39\nl
 288 & 12:29:22.03 &  7:53:51.0 & 536.3 & 225.0 & 21.03 & 1.72 &
-0.46$\pm$0.15 &    286$\pm$ 47 & 2&   286$\pm$ 47\nl
 305 & 12:29:55.59 &  7:53:54.2 & 394.8 & 162.4 & 20.52 & 1.96 & 
0.11$\pm$0.06 &    874$\pm$ 58 & 3&   874$\pm$ 58\nl
 337 & 12:29:57.61 &  7:54:00.5 & 399.1 & 158.0 & 20.69 & 1.39 &
-1.24$\pm$0.08 &    449$\pm$136 & 1&   449$\pm$136\nl
 342 & 12:29:27.65 &  7:54:00.9 & 473.2 & 218.7 & 21.53 & 1.61 &
-0.70$\pm$0.09 &    825$\pm$ 50 & 2&   825$\pm$ 50\nl
 430 & 12:29:35.07 &  7:54:14.6 & 401.2 & 207.5 & 20.92 & 1.29 &
-1.46$\pm$0.04 &   1167$\pm$164 & 1&  1167$\pm$164\nl
 455 & 12:29:30.49 &  7:54:19.4 & 432.9 & 215.8 & 20.71 & 1.31 &
-1.42$\pm$0.05 &    294$\pm$ 77 & 2&   294$\pm$ 77\nl
 460 & 12:29:57.78 &  7:54:20.1 & 382.0 & 156.6 & 21.78 & 1.50 &
-0.98$\pm$0.11 &   1800$\pm$134 & 1&  1800$\pm$134\nl
 463 & 12:29:50.42 &  7:54:21.0 & 352.1 & 173.0 & 19.93 & 1.59 &
-0.77$\pm$0.05 &    342$\pm$ 27 & Z&   342$\pm$ 27\nl
 522 & 12:29:53.54 &  7:54:31.1 & 350.9 & 165.3 & 21.02 & 1.67 &
-0.56$\pm$0.08 &    805$\pm$129 & 3&   805$\pm$129\nl
 564 & 12:29:56.82 &  7:54:37.9 & 360.0 & 157.5 & 21.06 & 1.57 &
-0.81$\pm$0.07 &   1088$\pm$ 83 & 1&  1088$\pm$ 83\nl
 637 & 12:29:42.59 &  7:54:49.5 & 329.3 & 192.9 & 19.95 & 1.54 &
-0.88$\pm$0.07 &    814$\pm$ 31 & Z&   814$\pm$ 31\nl
 647 & 12:29:37.12 &  7:54:50.8 & 355.2 & 205.9 & 21.15 & 1.34 &
-1.34$\pm$0.07 &   1101$\pm$ 39 & Z&  1101$\pm$ 39\nl
 676 & 12:29:52.36 &  7:54:56.0 & 322.6 & 167.2 & 20.72 & 1.46 &
-1.07$\pm$0.07 &   1162$\pm$167 & 3&  1289$\pm$ 53\nl
     &             &            &       &       &       &     
&                &   1304$\pm$ 56 & Z&             \nl
 677 & 12:29:52.89 &  7:54:56.0 & 324.3 & 165.9 & 22.03 & 1.82 &
-0.23$\pm$0.14 &   1090$\pm$154 & 1&  1090$\pm$154\nl
 714 & 12:30:07.42 &  7:55:01.1 & 427.7 & 136.4 & 20.32 & 1.55 &
-0.85$\pm$0.09 &   1061$\pm$ 39 & Z&  1061$\pm$ 39\nl
 744 & 12:29:53.46 &  7:55:05.8 & 317.2 & 163.9 & 19.73 & 1.38 &
-1.26$\pm$0.06 &    814$\pm$ 29 & Z&   814$\pm$ 29\nl
 830 & 12:29:58.72 &  7:55:17.9 & 336.5 & 150.4 & 21.24 & 2.22 & 
0.72$\pm$0.10 &   1313$\pm$105 & 1&  1313$\pm$105\nl
 876 & 12:30:00.68 &  7:55:23.9 & 346.8 & 145.8 & 19.63 & 1.46 &
-1.06$\pm$0.13 &   1487$\pm$ 24 & Z&  1487$\pm$ 24\nl
 888 & 12:29:53.50 &  7:55:25.0 & 298.9 & 162.8 & 20.58 & 1.42 &
-1.17$\pm$0.07 &   1088$\pm$172 & 3&  1088$\pm$172\nl
 929 & 12:29:55.43 &  7:55:29.4 & 304.6 & 157.4 & 21.36 & 1.93 & 
0.04$\pm$0.11 &   1268$\pm$143 & 1&  1268$\pm$143\nl
 952 & 12:29:38.07 &  7:55:31.4 & 312.6 & 206.8 & 21.16 & 1.52 &
-0.93$\pm$0.08 &    970$\pm$127 & 1&   970$\pm$127\nl
 995 & 12:29:18.62 &  7:55:35.2 & 510.2 & 237.4 & 20.90 & 1.54 &
-0.87$\pm$0.08 &    828$\pm$ 90 & Z&   828$\pm$ 90\nl
1047 & 12:29:33.50 &  7:55:39.2 & 342.2 & 217.6 & 21.14 & 1.89 &
-0.05$\pm$0.08 &   1083$\pm$ 30 & Z&  1091$\pm$ 28\nl
     &             &            &       &       &       &     
&                &   1147$\pm$ 77 & 2&             \nl
1087 & 12:29:49.75 &  7:55:43.0 & 269.5 & 173.0 & 19.64 & 1.44 &
-1.12$\pm$0.06 &   1070$\pm$ 29 & Z&  1070$\pm$ 29\nl
1095 & 12:29:39.43 &  7:55:43.6 & 292.8 & 204.3 & 22.00 & 1.22 &
-1.62$\pm$0.08 &   1772$\pm$127 & 1&  1772$\pm$127\nl
1110 & 12:29:44.59 &  7:55:45.8 & 268.3 & 189.4 & 19.88 & 1.37 &
-1.28$\pm$0.10 &   1626$\pm$ 24 & Z&  1626$\pm$ 24\nl
1174 & 12:29:55.89 &  7:55:51.3 & 287.4 & 154.5 & 21.80 & 2.05 & 
0.33$\pm$0.13 &    954$\pm$185 & 3&   954$\pm$185\nl
1193 & 12:29:57.99 &  7:55:53.7 & 300.1 & 148.9 & 21.00 & 1.88 &
-0.08$\pm$0.09 &    761$\pm$ 81 & 1&   761$\pm$ 81\nl
1207 & 12:29:58.63 &  7:55:54.9 & 304.1 & 147.2 & 21.34 & 1.25 &
-1.57$\pm$0.07 &    739$\pm$ 69 & Z&   739$\pm$ 69\nl
1234 & 12:29:19.39 &  7:55:57.4 & 488.8 & 238.9 & 20.81 & 1.47 &
-1.03$\pm$0.07 &   1059$\pm$ 64 & Z&  1075$\pm$ 36\nl
     &             &            &       &       &       &     
&                &   1082$\pm$ 44 & 2&             \nl
1255 & 12:29:52.92 &  7:55:59.9 & 263.0 & 162.4 & 19.71 & 1.35 &
-1.33$\pm$0.05 &    816$\pm$ 66 & Z&   816$\pm$ 66\nl
1300 & 12:29:54.90 &  7:56:04.4 & 269.3 & 156.1 & 21.95 & 2.02 & 
0.26$\pm$0.16 &    761$\pm$128 & 3&   761$\pm$128\nl
1315 & 12:29:43.94 &  7:56:06.4 & 249.9 & 192.4 & 20.68 & 1.42 &
-1.15$\pm$0.09 &   1370$\pm$ 76 & Z&  1497$\pm$ 61\nl
     &             &            &       &       &       &     
&                &   1735$\pm$103 & 1&             \nl
1369 & 12:29:59.20 &  7:56:11.7 & 295.0 & 144.1 & 21.21 & 1.40 &
-1.22$\pm$0.08 &   1235$\pm$117 & 1&  1235$\pm$117\nl
1411 & 12:30:11.66 &  7:56:15.8 & 428.3 & 123.3 & 21.61 & 1.30 &
-1.43$\pm$0.11 &    753$\pm$ 65 & 3&   753$\pm$ 65\nl
1423 & 12:29:35.31 &  7:56:16.7 & 296.1 & 217.9 & 20.94 & 1.57 &
-0.80$\pm$0.08 &   1768$\pm$ 89 & 1&  1653$\pm$ 27\nl
     &             &            &       &       &       &     
&                &   1641$\pm$ 29 & Z&             \nl
1448 & 12:29:29.46 &  7:56:19.5 & 354.2 & 229.3 & 20.79 & 1.34 &
-1.34$\pm$0.05 &   1353$\pm$ 26 & Z&  1353$\pm$ 26\nl
1475 & 12:29:40.45 &  7:56:22.5 & 251.1 & 204.8 & 21.15 & 1.46 &
-1.07$\pm$0.09 &    970$\pm$ 50 & 4&  1001$\pm$ 29\nl
     &             &            &       &       &       &     
&                &   1018$\pm$ 37 & Z&             \nl
1508 & 12:30:02.66 &  7:56:25.6 & 317.9 & 135.1 & 21.49 & 1.99 & 
0.18$\pm$0.11 &   1316$\pm$ 50 & 3&  1371$\pm$ 35\nl
     &             &            &       &       &       &     
&                &   1425$\pm$ 50 & 4&             \nl
1518 & 12:29:40.44 &  7:56:25.9 & 248.1 & 205.2 & 19.25 & 1.85 &
-0.15$\pm$0.10 &   1050$\pm$ 36 & S&  1050$\pm$ 36\nl
1570 & 12:29:39.07 &  7:56:31.0 & 252.9 & 209.8 & 20.98 & 1.58 &
-0.78$\pm$0.08 &   1034$\pm$ 61 & Z&  1034$\pm$ 61\nl
1587 & 12:29:59.34 &  7:56:32.9 & 279.4 & 141.2 & 21.16 & 1.12 &
-1.86$\pm$0.11 &    471$\pm$ 75 & Z&   471$\pm$ 75\nl
1650 & 12:29:55.84 &  7:56:39.1 & 244.7 & 149.8 & 20.85 & 1.95 & 
0.09$\pm$0.09 &    973$\pm$ 25 & 4&   985$\pm$ 21\nl
     &             &            &       &       &       &     
&                &    988$\pm$ 57 & 3&             \nl
     &             &            &       &       &       &     
&                &   1040$\pm$ 55 & Z&             \nl
1712 & 12:29:40.09 &  7:56:44.3 & 234.0 & 208.3 & 20.36 & 1.34 &
-1.35$\pm$0.10 &   1144$\pm$ 40 & S&  1144$\pm$ 40\nl
1731 & 12:30:01.35 &  7:56:46.7 & 289.2 & 134.8 & 20.71 & 1.82 &
-0.22$\pm$0.09 &   1241$\pm$ 25 & 4&  1250$\pm$ 19\nl
     &             &            &       &       &       &     
&                &   1246$\pm$ 40 & 3&             \nl
     &             &            &       &       &       &     
&                &   1294$\pm$ 51 & Z&             \nl
1749 & 12:29:47.41 &  7:56:48.1 & 202.4 & 180.6 & 20.92 & 1.98 & 
0.17$\pm$0.10 &   1407$\pm$ 88 & Z&  1407$\pm$ 88\nl
1764 & 12:29:44.70 &  7:56:49.1 & 205.8 & 191.9 & 20.82 & 1.72 &
-0.46$\pm$0.11 &    855$\pm$ 37 & Z&   855$\pm$ 37\nl
1782 & 12:29:37.28 &  7:56:50.4 & 251.4 & 217.3 & 21.91 & 1.99 & 
0.18$\pm$0.12 &    967$\pm$142 & 1&   986$\pm$ 97\nl
     &             &            &       &       &       &     
&                &   1002$\pm$135 & 1&             \nl
1798 & 12:29:45.21 &  7:56:51.5 & 202.0 & 189.9 & 20.69 & 1.98 & 
0.16$\pm$0.09 &    785$\pm$ 25 & 4&   795$\pm$ 19\nl
     &             &            &       &       &       &     
&                &    811$\pm$ 31 & Z&             \nl
1831 & 12:29:51.92 &  7:56:53.5 & 207.4 & 161.7 & 21.86 & 1.34 &
-1.35$\pm$0.09 &   1129$\pm$148 & 3&  1129$\pm$148\nl
1846 & 12:29:46.42 &  7:56:54.8 & 196.4 & 184.9 & 21.07 & 2.02 & 
0.26$\pm$0.11 &   1041$\pm$ 25 & 4&  1041$\pm$ 25\nl
1889 & 12:29:54.02 &  7:56:58.7 & 214.6 & 153.4 & 20.98 & 1.25 &
-1.55$\pm$0.06 &   1105$\pm$164 & 1&  1187$\pm$ 24\nl
     &             &            &       &       &       &     
&                &   1189$\pm$ 25 & 4&             \nl
1892 & 12:30:04.11 &  7:56:58.8 & 312.1 & 127.9 & 21.09 & 1.53 &
-0.91$\pm$0.11 &   1029$\pm$ 77 & 3&  1077$\pm$ 42\nl
     &             &            &       &       &       &     
&                &   1098$\pm$ 50 & 4&             \nl
1905 & 12:29:41.13 &  7:57:00.1 & 212.9 & 206.6 & 21.22 & 1.36 &
-1.29$\pm$0.11 &   1483$\pm$112 & 1&  1472$\pm$ 24\nl
     &             &            &       &       &       &     
&                &   1472$\pm$ 25 & 4&             \nl
2013 & 12:29:38.38 &  7:57:10.2 & 225.8 & 217.1 & 21.28 & 1.40 &
-1.21$\pm$0.11 &    517$\pm$ 25 & 4&   517$\pm$ 25\nl
2031 & 12:29:47.67 &  7:57:11.8 & 178.7 & 179.4 & 20.71 & 1.37 &
-1.28$\pm$0.08 &   1438$\pm$ 65 & M&  1380$\pm$ 22\nl
     &             &            &       &       &       &     
&                &   1366$\pm$ 25 & 4&             \nl
     &             &            &       &       &       &     
&                &   1426$\pm$ 71 & 3&             \nl
2045 & 12:29:39.05 &  7:57:13.1 & 217.6 & 215.4 & 20.94 & 1.77 &
-0.35$\pm$0.09 &    857$\pm$ 54 & S&   950$\pm$ 22\nl
     &             &            &       &       &       &     
&                &    970$\pm$ 25 & 4&             \nl
2060 & 12:29:39.71 &  7:57:13.9 & 211.4 & 213.4 & 20.62 & 1.29 &
-1.45$\pm$0.06 &   1217$\pm$ 65 & M&  1293$\pm$ 22\nl
     &             &            &       &       &       &     
&                &   1342$\pm$124 & 1&             \nl
     &             &            &       &       &       &     
&                &   1303$\pm$ 25 & 4&             \nl
2070 & 12:29:25.32 &  7:57:14.6 & 374.0 & 242.0 & 22.14 & 2.16 & 
0.58$\pm$0.15 &   1193$\pm$ 79 & 2&  1193$\pm$ 79\nl
2140 & 12:29:54.18 &  7:57:20.9 & 196.2 & 149.9 & 20.45 & 1.80 &
-0.28$\pm$0.10 &    730$\pm$ 53 & S&   770$\pm$ 26\nl
     &             &            &       &       &       &     
&                &    784$\pm$ 31 & Z&             \nl
2163 & 12:29:55.88 &  7:57:23.2 & 208.2 & 143.5 & 20.15 & 2.01 & 
0.22$\pm$0.05 &    402$\pm$ 43 & S&   402$\pm$ 43\nl
2178 & 12:29:37.39 &  7:57:24.6 & 224.1 & 222.3 & 21.51 & 1.19 &
-1.69$\pm$0.10 &    522$\pm$136 & 1&   773$\pm$ 24\nl
     &             &            &       &       &       &     
&                &    782$\pm$ 25 & 4&             \nl
2188 & 12:29:58.57 &  7:57:25.8 & 232.3 & 135.2 & 21.15 & 1.33 &
-1.36$\pm$0.11 &    623$\pm$102 & 1&   704$\pm$ 24\nl
     &             &            &       &       &       &     
&                &    709$\pm$ 25 & 4&             \nl
2195 & 12:29:30.72 &  7:57:26.1 & 299.1 & 236.7 & 21.33 & 1.14 &
-1.80$\pm$0.12 &   1241$\pm$ 56 & Z&  1241$\pm$ 56\nl
2303 & 12:30:00.66 &  7:57:34.7 & 249.5 & 128.7 & 21.91 & 1.33 &
-1.37$\pm$0.12 &    864$\pm$ 73 & 3&   864$\pm$ 73\nl
2306 & 12:29:57.69 &  7:57:34.9 & 216.7 & 135.9 & 20.35 & 1.63 &
-0.68$\pm$0.10 &    893$\pm$ 25 & 4&   893$\pm$ 25\nl
2341 & 12:29:32.76 &  7:57:38.3 & 267.1 & 235.3 & 20.76 & 1.91 &
-0.01$\pm$0.11 &   1001$\pm$ 68 & S&  1073$\pm$ 55\nl
     &             &            &       &       &       &     
&                &   1216$\pm$ 95 & 1&             \nl
2406 & 12:29:45.80 &  7:57:44.1 & 148.6 & 190.1 & 20.85 & 2.03 & 
0.27$\pm$0.11 &   1128$\pm$ 25 & 4&  1141$\pm$ 23\nl
     &             &            &       &       &       &     
&                &   1244$\pm$ 70 & S&             \nl
2420 & 12:29:41.00 &  7:57:44.7 & 175.2 & 213.7 & 20.95 & 2.08 & 
0.40$\pm$0.10 &    763$\pm$ 92 & Z&   763$\pm$ 92\nl
2421 & 12:29:52.18 &  7:57:45.0 & 161.0 & 154.7 & 21.09 & 1.43 &
-1.13$\pm$0.08 &   1726$\pm$ 25 & 4&  1729$\pm$ 24\nl
     &             &            &       &       &       &     
&                &   1801$\pm$117 & 3&             \nl
2452 & 12:29:57.82 &  7:57:48.1 & 208.8 & 133.0 & 21.49 & 1.40 &
-1.20$\pm$0.13 &   1828$\pm$ 81 & Z&  1828$\pm$ 81\nl
2482 & 12:29:42.71 &  7:57:50.0 & 157.7 & 207.1 & 21.58 & 2.08 & 
0.39$\pm$0.13 &    767$\pm$ 56 & S&   767$\pm$ 56\nl
2502 & 12:30:00.67 &  7:57:52.1 & 239.1 & 125.4 & 21.08 & 1.48 & 
0.01$\pm$0.08 &    930$\pm$ 25 & 4&   935$\pm$ 22\nl
     &             &            &       &       &       &     
&                &    955$\pm$ 49 & 3&             \nl
2528 & 12:29:49.33 &  7:57:53.7 & 139.4 & 169.0 & 20.34 & 1.46 &
-1.06$\pm$0.08 &    807$\pm$ 65 & M&   795$\pm$ 23\nl
     &             &            &       &       &       &     
&                &    794$\pm$ 25 & 4&             \nl
2543 & 12:29:52.87 &  7:57:54.8 & 157.1 & 149.8 & 20.27 & 1.36 &
-1.30$\pm$0.10 &   1199$\pm$ 48 & S&  1221$\pm$ 22\nl
     &             &            &       &       &       &     
&                &   1228$\pm$ 25 & 4&             \nl
2545 & 12:29:57.70 &  7:57:55.1 & 202.7 & 131.9 & 20.64 & 1.31 &
-1.41$\pm$0.08 &    414$\pm$ 53 & 1&   414$\pm$ 53\nl
2569 & 12:29:43.87 &  7:57:57.2 & 144.1 & 202.3 & 20.12 & 1.89 &
-0.06$\pm$0.12 &   1056$\pm$ 46 & S&  1068$\pm$ 22\nl
     &             &            &       &       &       &     
&                &   1072$\pm$ 25 & 4&             \nl
2622 & 12:29:52.65 &  7:58:02.2 & 149.1 & 149.4 & 21.09 & 1.65 &
-0.61$\pm$0.10 &    467$\pm$164 & 3&   467$\pm$164\nl
2634 & 12:29:39.63 &  7:58:03.3 & 173.1 & 222.8 & 19.70 & 1.56 &
-0.82$\pm$0.12 &   1014$\pm$ 57 & S&  1014$\pm$ 57\nl
2753 & 12:29:46.19 &  7:58:12.5 & 119.7 & 189.6 & 20.88 & 1.19 &
-1.70$\pm$0.09 &    945$\pm$100 & Z&   945$\pm$100\nl
2759 & 12:29:26.29 &  7:58:12.6 & 337.0 & 249.6 & 19.97 & 1.31 &
-1.42$\pm$0.04 &    654$\pm$ 92 & Z&   654$\pm$ 92\nl
2813 & 12:29:48.56 &  7:58:16.0 & 115.5 & 172.5 & 21.00 & 1.94 & 
0.07$\pm$0.11 &    363$\pm$ 25 & 4&   363$\pm$ 25\nl
2817 & 12:30:03.46 &  7:58:16.2 & 262.7 & 115.8 & 21.05 & 1.50 &
-0.98$\pm$0.10 &    665$\pm$ 47 & Z&   665$\pm$ 47\nl
2938 & 12:29:27.92 &  7:58:23.4 & 310.5 & 249.9 & 20.98 & 1.59 &
-0.75$\pm$0.06 &    843$\pm$ 87 & 2&   843$\pm$ 87\nl
2960 & 12:29:59.98 &  7:58:25.0 & 212.8 & 119.7 & 21.63 & 1.43 &
-1.13$\pm$0.11 &    637$\pm$125 & 1&   637$\pm$125\nl
3150 & 12:29:38.50 &  7:58:37.5 & 163.3 & 235.3 & 21.40 & 1.79 &
-0.29$\pm$0.12 &    952$\pm$ 42 & S&  1126$\pm$ 21\nl
     &             &            &       &       &       &     
&                &   1188$\pm$ 25 & 4&             \nl
3208 & 12:29:53.55 &  7:58:41.4 & 126.1 & 135.0 & 21.17 & 1.43 &
-1.14$\pm$0.10 &    677$\pm$ 96 & 1&   703$\pm$ 75\nl
     &             &            &       &       &       &     
&                &    745$\pm$122 & 3&             \nl
3250 & 12:29:27.74 &  7:58:44.2 & 306.6 & 253.7 & 21.20 & 1.23 &
-1.61$\pm$0.09 &   1117$\pm$ 56 & 2&  1117$\pm$ 56\nl
3289 & 12:29:37.79 &  7:58:46.2 & 167.6 & 239.8 & 21.48 & 1.90 &
-0.04$\pm$0.10 &   1476$\pm$101 & 1&  1476$\pm$101\nl
3307 & 12:30:02.45 &  7:58:46.9 & 236.8 & 110.7 & 20.25 & 1.53 &
-0.89$\pm$0.13 &   1834$\pm$ 65 & M&  1834$\pm$ 65\nl
3355 & 12:29:34.40 &  7:58:51.5 & 210.6 & 248.0 & 20.84 & 1.37 &
-1.28$\pm$0.10 &   1436$\pm$ 72 & 1&  1436$\pm$ 72\nl
3361 & 12:29:34.46 &  7:58:52.0 & 209.5 & 248.0 & 20.34 & 1.55 &
-0.85$\pm$0.10 &   1392$\pm$ 33 & Z&  1392$\pm$ 33\nl
3372 & 12:29:52.46 &  7:58:52.7 & 106.8 & 136.8 & 21.59 & 1.60 &
-0.74$\pm$0.10 &   1331$\pm$183 & 3&  1331$\pm$183\nl
3422 & 12:29:37.93 &  7:58:56.3 & 160.9 & 242.6 & 20.87 & 1.76 &
-0.36$\pm$0.12 &   1265$\pm$ 81 & 1&  1265$\pm$ 81\nl
3434 & 12:29:22.62 &  7:58:56.8 & 377.4 & 258.8 & 22.10 & 1.19 &
-1.70$\pm$0.10 &    255$\pm$ 75 & 2&   255$\pm$ 75\nl
3545 & 12:29:52.22 &  7:59:03.5 &  96.6 & 134.0 & 20.84 & 1.83 &
-0.19$\pm$0.12 &   1046$\pm$101 & 3&  1046$\pm$101\nl
3584 & 12:29:35.50 &  7:59:06.6 & 189.9 & 250.4 & 21.70 & 1.98 & 
0.16$\pm$0.11 &    952$\pm$158 & 1&   952$\pm$158\nl
3603 & 12:29:46.71 &  7:59:07.9 &  63.8 & 191.1 & 20.47 & 1.75 &
-0.39$\pm$0.09 &   1026$\pm$ 25 & 4&  1026$\pm$ 25\nl
3628 & 12:29:33.14 &  7:59:09.3 & 222.5 & 254.1 & 21.22 & 1.90 &
-0.02$\pm$0.09 &   1008$\pm$ 49 & S&  1008$\pm$ 49\nl
3635 & 12:29:56.57 &  7:59:10.1 & 147.2 & 114.3 & 20.86 & 1.34 &
-1.35$\pm$0.09 &    893$\pm$ 65 & 1&   907$\pm$ 53\nl
     &             &            &       &       &       &     
&                &    936$\pm$ 94 & Z&             \nl
3651 & 12:29:58.02 &  7:59:10.7 & 166.8 & 111.0 & 21.75 & 1.71 &
-0.47$\pm$0.12 &   1252$\pm$ 54 & 3&  1252$\pm$ 54\nl
3757 & 12:29:45.94 &  7:59:17.3 &  58.3 & 204.1 & 21.02 & 1.82 &
-0.23$\pm$0.11 &   1220$\pm$ 95 & Z&  1220$\pm$ 95\nl
3788 & 12:29:47.87 &  7:59:19.6 &  51.2 & 174.5 & 20.80 & 1.87 &
-0.10$\pm$0.11 &   1174$\pm$ 25 & 4&  1174$\pm$ 25\nl
3808 & 12:29:39.15 &  7:59:19.9 & 134.5 & 248.0 & 20.35 & 1.83 &
-0.21$\pm$0.12 &    832$\pm$ 35 & S&   832$\pm$ 35\nl
3900 & 12:29:49.53 &  7:59:25.7 &  53.7 & 146.6 & 21.04 & 1.89 &
-0.05$\pm$0.11 &    662$\pm$ 50 & 4&   662$\pm$ 50\nl
3909 & 12:30:06.11 &  7:59:26.1 & 279.4 &  99.2 & 21.43 & 1.36 &
-1.31$\pm$0.07 &   1253$\pm$ 90 & Z&  1253$\pm$ 90\nl
3980 & 12:29:35.60 &  7:59:30.8 & 181.8 & 257.4 & 21.15 & 1.28 &
-1.48$\pm$0.10 &   1112$\pm$ 45 & S&  1134$\pm$ 43\nl
     &             &            &       &       &       &     
&                &   1369$\pm$144 & 1&             \nl
3990 & 12:29:52.89 &  7:59:31.7 &  88.4 & 116.1 & 21.23 & 2.12 & 
0.49$\pm$0.12 &   1286$\pm$ 89 & 3&  1286$\pm$ 89\nl
4017 & 12:29:57.59 &  7:59:33.6 & 153.8 & 103.9 & 20.92 & 1.42 &
-1.17$\pm$0.09 &    949$\pm$ 96 & 1&   900$\pm$ 44\nl
     &             &            &       &       &       &     
&                &    888$\pm$ 50 & 4&             \nl
4062 & 12:29:45.38 &  7:59:36.9 &  46.4 & 223.7 & 20.77 & 2.01 & 
0.23$\pm$0.14 &    614$\pm$ 50 & 4&   614$\pm$ 50\nl
4125 & 12:29:39.16 &  7:59:41.6 & 127.8 & 257.0 & 22.11 & 1.72 &
-0.47$\pm$0.15 &   1435$\pm$169 & 1&  1435$\pm$169\nl
4144 & 12:30:02.01 &  7:59:43.1 & 216.7 &  97.3 & 20.74 & 1.33 &
-1.37$\pm$0.06 &    884$\pm$ 25 & 4&   884$\pm$ 25\nl
4168 & 12:29:40.18 &  7:59:44.8 & 112.2 & 256.8 & 20.36 & 1.68 &
-0.55$\pm$0.10 &   1384$\pm$ 44 & S&  1445$\pm$ 21\nl
     &             &            &       &       &       &     
&                &   1465$\pm$ 25 & 4&             \nl
4187 & 12:30:00.02 &  7:59:46.6 & 186.9 &  97.4 & 21.75 & 1.90 &
-0.03$\pm$0.10 &    981$\pm$156 & 1&   981$\pm$156\nl
4210 & 12:30:06.78 &  7:59:48.6 & 286.6 &  94.4 & 20.53 & 1.62 &
-0.69$\pm$0.05 &   1910$\pm$ 29 & Z&  1910$\pm$ 29\nl
4216 & 12:29:25.97 &  7:59:49.0 & 321.0 & 266.2 & 22.25 & 1.14 &
-1.83$\pm$0.10 &    591$\pm$ 70 & 2&   591$\pm$ 70\nl
4217 & 12:29:52.04 &  7:59:49.4 &  70.1 & 107.5 & 20.66 & 1.79 &
-0.29$\pm$0.09 &    887$\pm$135 & 1&  1067$\pm$ 23\nl
     &             &            &       &       &       &     
&                &   1040$\pm$ 76 & 3&             \nl
     &             &            &       &       &       &     
&                &   1077$\pm$ 25 & 4&             \nl
4296 & 12:30:04.68 &  7:59:54.1 & 255.1 &  93.7 & 20.79 & 1.27 &
-1.52$\pm$0.07 &   1212$\pm$ 25 & 4&  1214$\pm$ 23\nl
     &             &            &       &       &       &     
&                &   1234$\pm$ 67 & 3&             \nl
4297 & 12:29:37.74 &  7:59:54.1 & 146.5 & 263.6 & 21.41 & 1.49 &
-0.98$\pm$0.11 &   1435$\pm$163 & 1&  1435$\pm$163\nl
4332 & 12:30:01.46 &  7:59:56.4 & 207.3 &  93.9 & 21.95 & 1.16 &
-1.78$\pm$0.09 &    878$\pm$134 & 1&   878$\pm$134\nl
4351 & 12:29:49.03 &  7:59:57.5 &  25.6 & 120.4 & 20.38 & 1.37 &
-1.27$\pm$0.12 &    263$\pm$ 25 & 4&   263$\pm$ 25\nl
4386 & 12:29:51.70 &  8:00:00.2 &  62.6 &  99.5 & 19.83 & 1.94 & 
0.05$\pm$0.10 &   1197$\pm$ 33 & S&  1197$\pm$ 33\nl
4401 & 12:30:00.25 &  8:00:01.8 & 189.0 &  92.7 & 20.85 & 1.91 & 
0.00$\pm$0.08 &   1286$\pm$ 25 & 4&  1286$\pm$ 25\nl
4436 & 12:29:41.01 &  8:00:03.7 &  97.2 & 266.0 & 21.29 & 1.90 &
-0.04$\pm$0.12 &   1047$\pm$119 & 1&  1047$\pm$119\nl
4443 & 12:29:55.86 &  8:00:04.0 & 123.8 &  93.0 & 21.53 & 1.72 &
-0.47$\pm$0.12 &   1126$\pm$124 & 3&  1126$\pm$124\nl
4494 & 12:29:37.06 &  8:00:07.2 & 155.7 & 268.8 & 20.99 & 1.42 &
-1.15$\pm$0.10 &   1132$\pm$166 & 1&  1132$\pm$166\nl
4513 & 12:29:42.31 &  8:00:08.2 &  77.7 & 268.3 & 20.10 & 1.85 &
-0.14$\pm$0.13 &    908$\pm$ 80 & S&   991$\pm$ 23\nl
     &             &            &       &       &       &     
&                &   1000$\pm$ 25 & 4&             \nl
4541 & 12:29:54.22 &  8:00:10.3 &  99.3 &  90.1 & 20.83 & 1.56 &
-0.79$\pm$0.10 &    733$\pm$ 25 & 4&   733$\pm$ 25\nl
4582 & 12:29:38.23 &  8:00:13.0 & 138.2 & 271.1 & 21.66 & 1.93 & 
0.04$\pm$0.13 &    475$\pm$136 & 1&   475$\pm$136\nl
4628 & 12:29:26.21 &  8:00:16.3 & 316.8 & 271.1 & 20.66 & 1.29 &
-1.47$\pm$0.09 &    726$\pm$ 52 & 3&   726$\pm$ 52\nl
4663 & 12:29:59.27 &  8:00:18.9 & 174.3 &  91.6 & 20.38 & 1.87 &
-0.10$\pm$0.08 &   1194$\pm$ 25 & 4&  1194$\pm$ 25\nl
4682 & 12:29:55.55 &  8:00:20.2 & 119.4 &  85.4 & 21.42 & 1.55 &
-0.86$\pm$0.09 &    900$\pm$ 50 & 4&   900$\pm$ 50\nl
4721 & 12:29:51.42 &  8:00:23.2 &  59.0 &  77.6 & 21.32 & 1.77 &
-0.34$\pm$0.13 &    386$\pm$131 & 1&   386$\pm$131\nl
4731 & 12:29:42.09 &  8:00:23.5 &  82.0 & 279.2 & 19.96 & 1.43 &
-1.14$\pm$0.10 &    698$\pm$ 57 & S&   698$\pm$ 57\nl
4780 & 12:29:53.32 &  8:00:26.4 &  87.4 &  79.5 & 19.52 & 1.95 & 
0.09$\pm$0.10 &    971$\pm$ 45 & S&   971$\pm$ 45\nl
4834 & 12:29:44.50 &  8:00:30.1 &  49.2 & 293.5 & 20.24 & 1.47 &
-1.04$\pm$0.09 &   1221$\pm$ 50 & 4&  1221$\pm$ 50\nl
4852 & 12:29:56.33 &  8:00:31.1 & 132.2 &  81.1 & 21.13 & 2.04 & 
0.30$\pm$0.11 &   1127$\pm$ 25 & 4&  1127$\pm$ 25\nl
4862 & 12:29:42.19 &  8:00:31.7 &  82.2 & 285.0 & 21.07 & 1.73 &
-0.42$\pm$0.12 &   1028$\pm$108 & 1&  1028$\pm$108\nl
4864 & 12:29:57.01 &  8:00:31.9 & 142.3 &  81.4 & 20.63 & 1.99 & 
0.18$\pm$0.09 &    385$\pm$161 & 1&   312$\pm$ 24\nl
     &             &            &       &       &       &     
&                &    311$\pm$ 25 & 4&             \nl
4882 & 12:29:28.46 &  8:00:32.9 & 284.3 & 274.6 & 20.68 & 1.35 &
-1.32$\pm$0.07 &    564$\pm$ 62 & 3&   564$\pm$ 62\nl
4959 & 12:29:55.58 &  8:00:37.5 & 122.5 &  77.3 & 21.38 & 1.33 &
-1.38$\pm$0.10 &   1449$\pm$ 44 & S&  1449$\pm$ 44\nl
5003 & 12:29:43.05 &  8:00:40.2 &  73.0 & 294.0 & 20.72 & 1.43 &
-1.13$\pm$0.12 &    770$\pm$ 25 & 4&   770$\pm$ 25\nl
5018 & 12:29:43.78 &  8:00:41.4 &  63.8 & 299.0 & 20.70 & 2.04 & 
0.30$\pm$0.11 &    614$\pm$ 50 & 4&   614$\pm$ 50\nl
5090 & 12:29:41.97 &  8:00:45.6 &  89.9 & 293.1 & 19.83 & 1.61 &
-0.70$\pm$0.12 &    582$\pm$ 46 & S&   582$\pm$ 46\nl
5097 & 12:29:37.74 &  8:00:46.3 & 149.8 & 283.8 & 20.76 & 2.18 & 
0.62$\pm$0.11 &    864$\pm$ 70 & 1&   824$\pm$ 40\nl
     &             &            &       &       &       &     
&                &    804$\pm$ 50 & 4&             \nl
5182 & 12:29:58.80 &  8:00:50.4 & 172.0 &  76.6 & 20.94 & 1.57 &
-0.79$\pm$0.08 &   1292$\pm$ 84 & 1&  1292$\pm$ 84\nl
5213 & 12:29:40.13 &  8:00:52.0 & 117.6 & 290.7 & 20.89 & 1.64 &
-0.65$\pm$0.10 &    954$\pm$ 69 & 1&   954$\pm$ 69\nl
5217 & 12:29:52.92 &  8:00:52.5 &  90.3 &  62.3 & 20.60 & 1.32 &
-1.39$\pm$0.08 &    553$\pm$ 50 & 4&   553$\pm$ 50\nl
5281 & 12:29:37.49 &  8:00:56.7 & 156.2 & 287.3 & 21.28 & 1.38 &
-1.25$\pm$0.10 &   1400$\pm$131 & 1&  1400$\pm$131\nl
5282 & 12:30:03.57 &  8:00:56.9 & 242.5 &  79.0 & 21.90 & 1.95 & 
0.08$\pm$0.13 &   1356$\pm$ 68 & 3&  1356$\pm$ 68\nl
5400 & 12:29:43.70 &  8:01:03.6 &  77.9 & 313.0 & 20.70 & 1.32 &
-1.40$\pm$0.07 &    918$\pm$113 & 1&   918$\pm$113\nl
5456 & 12:29:45.03 &  8:01:06.9 &  67.6 & 326.6 & 19.26 & 1.39 &
-1.23$\pm$0.10 &    882$\pm$ 65 & M&   882$\pm$ 65\nl
5501 & 12:29:34.11 &  8:01:10.1 & 208.2 & 286.7 & 21.63 & 1.48 &
-1.01$\pm$0.12 &    799$\pm$158 & 1&   799$\pm$158\nl
5561 & 12:29:29.39 &  8:01:13.6 & 276.9 & 283.2 & 20.82 & 1.39 &
-1.22$\pm$0.08 &    903$\pm$ 48 & S&   903$\pm$ 48\nl
5564 & 12:30:07.38 &  8:01:14.2 & 301.6 &  77.8 & 21.26 & 1.37 &
-1.29$\pm$0.06 &    862$\pm$ 19 & Z&   862$\pm$ 19\nl
5629 & 12:29:44.44 &  8:01:17.7 &  81.4 & 325.6 & 21.09 & 1.36 &
-1.29$\pm$0.12 &    522$\pm$ 52 & S&   522$\pm$ 52\nl
5694 & 12:29:57.41 &  8:01:21.0 & 162.7 &  64.3 & 21.89 & 2.02 & 
0.25$\pm$0.10 &   1185$\pm$130 & 1&  1185$\pm$130\nl
5707 & 12:29:45.17 &  8:01:21.4 &  79.1 & 333.6 & 21.47 & 1.44 &
-1.11$\pm$0.12 &   1712$\pm$ 30 & Z&  1712$\pm$ 30\nl
5750 & 12:29:44.62 &  8:01:24.4 &  85.7 & 329.6 & 21.50 & 1.47 &
-1.05$\pm$0.10 &   1063$\pm$ 89 & Z&  1063$\pm$ 89\nl
5821 & 12:30:09.22 &  8:01:29.8 & 331.7 &  76.2 & 21.73 & 1.30 &
-1.44$\pm$0.07 &    928$\pm$ 59 & 3&   928$\pm$ 59\nl
5856 & 12:29:41.64 &  8:01:32.1 & 119.7 & 313.0 & 21.03 & 1.13 &
-1.84$\pm$0.13 &   1059$\pm$104 & 1&  1053$\pm$ 65\nl
     &             &            &       &       &       &     
&                &   1050$\pm$ 84 & Z&             \nl
6051 & 12:29:39.09 &  8:01:44.7 & 156.9 & 306.9 & 20.94 & 1.31 &
-1.41$\pm$0.07 &    973$\pm$ 88 & 1&   927$\pm$ 40\nl
     &             &            &       &       &       &     
&                &   1000$\pm$110 & 1&             \nl
     &             &            &       &       &       &     
&                &    897$\pm$ 50 & 4&             \nl
6092 & 12:29:52.17 &  8:01:47.4 & 118.8 &  35.4 & 20.92 & 1.54 &
-0.88$\pm$0.09 &    804$\pm$147 & 1&   804$\pm$147\nl
6108 & 12:29:57.19 &  8:01:48.4 & 173.6 &  55.7 & 21.49 & 1.42 &
-1.15$\pm$0.07 &    913$\pm$ 39 & Z&   913$\pm$ 39\nl
6164 & 12:29:44.79 &  8:01:52.6 & 110.0 & 338.2 & 19.79 & 1.65 &
-0.61$\pm$0.09 &    426$\pm$ 30 & S&   426$\pm$ 30\nl
6177 & 12:30:02.82 &  8:01:53.6 & 249.3 &  65.6 & 21.02 & 1.45 &
-1.10$\pm$0.06 &   1187$\pm$ 86 & 1&  1187$\pm$ 86\nl
6198 & 12:30:05.76 &  8:01:55.4 & 290.3 &  68.8 & 20.75 & 1.51 &
-0.94$\pm$0.07 &    306$\pm$ 53 & 3&   306$\pm$ 53\nl
6220 & 12:29:37.87 &  8:01:57.3 & 179.0 & 306.7 & 21.50 & 1.32 &
-1.39$\pm$0.12 &   1213$\pm$145 & 1&  1213$\pm$145\nl
6231 & 12:29:49.09 &  8:01:58.0 & 110.0 &  12.2 & 20.77 & 1.82 &
-0.22$\pm$0.09 &   1119$\pm$ 87 & 1&  1100$\pm$ 36\nl
     &             &            &       &       &       &     
&                &   1046$\pm$ 50 & Z&             \nl
     &             &            &       &       &       &     
&                &   1182$\pm$ 65 & M&             \nl
6284 & 12:29:57.92 &  8:02:02.4 & 190.5 &  54.1 & 19.44 & 1.57 &
-0.81$\pm$0.10 &    569$\pm$ 54 & S&   569$\pm$ 54\nl
6294 & 12:29:33.84 &  8:02:03.1 & 232.5 & 299.0 & 21.02 & 1.64 &
-0.63$\pm$0.09 &   1034$\pm$ 84 & S&  1034$\pm$ 84\nl
6344 & 12:29:40.14 &  8:02:07.0 & 160.1 & 316.7 & 20.89 & 2.01 & 
0.23$\pm$0.11 &   1266$\pm$ 60 & 1&  1257$\pm$ 34\nl
     &             &            &       &       &       &     
&                &   1337$\pm$ 79 & 1&             \nl
     &             &            &       &       &       &     
&                &   1220$\pm$ 50 & Z&             \nl
6357 & 12:29:52.67 &  8:02:07.4 & 139.6 &  33.1 & 20.47 & 1.26 &
-1.53$\pm$0.09 &    958$\pm$ 29 & Z&   958$\pm$ 29\nl
6388 & 12:29:55.68 &  8:02:10.1 & 170.0 &  45.4 & 20.18 & 1.36 &
-1.30$\pm$0.07 &   1212$\pm$ 24 & Z&  1212$\pm$ 24\nl
6394 & 12:29:35.57 &  8:02:10.4 & 214.4 & 304.0 & 21.40 & 1.42 &
-1.16$\pm$0.11 &    760$\pm$ 85 & Z&   760$\pm$ 85\nl
6427 & 12:29:45.01 &  8:02:13.0 & 128.1 & 343.0 & 21.11 & 1.79 &
-0.28$\pm$0.10 &   1141$\pm$ 50 & S&  1141$\pm$ 50\nl
6476 & 12:30:15.54 &  8:02:15.9 & 434.5 &  73.3 & 20.94 & 2.23 & 
0.74$\pm$0.12 &    966$\pm$ 36 & Z&   966$\pm$ 36\nl
6479 & 12:29:41.04 &  8:02:16.1 & 158.5 & 322.5 & 21.60 & 1.94 & 
0.07$\pm$0.10 &   1124$\pm$132 & 1&  1124$\pm$132\nl
6485 & 12:29:55.03 &  8:02:16.8 & 168.3 &  41.4 & 21.07 & 1.53 &
-0.89$\pm$0.09 &    510$\pm$ 82 & Z&   510$\pm$ 82\nl
6511 & 12:29:57.40 &  8:02:18.5 & 194.5 &  48.9 & 21.76 & 1.95 & 
0.08$\pm$0.09 &   1029$\pm$127 & 1&  1029$\pm$127\nl
6519 & 12:29:42.42 &  8:02:19.2 & 149.5 & 329.4 & 21.03 & 1.26 &
-1.54$\pm$0.09 &    347$\pm$ 79 & 1&   347$\pm$ 79\nl
6520 & 12:29:49.28 &  8:02:19.0 & 131.1 &  11.4 & 20.06 & 1.86 &
-0.14$\pm$0.10 &    607$\pm$ 57 & S&   607$\pm$ 57\nl
6564 & 12:29:43.31 &  8:02:22.7 & 146.4 & 334.6 & 20.03 & 1.34 &
-1.34$\pm$0.11 &   1077$\pm$ 31 & S&  1077$\pm$ 31\nl
6615 & 12:30:14.00 &  8:02:27.0 & 416.1 &  70.9 & 20.48 & 1.47 &
-1.05$\pm$0.11 &   1923$\pm$ 49 & Z&  1923$\pm$ 49\nl
6647 & 12:29:38.13 &  8:02:29.2 & 196.9 & 314.8 & 21.33 & 1.30 &
-1.44$\pm$0.09 &    672$\pm$125 & 1&   672$\pm$125\nl
6696 & 12:29:53.98 &  8:02:34.0 & 172.5 &  33.7 & 20.08 & 1.59 &
-0.76$\pm$0.09 &    550$\pm$ 52 & S&   559$\pm$ 21\nl
     &             &            &       &       &       &     
&                &    561$\pm$ 23 & Z&             \nl
6701 & 12:29:26.20 &  8:02:33.9 & 347.9 & 294.4 & 20.97 & 2.00 & 
0.19$\pm$0.06 &   1092$\pm$141 & Z&  1303$\pm$ 50\nl
     &             &            &       &       &       &     
&                &   1334$\pm$ 54 & 3&             \nl
6721 & 12:29:39.90 &  8:02:35.9 & 184.5 & 322.1 & 21.09 & 1.73 &
-0.44$\pm$0.08 &   1180$\pm$ 45 & S&  1209$\pm$ 40\nl
     &             &            &       &       &       &     
&                &   1329$\pm$ 91 & 1&             \nl
6748 & 12:29:40.78 &  8:02:37.8 & 178.3 & 325.8 & 20.21 & 1.53 &
-0.90$\pm$0.09 &    817$\pm$ 20 & Z&   817$\pm$ 20\nl
6872 & 12:29:41.75 &  8:02:46.1 & 177.8 & 331.1 & 20.15 & 1.46 &
-1.08$\pm$0.10 &    870$\pm$ 41 & S&   870$\pm$ 41\nl
6905 & 12:29:29.53 &  8:02:48.9 & 310.8 & 300.7 & 22.18 & 1.67 &
-0.58$\pm$0.12 &   1065$\pm$ 69 & 3&  1065$\pm$ 69\nl
6989 & 12:29:54.53 &  8:02:56.2 & 195.5 &  32.1 & 20.61 & 1.75 &
-0.38$\pm$0.07 &   1009$\pm$ 24 & Z&  1028$\pm$ 21\nl
     &             &            &       &       &       &     
&                &   1154$\pm$ 87 & 1&             \nl
     &             &            &       &       &       &     
&                &   1071$\pm$ 50 & S&             \nl
7028 & 12:30:12.03 &  8:02:59.2 & 401.1 &  65.2 & 21.40 & 1.38 &
-1.24$\pm$0.07 &   1548$\pm$ 39 & Z&  1548$\pm$ 39\nl
7043 & 12:29:29.23 &  8:02:59.7 & 320.2 & 301.9 & 20.47 & 1.78 &
-0.31$\pm$0.08 &    808$\pm$ 67 & Z&   808$\pm$ 67\nl
7095 & 12:30:01.57 &  8:03:05.4 & 272.1 &  50.0 & 21.43 & 1.56 &
-0.83$\pm$0.08 &   1285$\pm$ 80 & Z&  1285$\pm$ 80\nl
7110 & 12:29:41.98 &  8:03:06.6 & 194.6 & 334.9 & 21.82 & 1.35 &
-1.32$\pm$0.09 &   1242$\pm$138 & 1&  1242$\pm$138\nl
7157 & 12:29:36.85 &  8:03:11.0 & 240.4 & 318.7 & 21.52 & 1.57 &
-0.81$\pm$0.10 &    804$\pm$ 90 & 1&   804$\pm$ 90\nl
7197 & 12:29:40.98 &  8:03:13.6 & 207.5 & 332.0 & 20.94 & 1.50 &
-0.96$\pm$0.11 &    782$\pm$ 50 & S&   782$\pm$ 50\nl
7281 & 12:29:53.17 &  8:03:19.4 & 206.5 &  23.9 & 21.43 & 1.25 &
-1.55$\pm$0.10 &    389$\pm$123 & Z&   389$\pm$123\nl
7340 & 12:29:49.93 &  8:03:24.5 & 197.2 &  10.4 & 20.91 & 1.77 &
-0.33$\pm$0.08 &   1067$\pm$ 29 & Z&  1079$\pm$ 28\nl
     &             &            &       &       &       &     
&                &   1308$\pm$124 & S&             \nl
7364 & 12:29:31.27 &  8:03:26.8 & 311.3 & 309.1 & 21.36 & 1.36 &
-1.31$\pm$0.08 &   1522$\pm$100 & Z&  1522$\pm$100\nl
7382 & 12:29:41.39 &  8:03:29.0 & 218.4 & 335.3 & 21.12 & 1.56 &
-0.83$\pm$0.09 &    860$\pm$131 & 1&   860$\pm$131\nl
7390 & 12:29:40.01 &  8:03:29.8 & 228.6 & 330.7 & 21.48 & 1.73 &
-0.43$\pm$0.09 &    301$\pm$ 89 & 1&   301$\pm$ 89\nl
7399 & 12:29:33.87 &  8:03:30.5 & 285.0 & 314.6 & 20.35 & 1.40 &
-1.20$\pm$0.07 &   1005$\pm$ 44 & S&  1005$\pm$ 44\nl
7430 & 12:29:25.60 &  8:03:31.8 & 383.1 & 301.7 & 20.76 & 1.41 &
-1.18$\pm$0.08 &    862$\pm$ 66 & Z&   895$\pm$ 48\nl
     &             &            &       &       &       &     
&                &    934$\pm$ 70 & 3&             \nl
7449 & 12:29:19.99 &  8:03:33.0 & 456.5 & 296.4 & 20.75 & 1.65 &
-0.62$\pm$0.05 &    724$\pm$ 65 & Z&   724$\pm$ 65\nl
7458 & 12:29:44.76 &  8:03:34.1 & 207.7 & 348.6 & 20.75 & 1.84 &
-0.18$\pm$0.09 &    807$\pm$ 57 & S&   807$\pm$ 57\nl
7478 & 12:29:51.98 &  8:03:35.8 & 215.7 &  17.8 & 21.10 & 1.47 &
-1.05$\pm$0.08 &    791$\pm$154 & 1&   791$\pm$154\nl
7531 & 12:29:25.10 &  8:03:38.8 & 393.1 & 302.0 & 19.71 & 2.07 & 
0.38$\pm$0.07 &    818$\pm$ 72 & Z&   818$\pm$ 72\nl
7548 & 12:29:41.51 &  8:03:40.7 & 228.5 & 337.0 & 20.57 & 1.62 &
-0.68$\pm$0.08 &   1110$\pm$ 76 & 1&  1136$\pm$ 59\nl
     &             &            &       &       &       &     
&                &   1175$\pm$ 94 & 1&             \nl
7616 & 12:29:49.08 &  8:03:45.7 & 216.4 &   6.1 & 21.35 & 1.49 &
-0.99$\pm$0.08 &    600$\pm$ 90 & Z&   600$\pm$ 90\nl
7638 & 12:29:54.52 &  8:03:47.8 & 240.8 &  25.6 & 21.19 & 2.05 & 
0.32$\pm$0.09 &   1219$\pm$127 & 1&  1219$\pm$127\nl
7659 & 12:29:43.09 &  8:03:49.5 & 228.7 & 343.3 & 19.87 & 1.34 &
-1.34$\pm$0.10 &   1520$\pm$ 44 & Z&  1539$\pm$ 34\nl
     &             &            &       &       &       &     
&                &   1571$\pm$ 56 & S&             \nl
7702 & 12:30:11.38 &  8:03:53.1 & 418.3 &  57.9 & 21.19 & 1.88 &
-0.07$\pm$0.09 &   1388$\pm$ 58 & Z&  1388$\pm$ 58\nl
7746 & 12:29:59.53 &  8:03:57.1 & 288.2 &  38.2 & 21.29 & 1.42 &
-1.15$\pm$0.07 &    712$\pm$ 68 & Z&   755$\pm$ 60\nl
     &             &            &       &       &       &     
&                &    932$\pm$137 & 1&             \nl
7784 & 12:29:55.74 &  8:03:60.0 & 259.8 &  28.0 & 19.20 & 1.52 &
-0.92$\pm$0.10 &    868$\pm$ 51 & S&   868$\pm$ 51\nl
7798 & 12:30:05.08 &  8:04:01.1 & 347.9 &  48.5 & 20.96 & 1.39 &
-1.23$\pm$0.07 &   1340$\pm$ 39 & Z&  1340$\pm$ 39\nl
7872 & 12:29:47.24 &  8:04:07.5 & 237.0 & 359.0 & 20.33 & 1.45 &
-1.08$\pm$0.08 &    908$\pm$ 77 & Z&   908$\pm$ 77\nl
7883 & 12:30:01.27 &  8:04:09.3 & 314.1 &  40.5 & 21.32 & 1.33 &
-1.36$\pm$0.08 &   1079$\pm$108 & 1&  1079$\pm$108\nl
7886 & 12:29:28.54 &  8:04:09.4 & 369.7 & 310.3 & 20.62 & 1.58 &
-0.78$\pm$0.07 &   1208$\pm$ 57 & 3&  1229$\pm$ 28\nl
     &             &            &       &       &       &     
&                &   1236$\pm$ 33 & Z&             \nl
7889 & 12:29:58.01 &  8:04:09.1 & 284.8 &  33.1 & 18.84 & 1.58 &
-0.78$\pm$0.10 &    770$\pm$ 65 & M&   770$\pm$ 65\nl
7894 & 12:29:34.53 &  8:04:09.6 & 302.0 & 320.3 & 21.61 & 1.73 &
-0.43$\pm$0.11 &    730$\pm$ 81 & S&   730$\pm$ 81\nl
7914 & 12:29:50.60 &  8:04:11.9 & 245.7 &  10.7 & 21.20 & 1.33 &
-1.37$\pm$0.08 &   1101$\pm$ 29 & Z&  1101$\pm$ 29\nl
7938 & 12:29:44.44 &  8:04:13.4 & 247.2 & 349.3 & 20.92 & 1.44 &
-1.12$\pm$0.08 &   1298$\pm$ 96 & 1&  1261$\pm$ 44\nl
     &             &            &       &       &       &     
&                &   1251$\pm$ 50 & S&             \nl
7945 & 12:29:24.47 &  8:04:13.5 & 420.0 & 305.4 & 19.79 & 1.59 &
-0.76$\pm$0.05 &    651$\pm$ 33 & Z&   651$\pm$ 33\nl
8000 & 12:30:00.95 &  8:04:18.9 & 318.4 &  38.8 & 21.08 & 1.49 &
-1.00$\pm$0.05 &    368$\pm$ 40 & Z&   399$\pm$ 36\nl
     &             &            &       &       &       &     
&                &    544$\pm$ 86 & 1&             \nl
8090 & 12:29:45.67 &  8:04:26.4 & 257.4 & 353.8 & 20.51 & 1.46 &
-1.06$\pm$0.11 &    903$\pm$ 66 & S&   903$\pm$ 66\nl
8143 & 12:30:04.26 &  8:04:30.4 & 359.4 &  43.7 & 20.96 & 1.52 &
-0.91$\pm$0.05 &    672$\pm$109 & Z&   672$\pm$109\nl
8164 & 12:29:58.85 &  8:04:32.5 & 311.2 &  32.7 & 21.32 & 2.04 & 
0.29$\pm$0.09 &    738$\pm$ 40 & Z&   769$\pm$ 36\nl
     &             &            &       &       &       &     
&                &    916$\pm$ 86 & 1&             \nl
8165 & 12:29:56.09 &  8:04:32.4 & 291.1 &  25.9 & 20.22 & 1.39 &
-1.24$\pm$0.06 &   1027$\pm$ 47 & S&  1027$\pm$ 47\nl
8210 & 12:29:26.69 &  8:04:35.4 & 407.5 & 310.6 & 20.43 & 1.32 &
-1.40$\pm$0.07 &    576$\pm$ 88 & Z&   576$\pm$ 88\nl
8228 & 12:29:41.74 &  8:04:37.6 & 280.7 & 342.2 & 21.48 & 1.77 &
-0.33$\pm$0.11 &    910$\pm$168 & 1&   910$\pm$168\nl
8254 & 12:29:26.30 &  8:04:40.3 & 415.0 & 310.6 & 21.40 & 1.44 &
-1.12$\pm$0.07 &   1025$\pm$ 77 & 3&  1025$\pm$ 77\nl
8273 & 12:29:33.64 &  8:04:41.3 & 340.5 & 322.7 & 20.60 & 1.43 &
-1.14$\pm$0.11 &    784$\pm$ 90 & Z&   784$\pm$ 90\nl
8332 & 12:29:43.83 &  8:04:46.8 & 281.8 & 348.8 & 21.09 & 1.41 &
-1.19$\pm$0.08 &   1109$\pm$ 99 & 1&  1167$\pm$ 70\nl
     &             &            &       &       &       &     
&                &   1226$\pm$100 & Z&             \nl
8353 & 12:29:41.24 &  8:04:48.2 & 293.1 & 341.4 & 20.03 & 1.98 & 
0.15$\pm$0.10 &    928$\pm$ 40 & S&   928$\pm$ 40\nl
8357 & 12:29:36.87 &  8:04:48.3 & 319.8 & 330.3 & 20.26 & 1.45 &
-1.10$\pm$0.07 &    981$\pm$ 61 & Z&   981$\pm$ 61\nl
8384 & 12:29:47.83 &  8:04:49.9 & 279.4 &   0.9 & 21.39 & 1.41 &
-1.18$\pm$0.07 &    768$\pm$ 54 & S&   768$\pm$ 54\nl
8409 & 12:29:22.37 &  8:04:51.6 & 467.7 & 307.0 & 21.18 & 1.39 &
-1.23$\pm$0.08 &   1089$\pm$ 62 & 3&  1089$\pm$ 62\nl
8596 & 12:30:06.88 &  8:05:10.3 & 415.2 &  43.8 & 19.73 & 1.31 &
-1.42$\pm$0.11 &    888$\pm$ 31 & Z&   888$\pm$ 31\nl
8606 & 12:29:24.43 &  8:05:10.7 & 456.0 & 311.2 & 21.13 & 1.80 &
-0.26$\pm$0.06 &   1380$\pm$ 67 & 3&  1380$\pm$ 67\nl
8630 & 12:29:38.30 &  8:05:12.4 & 331.7 & 335.6 & 22.22 & 1.57 &
-0.82$\pm$0.10 &   1082$\pm$141 & 1&  1082$\pm$141\nl
8653 & 12:29:16.38 &  8:05:14.7 & 553.8 & 303.4 & 20.50 & 1.26 &
-1.53$\pm$0.09 &    744$\pm$ 44 & Z&   744$\pm$ 44\nl
8712 & 12:30:13.50 &  8:05:20.4 & 494.7 &  51.2 & 20.93 & 1.49 &
-0.98$\pm$0.06 &    817$\pm$ 63 & Z&   817$\pm$ 63\nl
8740 & 12:29:42.30 &  8:05:23.2 & 322.2 & 346.1 & 21.27 & 2.14 & 
0.53$\pm$0.07 &    380$\pm$124 & 1&   688$\pm$ 80\nl
     &             &            &       &       &       &     
&                &    913$\pm$106 & Z&             \nl
8890 & 12:29:48.29 &  8:05:39.5 & 329.2 &   2.0 & 20.41 & 1.88 &
-0.08$\pm$0.06 &    870$\pm$ 65 & S&   870$\pm$ 65\nl
8919 & 12:29:34.36 &  8:05:41.4 & 384.4 & 329.4 & 19.87 & 1.45 &
-1.08$\pm$0.13 &   1014$\pm$ 65 & Z&  1014$\pm$ 65\nl
9009 & 12:29:44.80 &  8:05:49.8 & 341.8 & 353.2 & 21.28 & 1.99 & 
0.18$\pm$0.08 &   1218$\pm$120 & 1&  1218$\pm$120\nl
9087 & 12:29:58.26 &  8:05:59.2 & 383.3 &  24.6 & 22.25 & 1.97 & 
0.12$\pm$0.13 &   1262$\pm$157 & 1&  1262$\pm$157\nl
9145 & 12:30:14.87 &  8:06:04.8 & 538.9 &  48.9 & 19.79 & 1.76 &
-0.35$\pm$0.14 &    973$\pm$ 38 & Z&   973$\pm$ 38\nl
9360 & 12:29:19.15 &  8:06:32.0 & 568.6 & 312.2 & 21.00 & 1.67 &
-0.58$\pm$0.09 &   1191$\pm$ 99 & Z&  1191$\pm$ 99\nl
9414 & 12:29:38.63 &  8:06:39.9 & 411.3 & 341.3 & 20.82 & 1.74 &
-0.41$\pm$0.06 &    664$\pm$111 & 1&   813$\pm$ 37\nl
     &             &            &       &       &       &     
&                &    832$\pm$ 40 & Z&             \nl
9527 & 12:29:32.40 &  8:06:57.4 & 464.8 & 331.1 & 20.91 & 1.58 &
-0.78$\pm$0.10 &    941$\pm$ 61 & Z&   941$\pm$ 61\nl
9666 & 12:29:51.39 &  8:07:16.8 & 430.2 &   7.7 & 20.04 & 1.74 &
-0.42$\pm$0.07 &    811$\pm$ 71 & Z&   811$\pm$ 71\nl
9991 & 12:29:58.92 &  7:58:01.0 & 212.9 & 127.5 & 19.41 & 1.27 &
-1.51$\pm$0.10 &   1156$\pm$ 65 & M&  1156$\pm$ 65\nl
9992 & 12:29:48.34 &  8:00:42.2 &  33.9 &  20.5 & 19.99 & 1.47 &
-1.04$\pm$0.10 &    795$\pm$ 65 & M&   795$\pm$ 65\nl
\multicolumn{11}{c}{} \nl
\multicolumn{11}{c}{\underbar{Foreground Stars and Background Galaxies}}
\nl
\multicolumn{11}{c}{} \nl
 902 & 12:29:30.17 &  7:55:27.1 & 383.2 & 222.3 & 20.55 & 1.15
&                &    -19$\pm$ 65 & 2&   -19$\pm$ 65\nl
1608 & 12:29:27.99 &  7:56:35.0 & 361.6 & 233.5 & 21.39 & 2.24
&                &    -40$\pm$ 48 & 2&   -40$\pm$ 48\nl
1824 & 12:29:33.92 &  7:56:52.9 & 282.9 & 225.7 & 21.61 & 1.24
&                &    242$\pm$ 63 & 2&   242$\pm$ 63\nl
2071 & 12:29:57.92 &  7:57:14.9 & 233.7 & 138.7 & 21.03 & 1.38
&                &    -25$\pm$ 73 & 1&   -25$\pm$ 73\nl
2668 & 12:30:10.37 &  7:58:06.5 & 361.1 & 110.1 & 20.72 & 1.10
&                &    228$\pm$ 72 & 3&   228$\pm$ 72\nl
2860 & 12:29:54.56 &  7:58:18.4 & 153.1 & 137.1 & 20.27 & 1.21
&                &    -10$\pm$ 15 & 4&   -10$\pm$ 15\nl
4497 & 12:30:00.58 &  8:00:07.4 & 193.7 &  91.0 & 22.18 & 1.33
&                &   6960$\pm$146 & 1&  6960$\pm$146\nl
5323 & 12:29:48.83 &  8:00:59.3 &  52.4 &  21.4 & 20.33 & 0.82
&                &   1357$\pm$ 65 & M&  1357$\pm$ 65\nl
8096 & 12:29:20.93 &  8:04:26.8 & 471.1 & 303.0 & 21.81 & 1.68
&                &     75$\pm$ 51 & 3&    75$\pm$ 51\nl
9086 & 12:29:27.33 &  8:05:59.0 & 459.9 & 319.3 & 21.80 & 1.30
&                &     10$\pm$ 54 & 3&    10$\pm$ 54\nl
9228 & 12:29:41.62 &  8:06:16.2 & 376.1 & 346.5 & 21.02 & 1.61
&                &    155$\pm$172 & 1&   155$\pm$172\nl
\multicolumn{11}{c}{} \nl
\multicolumn{11}{c}{\underbar{Discrepant Objects\tablenotemark{2}}} \nl
\multicolumn{11}{c}{} \nl
1982 & 12:29:41.67 &  7:57:07.4 & 202.8 & 205.5 & 20.89 & 1.01 &
-2.02$\pm$0.11 &    648$\pm$ 43 & Z&  2558$\pm$ 41\nl
     &             &            &       &       &       &     
&                &  25800$\pm$150 & 4&             \nl
2256 & 12:29:47.53 &  7:57:31.4 & 159.1 & 180.1 & 21.24 & 1.90 & 
0.08$\pm$0.13 &    242$\pm$ 61 & 3&   830$\pm$ 25\nl
     &             &            &       &       &       &     
&                &    954$\pm$ 28 & Z&             \nl
\enddata
\tablenotetext{1}{Key to Source codes: (M) Mould \etal (1990); (S)
Sharples \etal (1998); (Z) Zepf \etal (2000);
(1-4) as given in Table~\ref{tab1}.}
\tablenotetext{2}{Omitted from final GC sample.}
\end{deluxetable}
 
\clearpage
 
\begin{deluxetable}{cccrrrrr}
\tablewidth{0pc}
\tablenum{3}
\tablecaption{Kinematics of the M49 Globular Cluster
              System\tablenotemark{1}\label{tab3}}
\tablehead{
\colhead{$R$} &
\colhead{$\langle R\rangle$} &
\colhead{$N$} &
\colhead{$\overline{v_p}$} &
\colhead{$\sigma_p$} &
\colhead{$\Theta_0$\tablenotemark{2}} &
\colhead{$\Omega R$\tablenotemark{~2}} &
\colhead{${\sigma}_{p,r}$\tablenotemark{2}} \nl
\colhead{(arcsec)} &
\colhead{(arcsec)} &
\colhead{} &
\colhead{(km s$^{-1}$)} &
\colhead{(km s$^{-1}$)} &
\colhead{(deg)} &
\colhead{(km s$^{-1}$)} &
\colhead{(km s$^{-1}$)} }
\startdata
\multicolumn{8}{c}{} \nl
\multicolumn{8}{c}{Full Sample: 263 Clusters with 1.00 $\le$ $(C-T_1)$
$\le$ 2.25} \nl
\multicolumn{8}{c}{} \nl
~25--570 & 234 & 263 & ~973$^{+22}_{-23}$ & 313$^{+27}_{-9}$ &
   105$^{+45}_{-45}$ &  53$^{+52}_{-25}$ & 312$^{+27}_{-8}$ \nl
\multicolumn{8}{c}{} \nl
~25--150 & 110 & ~58 & ~910$^{+43}_{-54}$ & 328$^{+33}_{-32}$ &
   $-$168$^{+41}_{-42}$ & 115$^{+103}_{-60}$ & 335$^{+36}_{-36}$ \nl  
150--250 & 207 & ~87 & 1025$^{+35}_{-38}$ & 306$^{+35}_{-21}$ &
   96$^{+53}_{-48}$ &  55$^{+81}_{-47}$ & 302$^{+33}_{-22}$ \nl 
250--350 & 300 & ~69 & ~993$^{+39}_{-46}$ & 294$^{+41}_{-26}$ &
   ~74$^{+24}_{-25}$ & 223$^{+80}_{-55}$ & 307$^{+44}_{-36}$ \nl
350--570 & 418 & ~49 & ~925$^{+48}_{-51}$ & 329$^{+50}_{-48}$ &
   $-$56$^{+44}_{-38}$ & 123$^{+135}_{-78}$ & 331$^{+43}_{-48}$ \nl
\multicolumn{8}{c}{} \nl
\multicolumn{8}{c}{} \nl
\multicolumn{8}{c}{Metal-poor Sample: 158 Clusters with 1.00 $\le$
$(C-T_1)$ $<$ 1.625} \nl
\multicolumn{8}{c}{} \nl
~25--570 & 251 & 158 & ~957$^{+32}_{-28}$ & 345$^{+34}_{-18}$ &
   100$^{+37}_{-40}$ & ~93$^{+69}_{-37}$ & 342$^{+33}_{-18}$ \nl
\multicolumn{8}{c}{} \nl
~25--150 & 107 & ~26 & ~869$^{+72}_{-64}$ & 326$^{+62}_{-70}$ &
   50$^{+45}_{-35}$ &  $-$180$^{+114}_{-164}$ & 353$^{+64}_{-89}$ \nl
150--250 & 209 & ~53 & 1017$^{+55}_{-47}$ & 350$^{+39}_{-32}$ &
   96$^{+31}_{-35}$ & 162$^{+107}_{-66}$ & 324$^{+41}_{-38}$ \nl
250--350 & 296 & ~48 & ~985$^{+51}_{-58}$ & 333$^{+48}_{-47}$ &
   73$^{+30}_{-26}$ & 233$^{+116}_{-81}$ & 352$^{+42}_{-53}$ \nl  
350--570 & 415 & ~31 & ~878$^{+77}_{-58}$ & 359$^{+68}_{-69}$ &
   $-$58$^{+46}_{-50}$ & 127$^{+188}_{-143}$ & 367$^{+52}_{-70}$ \nl
\multicolumn{8}{c}{} \nl
\multicolumn{8}{c}{} \nl
\multicolumn{8}{c}{Metal-rich Sample: 105 Clusters with 1.625 $\le$
$(C-T_1)$ $\le$ 2.25} \nl
\multicolumn{8}{c}{} \nl
~25--570 & 202 & 105 & ~999$^{+31}_{-29}$ & 265$^{+30}_{-14}$ &
   195$^{+56}_{-58}$ &  12$^{+76}_{-74}$ & 265$^{+34}_{-13}$ \nl
\multicolumn{8}{c}{} \nl
~25--150 & 110 & ~32 & ~946$^{+98}_{-81}$ & 334$^{+33}_{-62}$ &
   153$^{+55}_{-54}$ & ~93$^{+141}_{-136}$ & 329$^{+35}_{-59}$ \nl
150--250 & 196 & ~34 & 1047$^{+43}_{-45}$ & 231$^{+46}_{-35}$ &
   97$^{+41}_{-35}$ & $-$86$^{+22}_{-61}$ & 199$^{+53}_{-28}$ \nl
250--350 & 311 & ~21 & 1008$^{+58}_{-59}$ & 208$^{+44}_{-16}$ &
    ... & ... & ... \nl
350--570 & 435 & ~18 & ~993$^{+61}_{-95}$ & 274$^{+57}_{-56}$ &
    130$^{+46}_{-46}$ & $-$114$^{+95}_{-152}$ & 277$^{+38}_{-74}$ \nl
\enddata
\tablenotetext{1}{All uncertainties correspond to 68\% (1-$\sigma$)
confidence
intervals.}
\tablenotetext{2}{Systemic velocity held fixed in fit of sine curve:
$v_{\rm sys}$ $\equiv$ 997 km s$^{-1}$.}
\end{deluxetable}
 
\clearpage
 
\begin{deluxetable}{lccccccccccc}
\scriptsize
\tablecolumns{12}
\tablenum{4}
\tablewidth{0pc}
\tablecaption{Global Kinematic Properties of
              GCs in M87 and M49\tablenotemark{1}\label{tab4}}
\tablehead{
 & \multicolumn{5}{c}{M87} & &
\multicolumn{5}{c}{M49\tablenotemark{2}} \\
 \cline{2-6} \cline{8-12} \\
 & $N$ & ${\sigma}_{p,r}$ & ${\Omega}R$ & ${\Theta_0}$\tablenotemark{3} &
${\Omega}R/{\sigma}_{p,r}$ & & $N$ & ${\sigma}_{p,r}$ & ${\Omega}R$ &
${\Theta_0}$\tablenotemark{4} & ${\Omega}R/{\sigma}_{p,r}$ \\
 & &(km~s$^{-1}$)&(km~s$^{-1}$)&(deg)& & & &(km~s$^{-1}$)&(km~s$^{-1}$)&(deg)& 
}
\startdata
All GCs        & 278 & 383$^{+31}_{-7}$ & 171$^{+39}_{-30}$ &
70$^{+11}_{-14}$ & 0.45$^{+0.09}_{-0.09}$ & & 253 & 316$^{+27}_{-8}$ &
48$^{+52}_{-26}$ & 109$^{+44}_{-47}$ & 0.15$^{+0.15}_{-0.08}$ \\
Metal-Poor GCs & 161 & 397$^{+36}_{-14}$ & 186$^{+58}_{-41}$ &
61$^{+17}_{-18}$ & 0.47$^{+0.13}_{-0.11}$ & & 158 & 342$^{+33}_{-18}$ &
93$^{+69}_{-37}$ & 100$^{+37}_{-40}$ & 0.27$^{+0.19}_{-0.11}$ \\
......$R \le 2R_{\rm eff}$\tablenotemark{5} &  46 & 345$^{+51}_{-42}$ &
81$^{+118}_{-85}$ & $-$14$^{+47}_{-46}$ & 0.23$^{+0.34}_{-0.25}$ & &    
&                   &                  &                  
&                        \\
Metal-Rich GCs & 117 & 365$^{+38}_{-18}$ & 155$^{+53}_{-37}$ &
81$^{+17}_{-20}$ & 0.43$^{+0.14}_{-0.12}$ & & 95 & 265$^{+34}_{-13}$ &
 $-$26$^{+64}_{-79}$ & 45$^{+55}_{-54}$ & 0.10$^{+0.27}_{-0.25}$ \\
\enddata
\tablenotetext{1}{All uncertainties correspond to 68\% confidence intervals.}
\tablenotetext{2}{The 10 metal-rich clusters between $302\arcsec\leq R\leq
                  337\arcsec$ in M49 have been omitted.}
\tablenotetext{3}{The rotation axis of the galaxy has position angle
                  $-23^{\circ}\pm10^{\circ}$ inside
                  $R\lae R_{\rm eff}$ (Davies \& Birkinshaw 1988).}
\tablenotetext{4}{The rotation axis of the galaxy has position angle
                  $59^{\circ}\pm3^{\circ}$ 
                  (Davies \& Birkinshaw 1988).}
\tablenotetext{5}{$R_{\rm eff} = 96\arcsec \simeq7$ kpc for M87 (de Vaucouleurs
                  \& Nieto 1978).}
\end{deluxetable}

\clearpage
 
\plotone{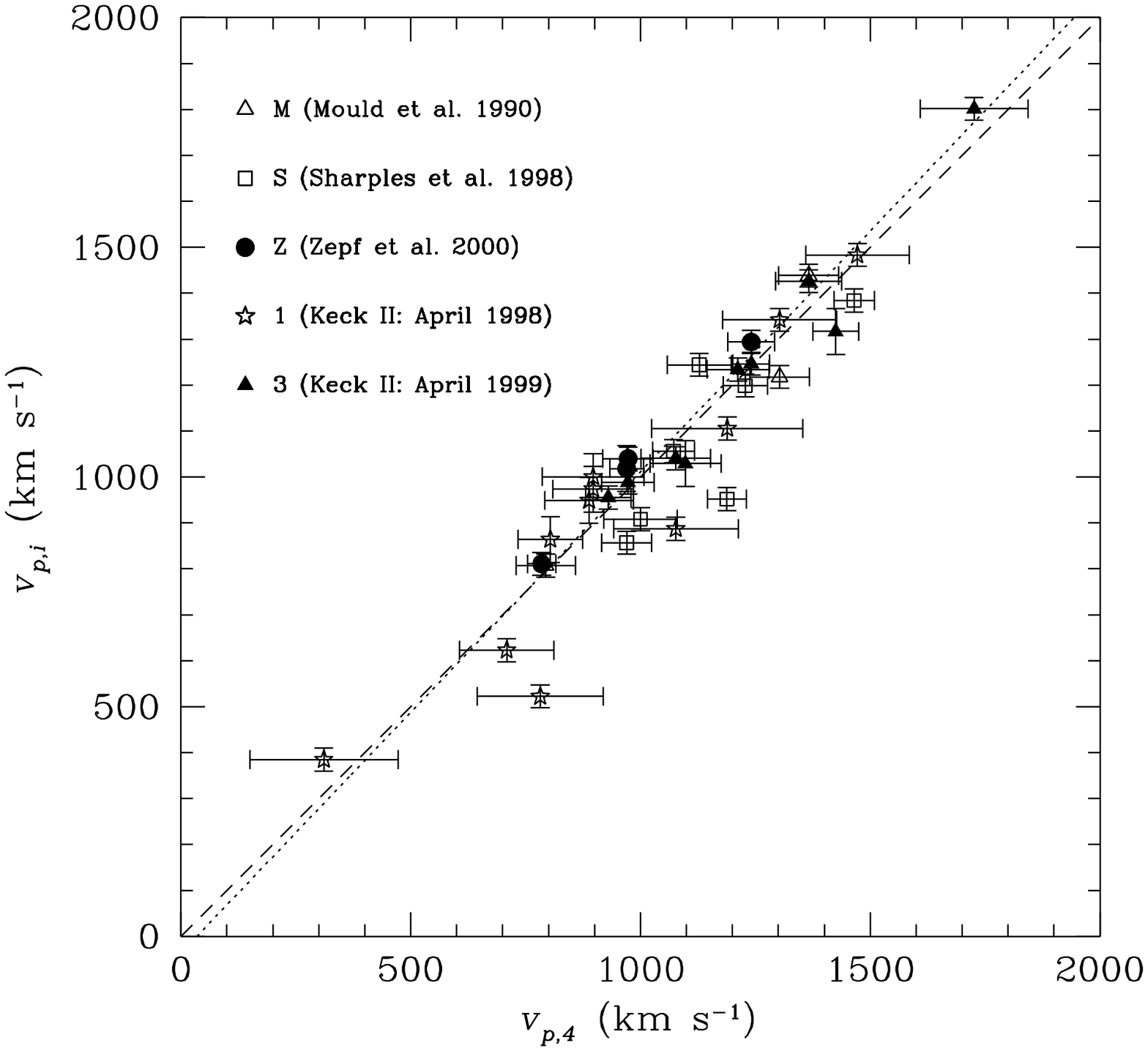}
 
\figcaption[fig01.eps]{\scriptsize Radial velocities measured for M49 globular
clusters from spectra obtained in April 2000 (i.e., run \#4) plotted against
independent velocity measurements from other datasets. The dashed line shows the
one-to-one relation, while the dotted line shows the least-squares line of best fit.
\label{fig01}}
 
\clearpage
 
\plotone{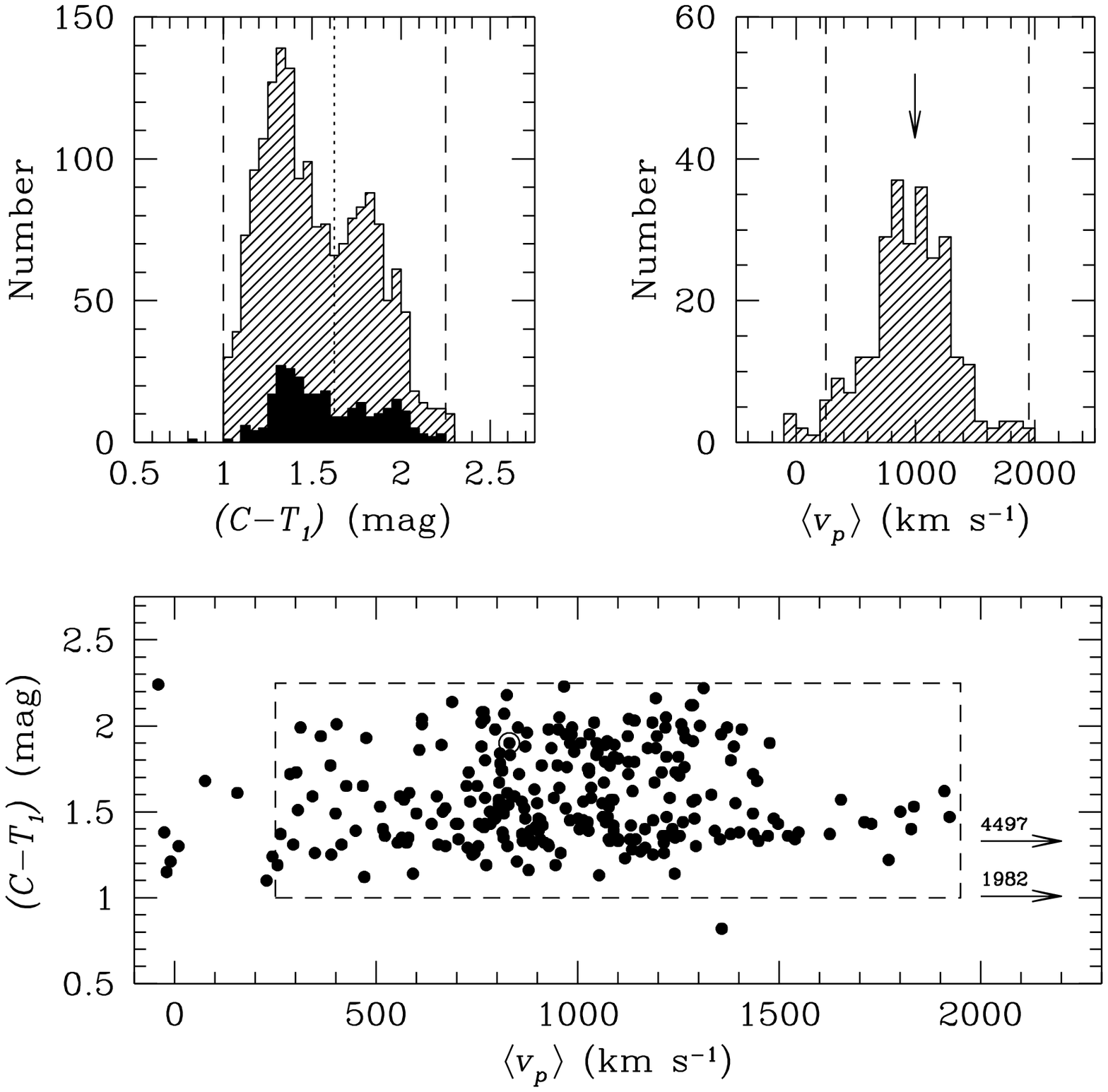}
 
\figcaption[fig02.eps]{\scriptsize {\it (Upper Left Panel)} $(C-T_1)$ colors for all 1774 g
lobular cluster candidates from the photometric survey of Geisler, Lee \& Kim 
(1996) are shown
as the dashed histogram. The solid histogram shows the color distribution
for the 276 globular cluster candidates with measured radial velocities.
The dashed lines indicate our adopted selection criteria on color:
1.00 $\le$ $(C-T_1)$ $\le$ 2.25 mag, while the dotted line shows the
color selection used to isolate metal-rich and metal-poor subsamples:
$(C-T_1)$ = 1.625.
{\it (Upper Right Panel)} Radial velocity
histogram for the 276 candidate globular clusters.
The vertical lines show our adopted selection criteria on velocity:
$250\le v \le 1950$ km s$^{-1}$.
{\it (Lower Panel)} Color versus radial velocity for our initial sample of 276 
objects (two objects which fall outside the plotted
region --- \#1982 and \#4497 --- are indicated by the arrows).
The dotted region shows the joint selection criteria on color and radial velocity.
The lone circled point in this box refers to object 2256, which was discarded from
the dynamical analysis due to discrepant radial velocity measurements.  Excluding
this object, the dashed box contains a total of 263 globular clusters.
\label{fig02}}
 
\clearpage
 
\plotone{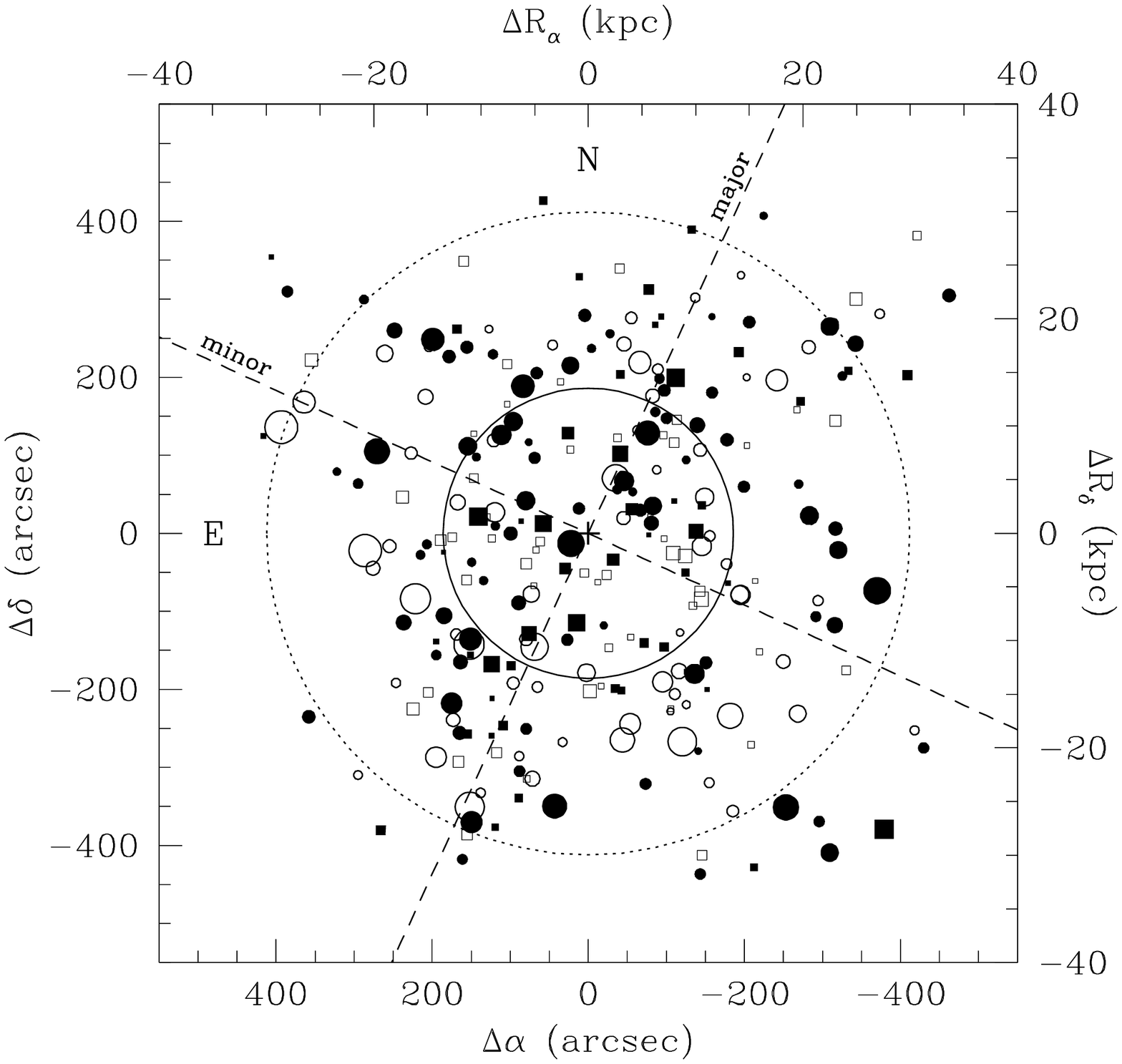}
 
\figcaption[fig03.eps]{\scriptsize Spatial distribution of the 263 globular
clusters with measured velocities. Circles and squares indicate metal-poor
and metal-rich globular clusters, respectively (see text for details).  For both
subsamples, open and filled symbols indicate objects having
positive and negative velocity residuals,
${\Delta}v_p = \langle{v_p}\rangle - v_{\rm gal}$,
respectively, where $v_{\rm gal} = 997$ km s$^{-1}$. The size of the symbol is
proportional to the absolute value of the velocity residual.  The center of the galaxy
is marked by the cross, while the solid circle shows our estimate for the galaxy's
effective radius, $R_{\rm eff} = 3\farcm1\simeq~13.5$~kpc, based on a fit (see \S4.2) 
of the surface photometry of Kim \etal (2000). The diagonal lines show
the photometric major and minor axes of the galaxy as determined by
Kim \etal (2000). Exterior to the large, dotted circle at $R = 30$ kpc,
our sample shows an apparent dearth of globular clusters with 
$\langle{v_p}\rangle \gae 1500$ km s$^{-1}$. Conclusions
about the kinematics of the globular cluster system beyond this point should be
considered provisional.
\label{fig03}}
 
\clearpage
 
\plotone{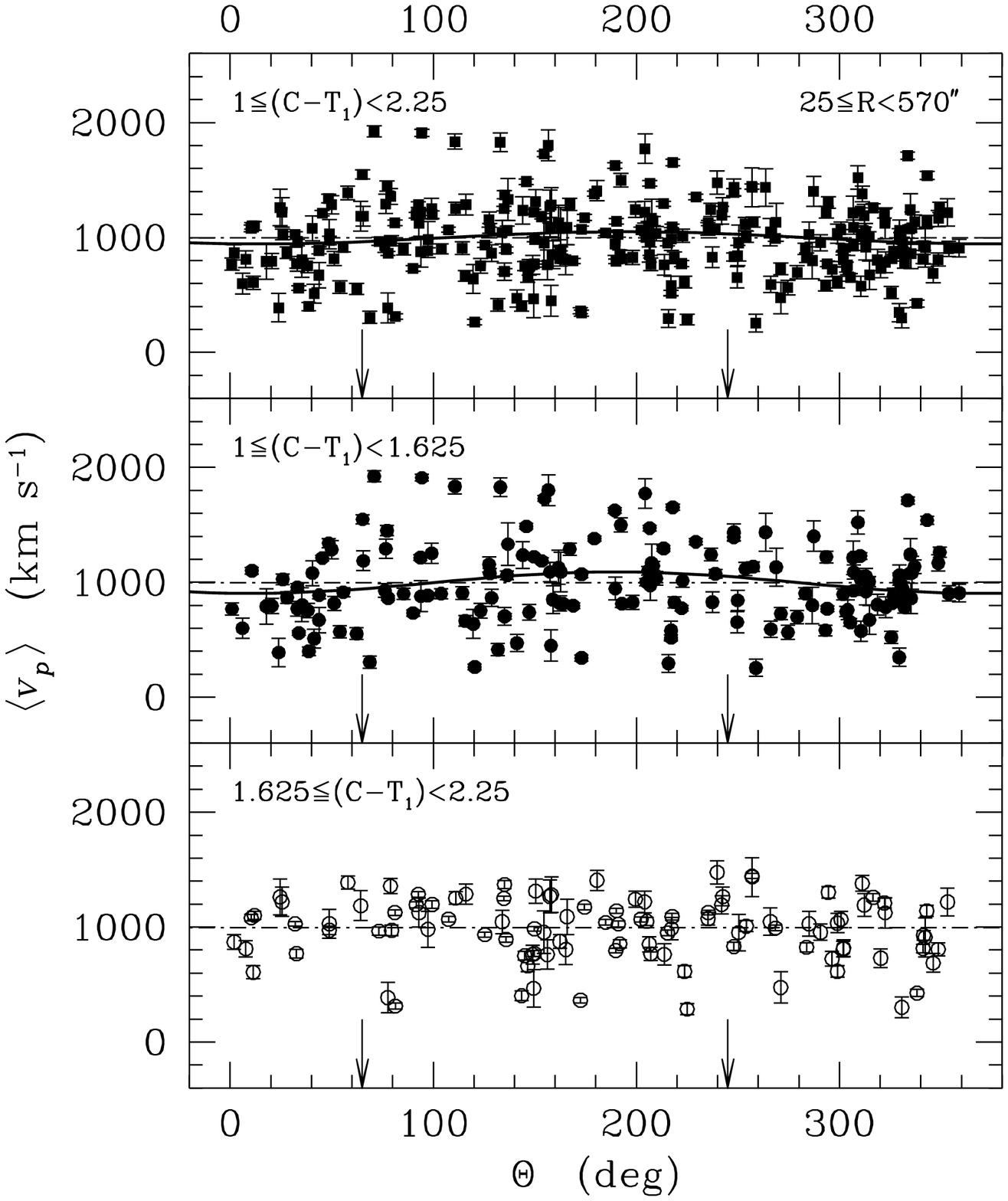}
 
\figcaption[fig04.eps]{\scriptsize {\it (Upper Panel)} Radial velocity,
$\langle v_p \rangle$, plotted against position angle, $\Theta$, for the
full sample of 263 globular clusters. {\it (Middle Panel)}
$\langle v_p \rangle$ versus $\Theta$ for the sample of 158 metal-poor
clusters. {\it (Lower Panel)} $\langle v_p \rangle$ versus
$\Theta$ for the sample of 105 metal-rich globular clusters. In all three
panels, the broken horizontal line indicates the galaxy's velocity,
$v_{\rm gal}=997$ \kms; in the top two, the best-fit sine functions from Table
\ref{tab3} are shown as bold, solid curves. The position angles of the
photometric minor axis of M49 (Kim \etal 2000) are indicated by the vertical
arrows at $\Theta = 65^{\circ}$ and $245^{\circ}$.
\label{fig04}}
 
\clearpage
 
\plotone{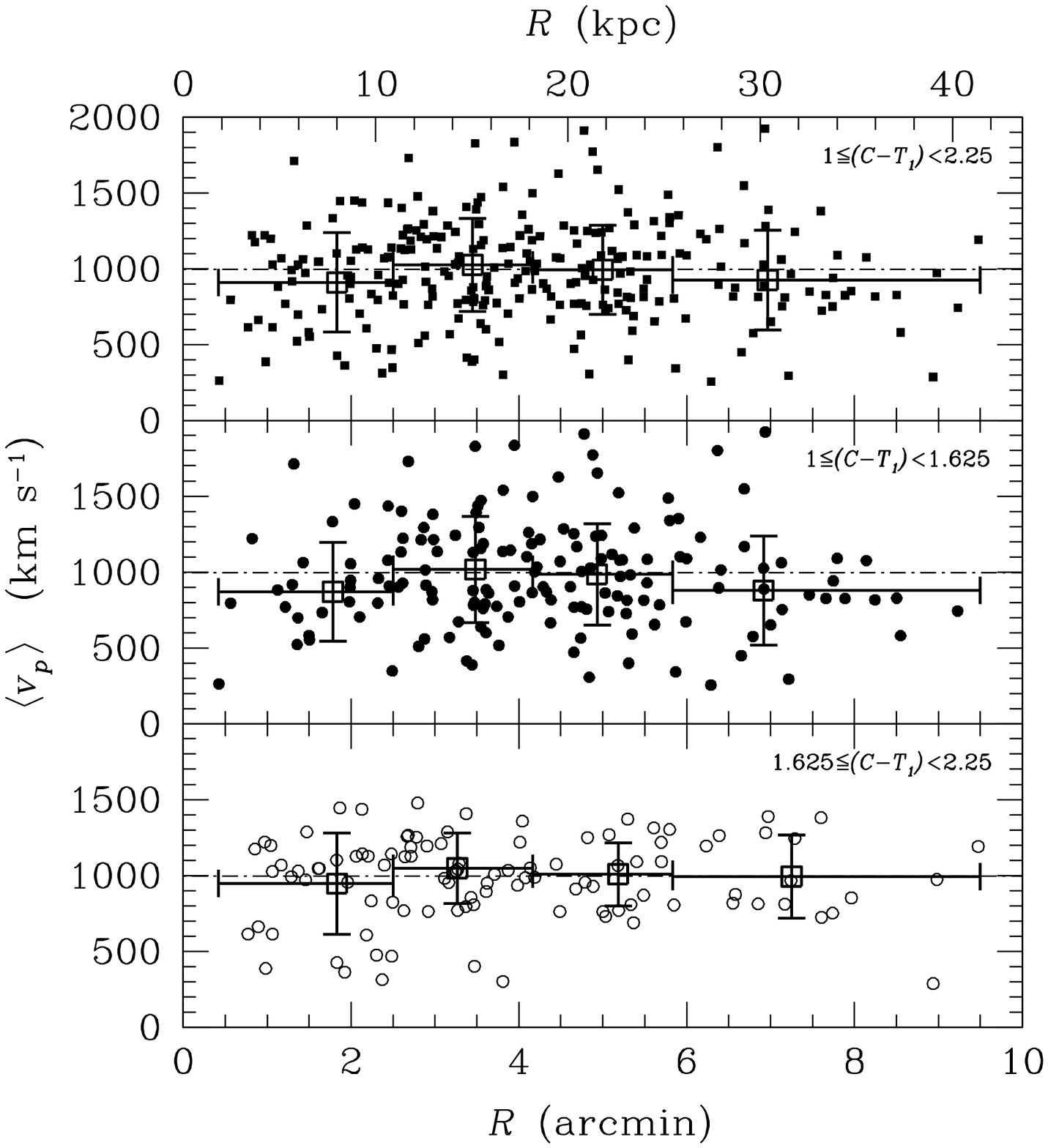}
 
\figcaption[fig05.eps]{\scriptsize {\it (Upper Panel)} Radial velocity,
$\langle v_p \rangle$, plotted against projected galactocentric distance,
$R$, for the full sample of 263 globular clusters. The broken horizontal line
shows the velocity of M49. {\it (Middle Panel)} $\langle v_p \rangle$ versus
$R$ for the sample of 158 metal-poor clusters. {\it (Lower Panel)}
$\langle v_p \rangle$ versus $R$ for the sample of 105 metal-rich globular
clusters. In each panel, the large, open squares are at the median $R$ and
average velocity, $\overline{v_p}$, of the globular clusters in the four coarse radial bins of Table
\ref{tab3}. Horizontal errorbars define the limiting radii of the bins, and
vertical errorbars represent the velocity dispersion, $\sigma_p$, in each bin.
\label{fig05}}
 
\clearpage
 
\plotone{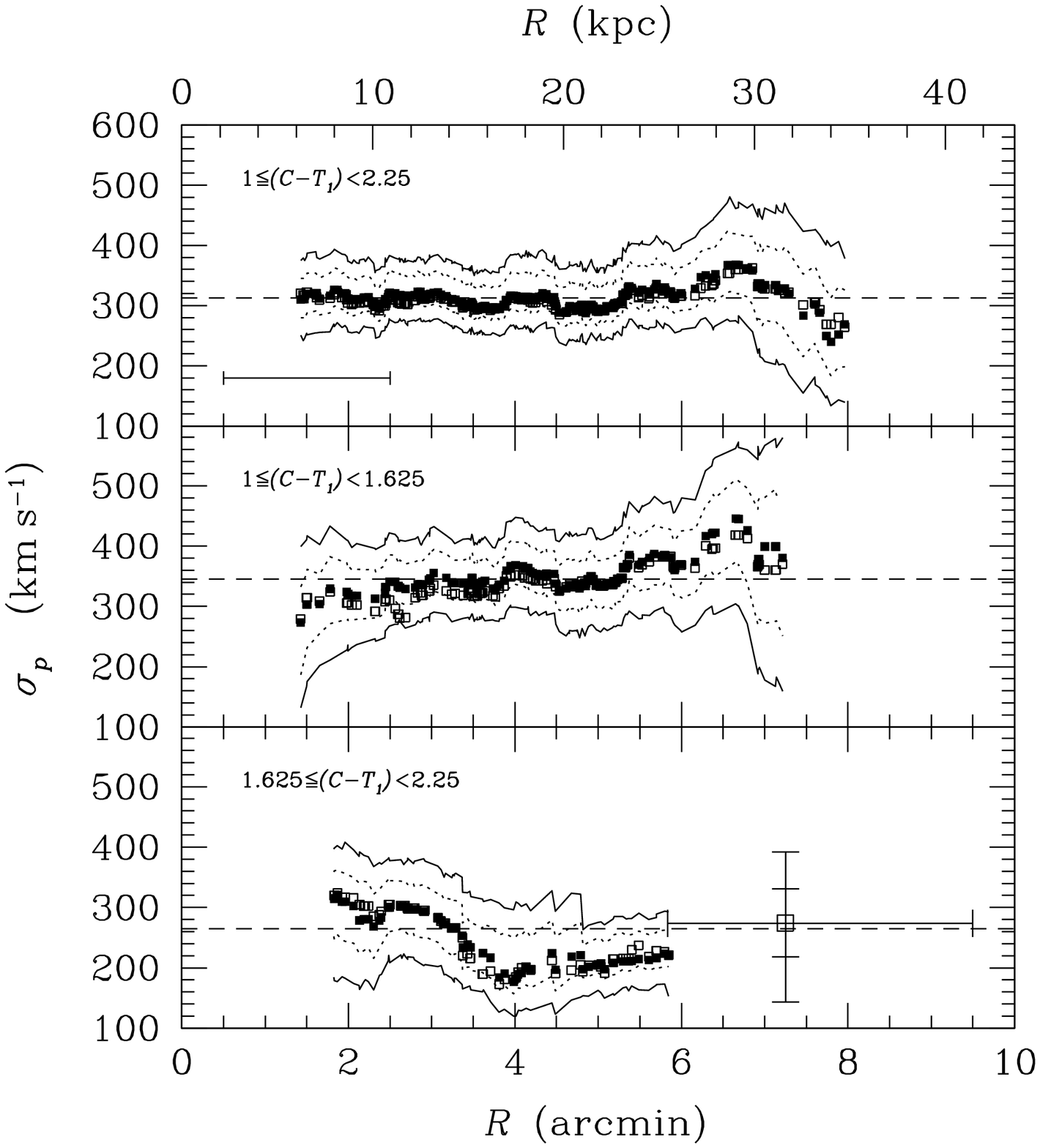}
 
\figcaption[fig06.eps]{\scriptsize {\it (Upper Panel)} Biweight estimates of velocity
dispersion as a function of galactocentric radius in the full sample of
263 M49 globular clusters. Solid squares represent the dispersion, $\sigma_p$,
about the average globular cluster velocity at each point; open squares show the dispersion,
$\sigma_{p,r}$, about the best-fit sine curve describing the variation of
$\langle v_p\rangle$ with $\Theta$ at each radius. Dotted and solid lines
around the points delineate the 68\% and 95\% confidence intervals on the
dispersion estimates. The dashed horizontal line shows the global value of
the velocity dispersion for the sample, and the horizontal errorbar shows the
width ($2\arcmin\simeq8.7$ kpc) of the sliding radial bin used to compute
this smoothed profile. {\it (Middle Panel)} As above, but for the 158
metal-poor globular clusters. {\it (Lower Panel)} As above, but for the
105 metal-rich clusters. Also shown is a large open square representing the 
single bin containing the 18 metal-rich globular clusters beyond $R = 350\arcsec\simeq25.5$ kpc.
\label{fig06}}
 
\clearpage
 
\plotone{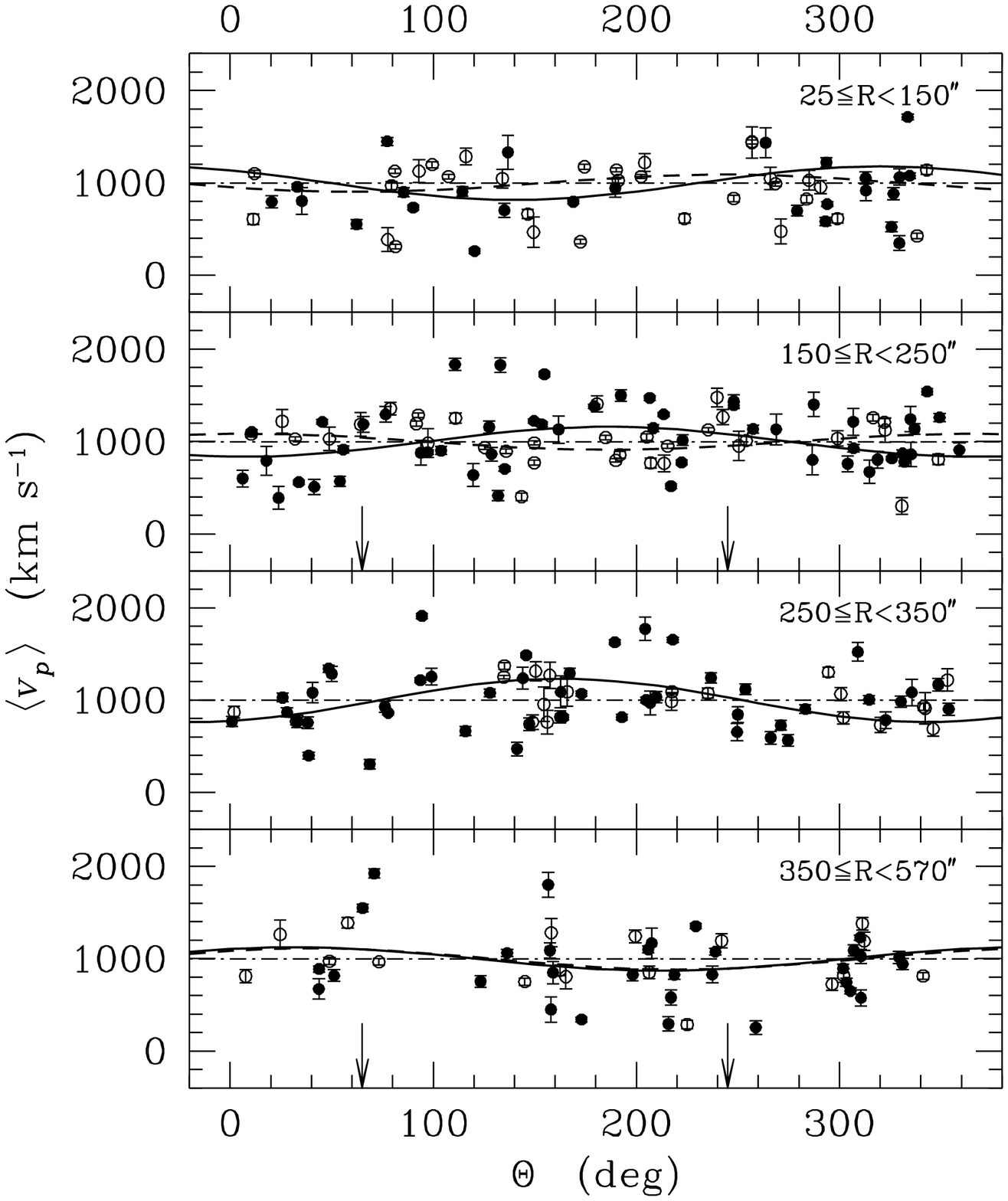}
 
\figcaption[fig07.eps]{\scriptsize Radial velocity versus projected position angle for
metal-poor and metal-rich globular clusters (filled and open points) in each of the four
coarse radial bins whose kinematics are summarized in Table \ref{tab3}.
The broken, horizontal lines in each panel represent the systemic velocity of
M49. The bold solid and dashed curves are the sine curves that best fit the
data for the metal-poor and metal-rich globular clusters (with parameters taken 
from Table~\ref{tab3}). Vertical
arrows in the second and bottom panels mark the position angle of the
galaxy's photometric minor axis ($\Theta=65^{\circ}$ and $245^{\circ}$).
\label{fig07}}
 
\clearpage
 
\plotone{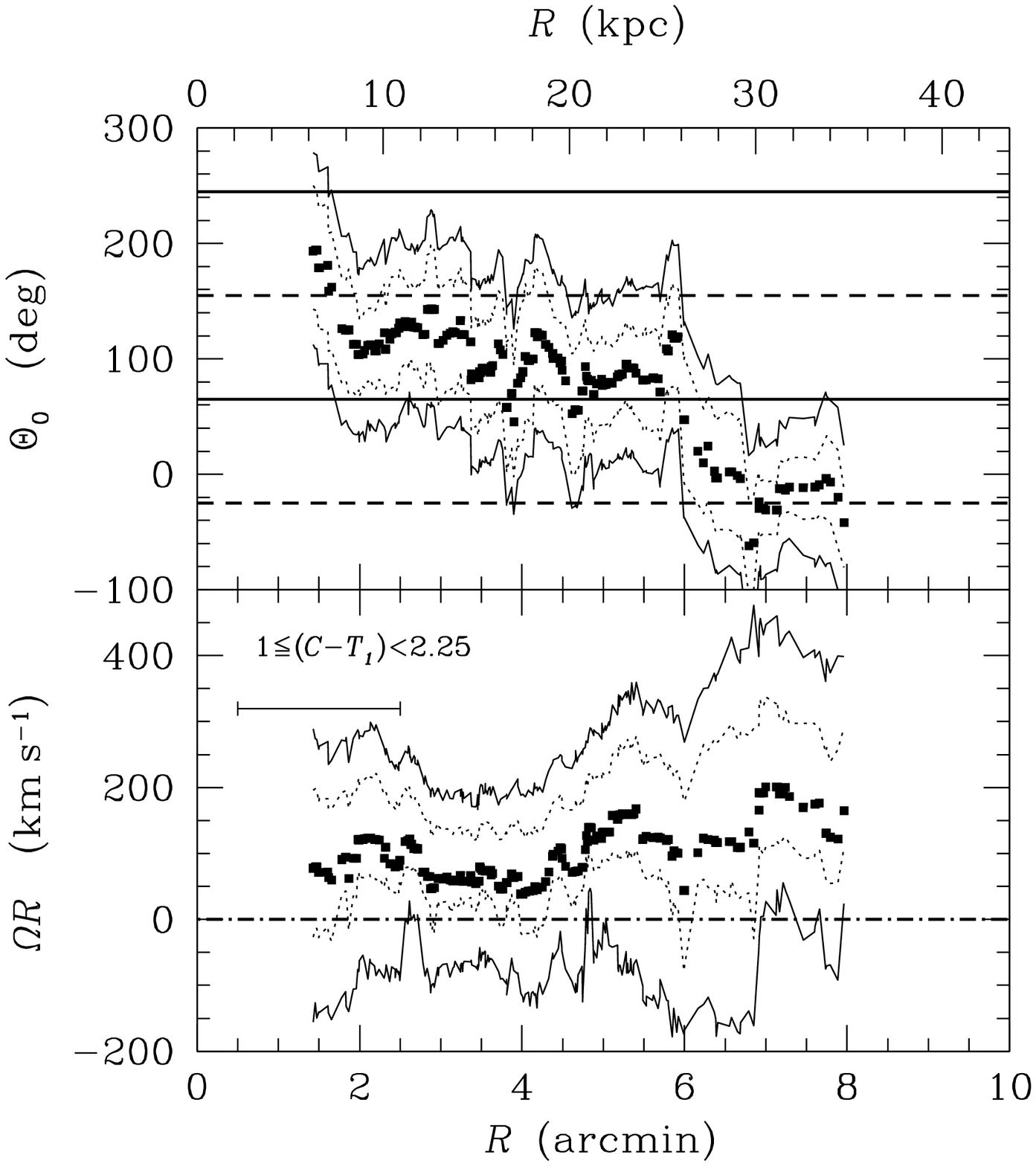}
 
\figcaption[fig08.eps]{\scriptsize {\it (Upper Panel)} Projected position angle of
the rotation axis of the globular cluster system of M49 (all metallicities), as a function
of galactocentric radius. The bold, solid lines at $\Theta_0=65^{\circ}$ and
$245^{\circ}$ represent the minor axis of the galaxy light; bold, dashed
lines at $\Theta_0=-25^{\circ}$ and $155^{\circ}$, the photometric major axis.
68\% and 95\% confidence bands for the fitted $\Theta_0$ are shown as dotted
and solid lines around the points. {\it (Lower Panel)} Projected amplitude of
rotation as a function of radius for the full globular cluster sample. The 68\% and 95\%
confidence bands are indicated as in the upper panel, and the horizontal
errorbar has a width of $2\arcmin$: the size of the sliding radial bin used
to construct the profiles.
\label{fig08}}
 
\clearpage
 
\plotone{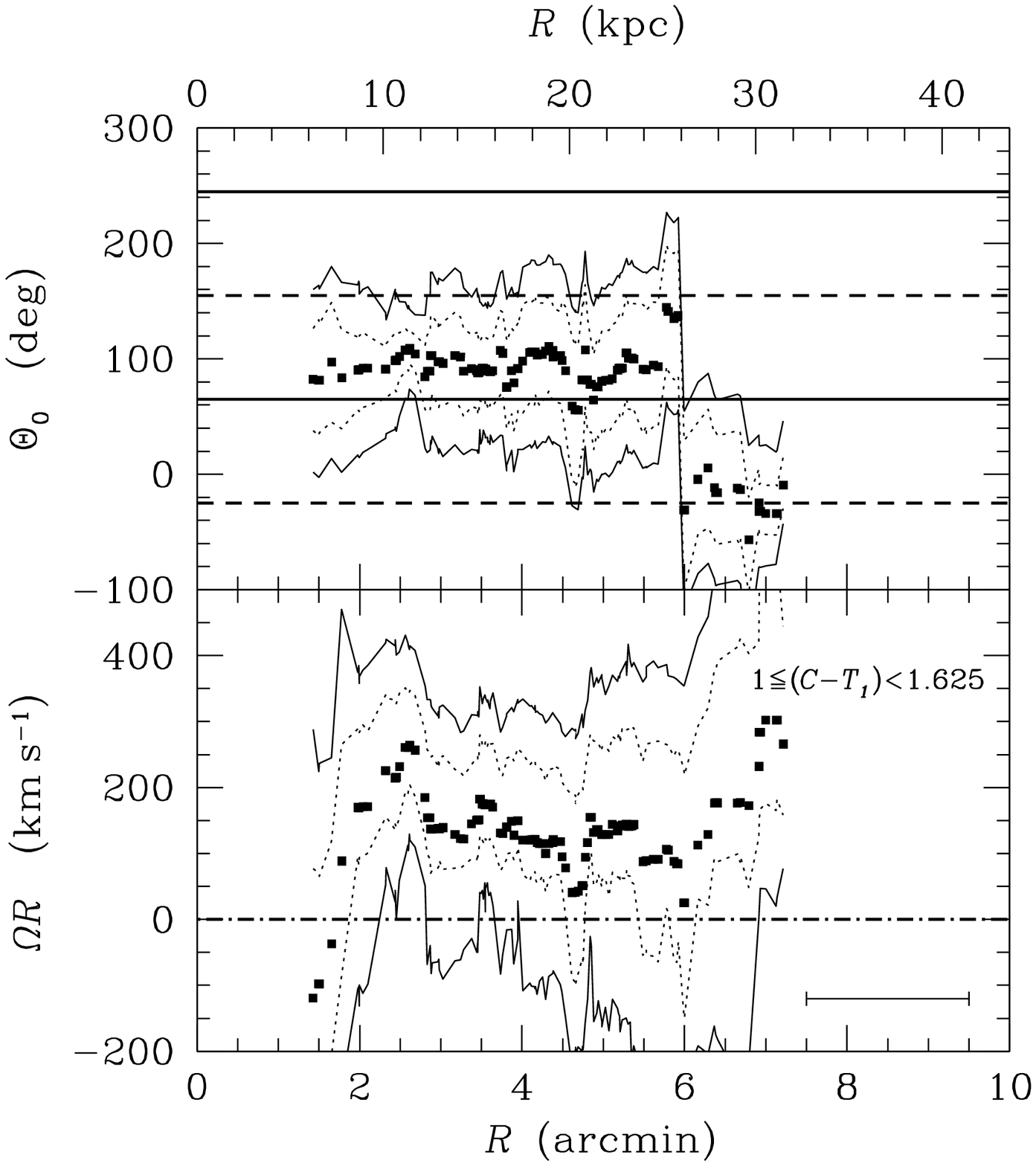}
 
\figcaption[fig09.eps]{\scriptsize Same as Figure~\ref{fig08}, but for the metal-poor globular
clusters.
\label{fig09}}
 
\clearpage
 
\plotone{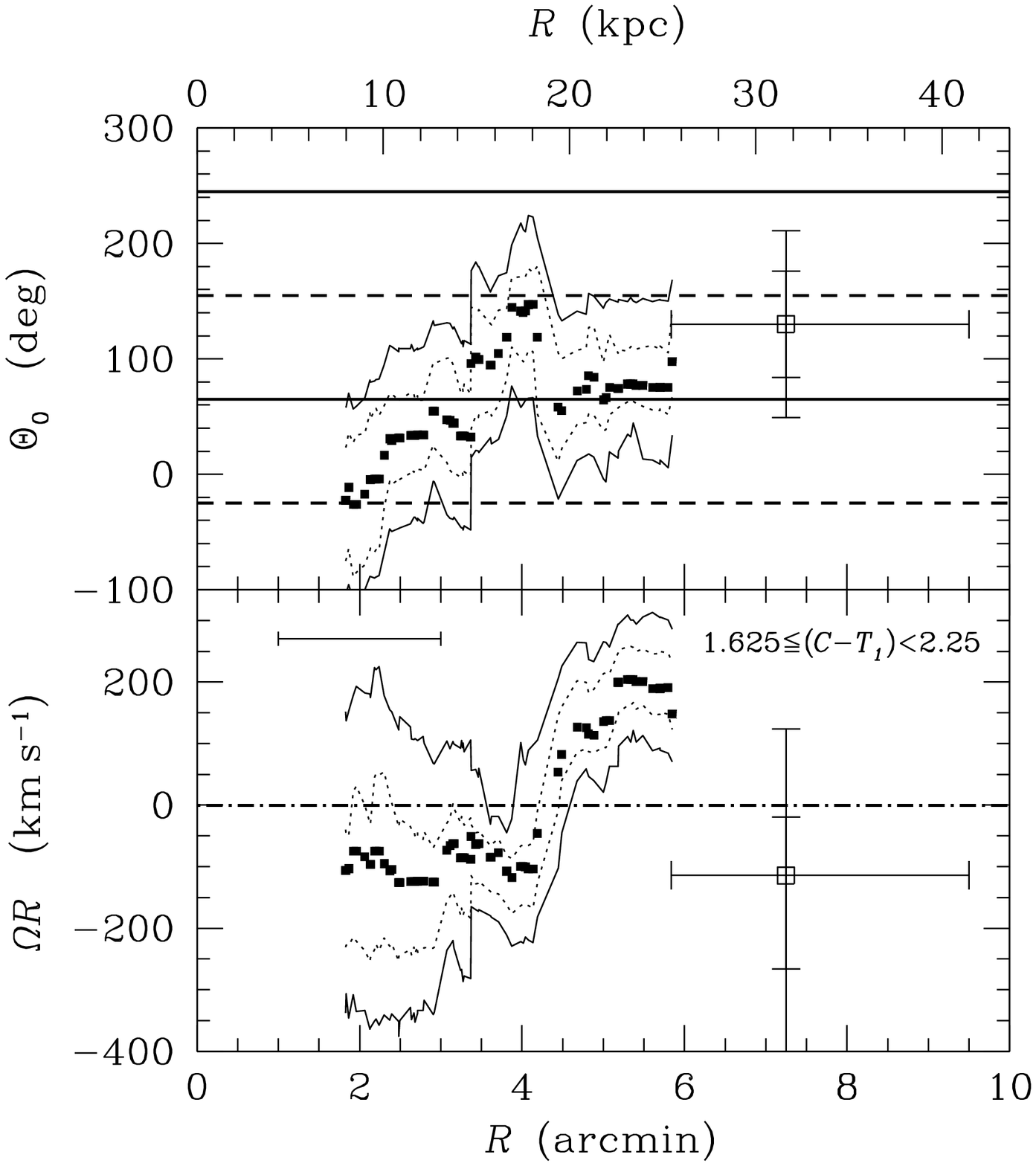}
 
\figcaption[fig10.eps]{\scriptsize Same as Figure~\ref{fig08}, but for the metal-rich globular
clusters.
Also shown, as the large open squares, are the best-fit $\Theta_0$ and
$\Omega R$, with 1- and 2-$\sigma$ errorbars, in the wide, outermost radial
bin of Table \ref{tab3}.
\label{fig10}}
 
\clearpage
 
\plotone{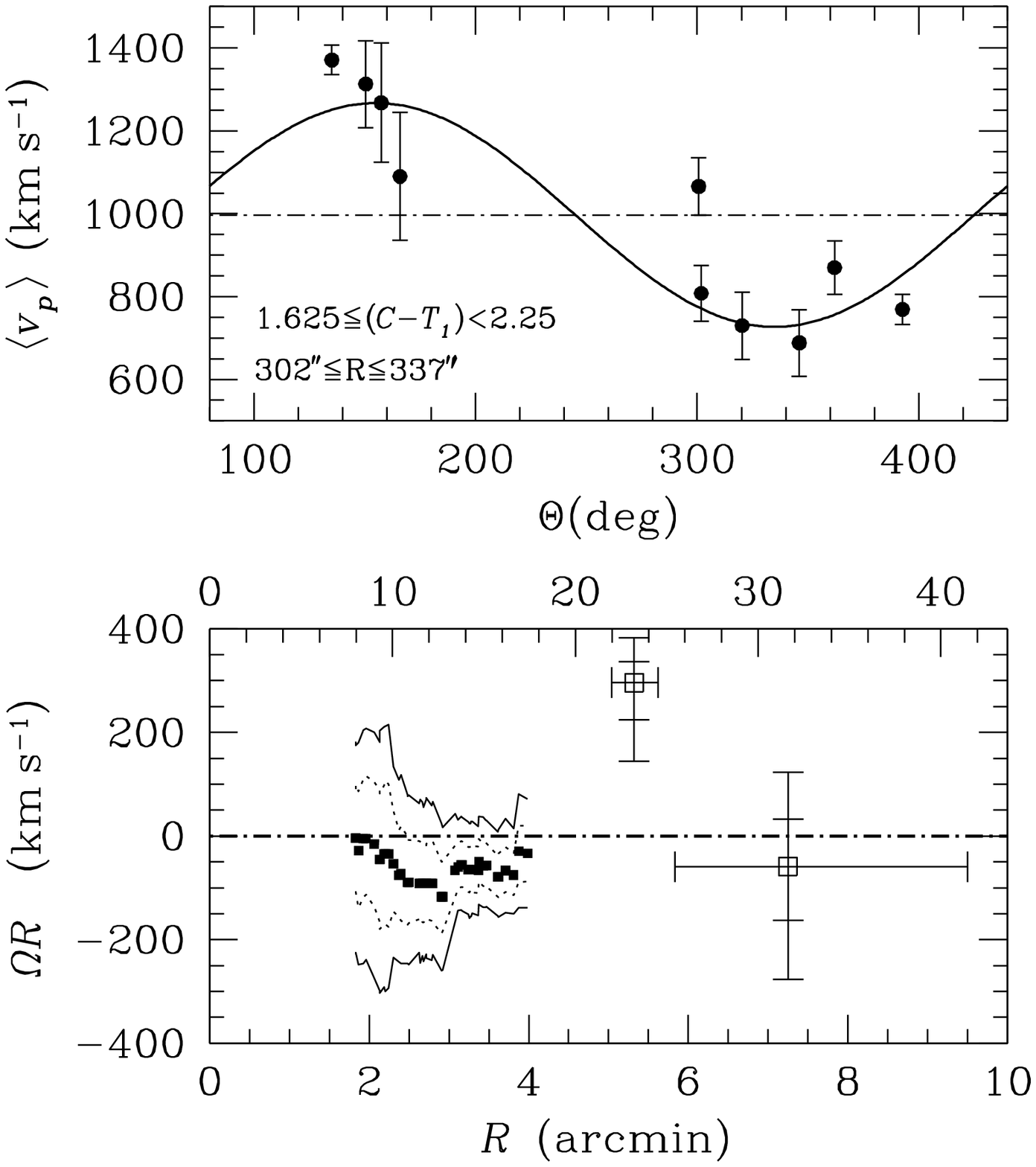}
 
\figcaption[fig11.eps]{\scriptsize {\it Upper Panel)} Detection of strong rotation from
10 metal-rich globular clusters within in a narrow range of galactocentric radius,
$22\leq R\leq 24.5$ kpc. The sine curve drawn has a zero at
the photometric minor axis of the galaxy, {\it i.e.}, $\Theta_0=245^\circ$
(Kim \etal 2000), and an amplitude of 296 \kms\ (with 1-$\sigma$
uncertainties of $+50$ \kms\ and $-70$ \kms; and 2-$\sigma$ uncertainties
of $+90$ \kms\ and $-150$ \kms).
{\it (Lower Panel)} Projected rotation amplitude as a function of
galactocentric radius in the metal-rich globular cluster system, for fits of the sine
curve in equation (\ref{eq5}) with both the systemic velocity held fixed
(at the velocity of M49 itself) and the position angle of the rotation
axis forced to coincide with the photometric minor axis of the galaxy.
The smoothing process of Figure~\ref{fig10} is used only on those globular 
clusters with
$R<302\arcsec=22$ kpc. The large point at $R=5\farcm3$ and $\Omega R=296$
\kms\ represents the 10 globular clusters isolated in the upper panel, while that at
$R=7\farcm25$ and $\Omega R=-59$ \kms\ represents the sine fit with
$v_{\rm sys}$ and $\Theta_0$ both fixed in the outermost radial bin,
$350\arcsec\leq R<570\arcsec$.
\label{fig11}}
 
\clearpage
 
\plotone{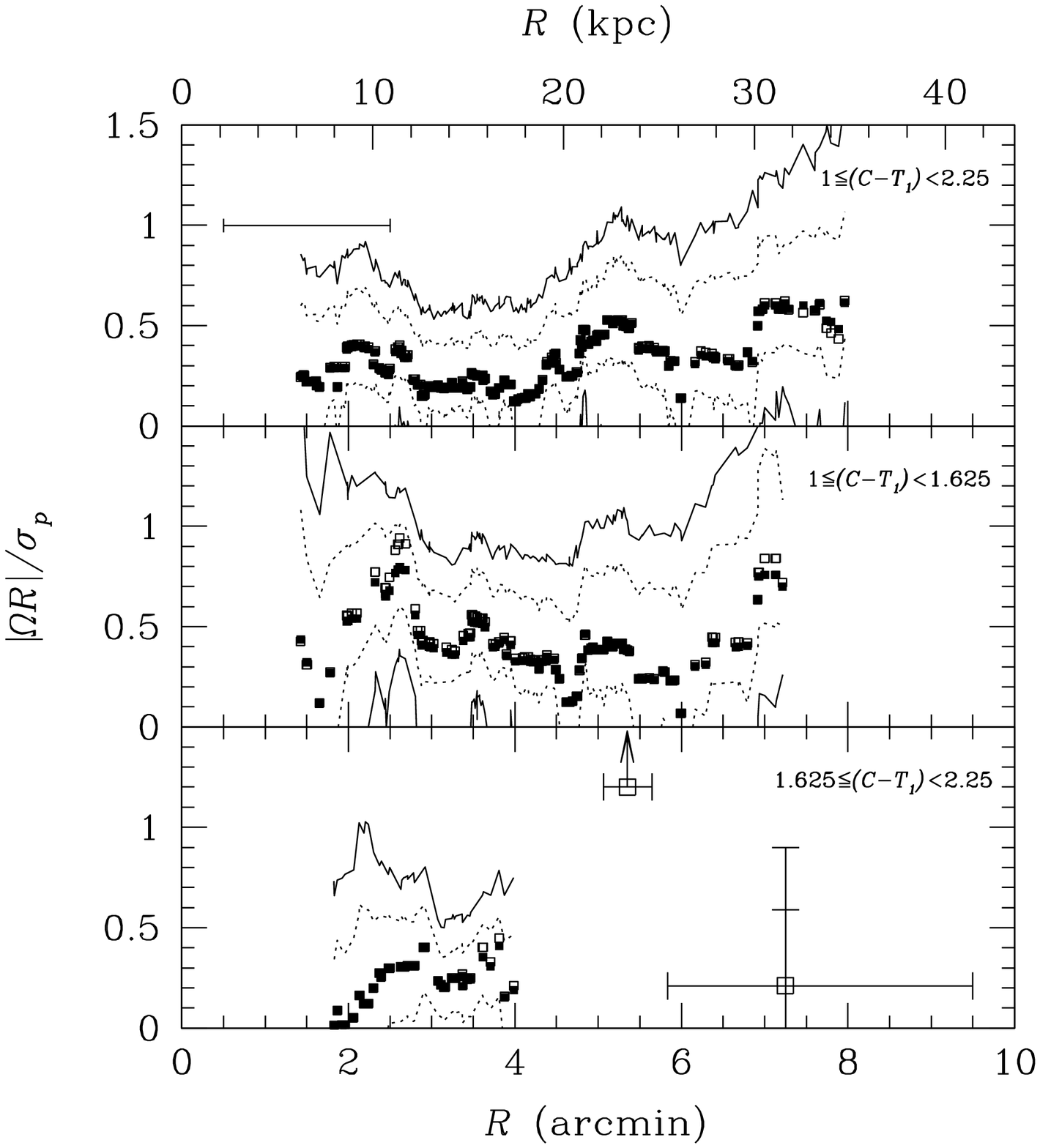}
 
\figcaption[fig12.eps]{\scriptsize {\it (Upper Panel)} Ratio of the projected
rotation amplitude to the line-of-sight velocity dispersion for M49 globular
clusters, plotted as a function of distance from the galaxy center. Filled
points take the ratio relative to the dispersion, $\sigma_p$, about the
average globular cluster velocity; open points use the dispersion, $\sigma_{p, r}$, about
the best fit of equation (\ref{eq5}) at every radius. The horizontal errorbar
shows the 2\arcmin\ width of the sliding radial bin used to ``smooth'' the
data. {\it (Middle Panel)} As above, but for the metal-poor globular clusters.
{\it (Lower Panel)} As above, but for the metal-rich globular clusters. The
rotation amplitude in question is taken from the lower panel of
Figure~\ref{fig11} rather than that in Figure~\ref{fig10}. The large open square
represents the 10 rapidly rotating, metal-rich globular clusters in the upper panel of
Figure~\ref{fig11}. For these objects, $|\Omega R|/\sigma_p=1.2$; the
corresponding rotation-corrected value is $|\Omega R|/\sigma_{p,r}=2.9$.
\label{fig12}}
 
\clearpage
 
\plotone{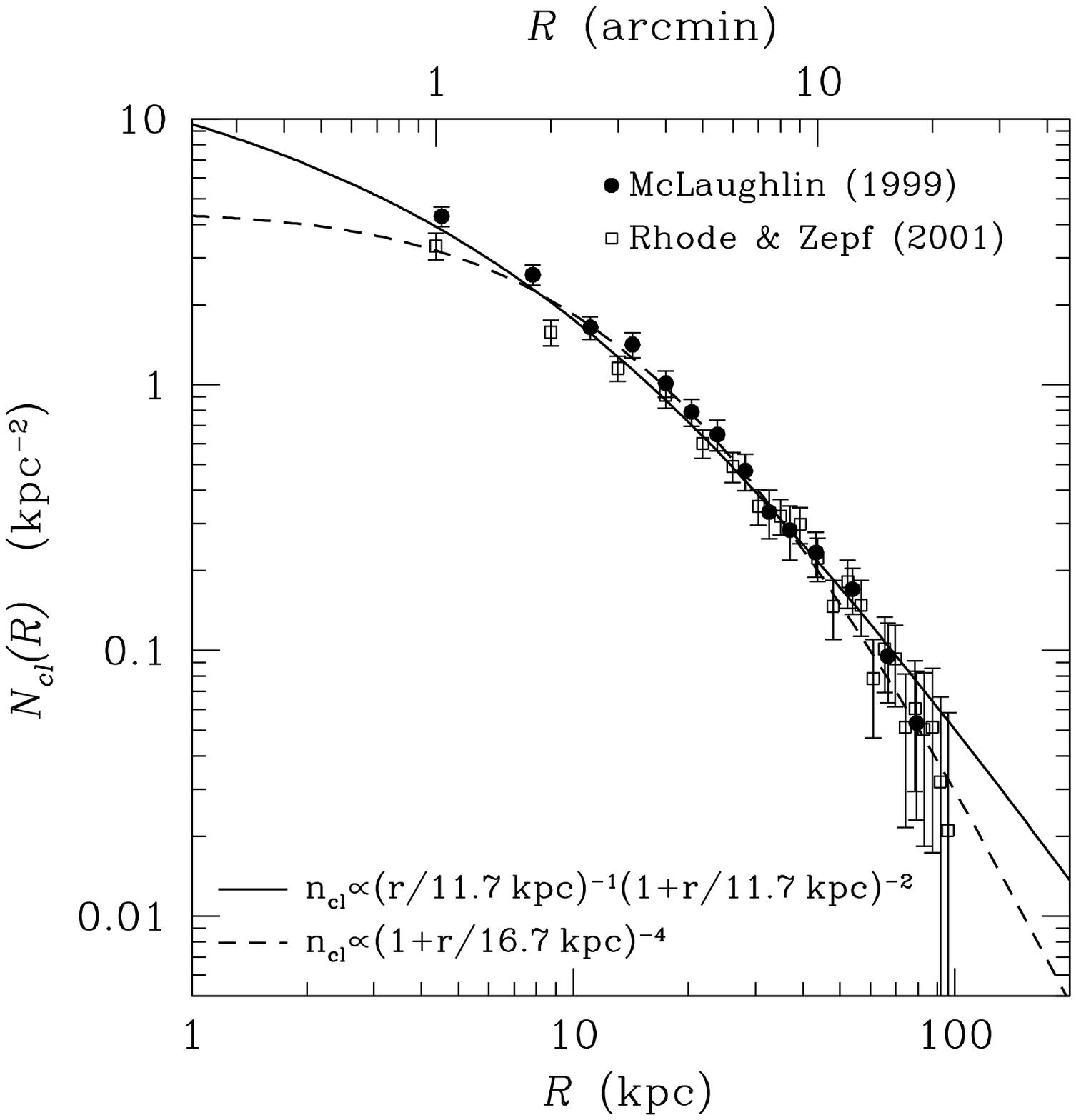}
 
\figcaption[fig13.eps]{\scriptsize Projected number-density profile for the M49
globular cluster system. Filled circles are data compiled by McLaughlin
(1999a); open squares come from the independent study of Rhode \& Zepf
(2001). Both profiles already include all scalings and corrections to
account for background and foreground contamination, and for photometric
incompleteness in the relevant globular cluster number counts. Solid line shows the best
fit of a projected three-dimensional law of the NFW form, as indicated on
the graph. The broken curve is the best-fitting projection of one of the family
of density models developed by Dehnen (1993). (See also eqs.~[\ref{eq9}]
and [\ref{eq10}] in the text.)
\label{fig13}}
 
\clearpage
 
\plotone{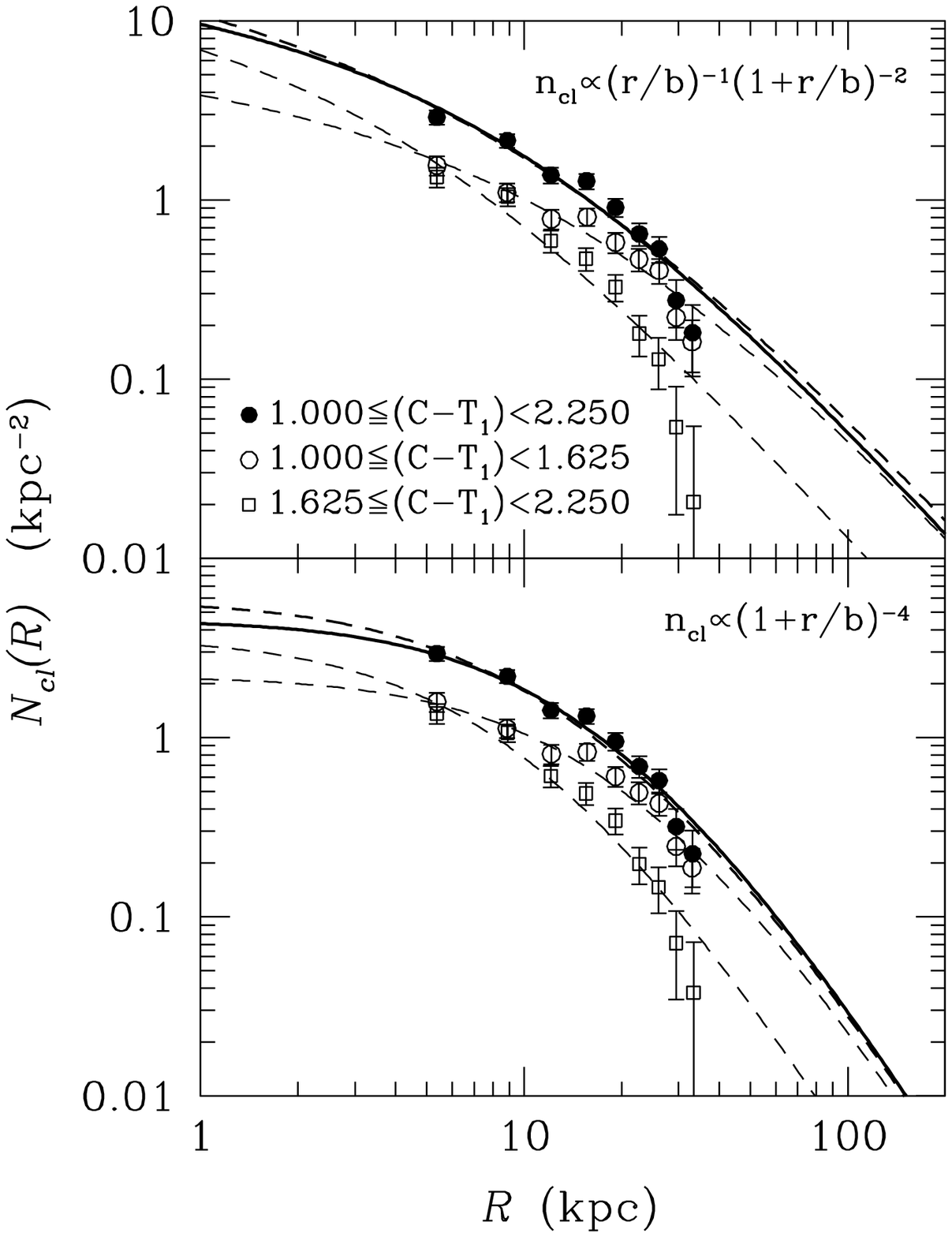}
 
\figcaption[fig14.eps]{\scriptsize Fits to the projected number-density profiles of
metal-poor and metal-rich globular clusters in M49 (open circles and squares), defined
by us from the catalog of Lee \etal (1998), for two assumed functional
forms of the three-dimensional $n_{\rm cl}(r)$ (see eqs.~[\ref{eq11}]
and [\ref{eq12}]). See the text for a description of the various curves.
\label{fig14}}
 
\clearpage

\plotone{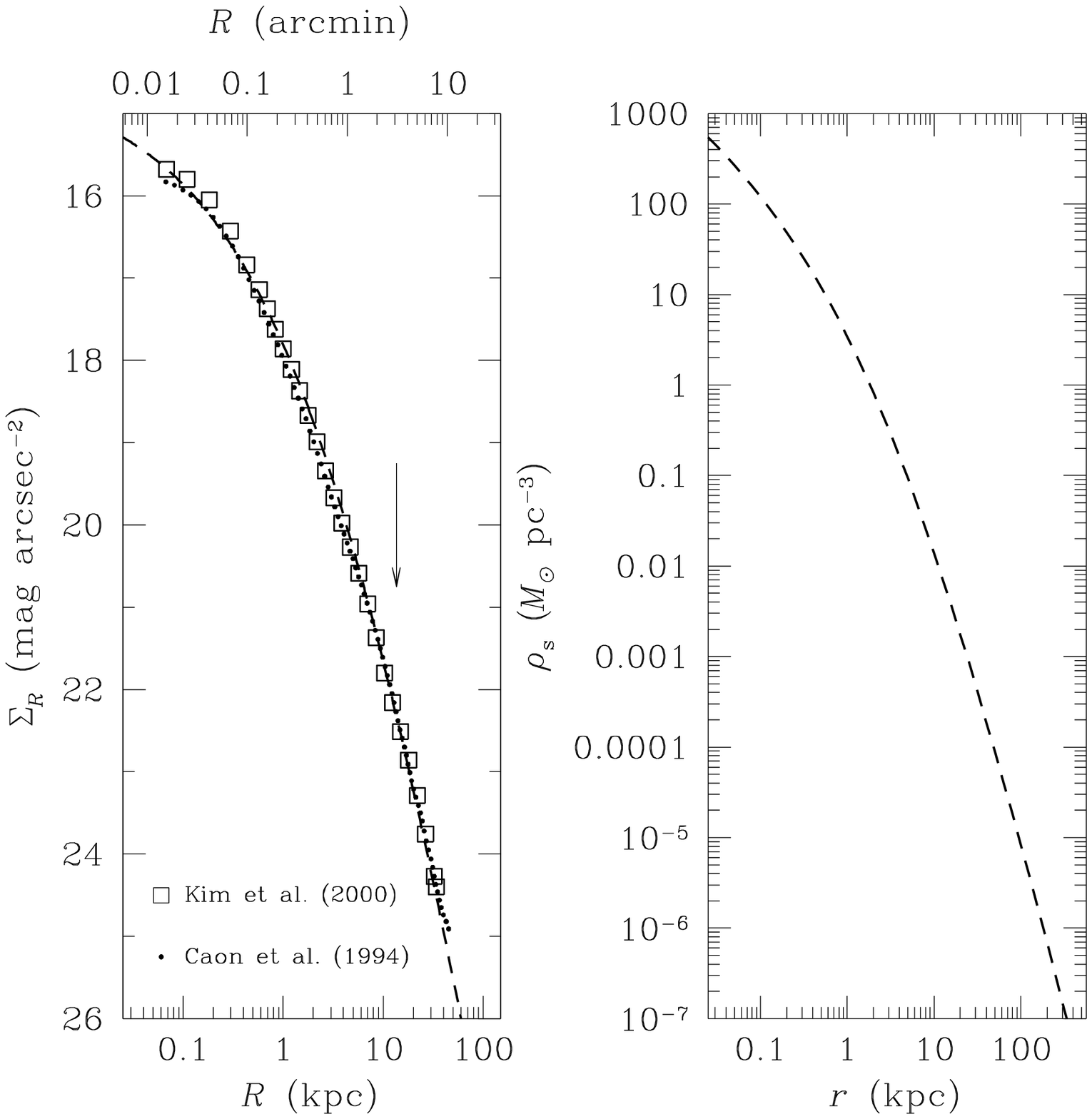}
 
\figcaption[fig15.eps]{\scriptsize {\it (Left Panel)} $R$-band surface brightness
profile for M49 (open squares) derived from the $CT_1$ photometry of
Kim \etal (2000). The dashed curve is the best fit, in projection, of the
model in equation \ref{eq13}. Dots show the $B$-band surface brightness
profile of Caon \etal (1994) after shifting by the mean galaxy color of
$\langle B-R \rangle = 1.52$. The arrow shows  our estimate for the galaxy's
effective radius, $R_{\rm eff}$ = 3\farcm1.
{\it (Right Panel)} Three-dimensional stellar mass density profile
corresponding to the model shown in the previous panel, for a spatially
constant stellar mass-to-light ratio of $\Upsilon_0 = 5.7 M_\odot\,
L_{R,\odot}^{-1}$ in the $R$-band (see \S4.3).
\label{fig15}}
 
\clearpage
 
\plotone{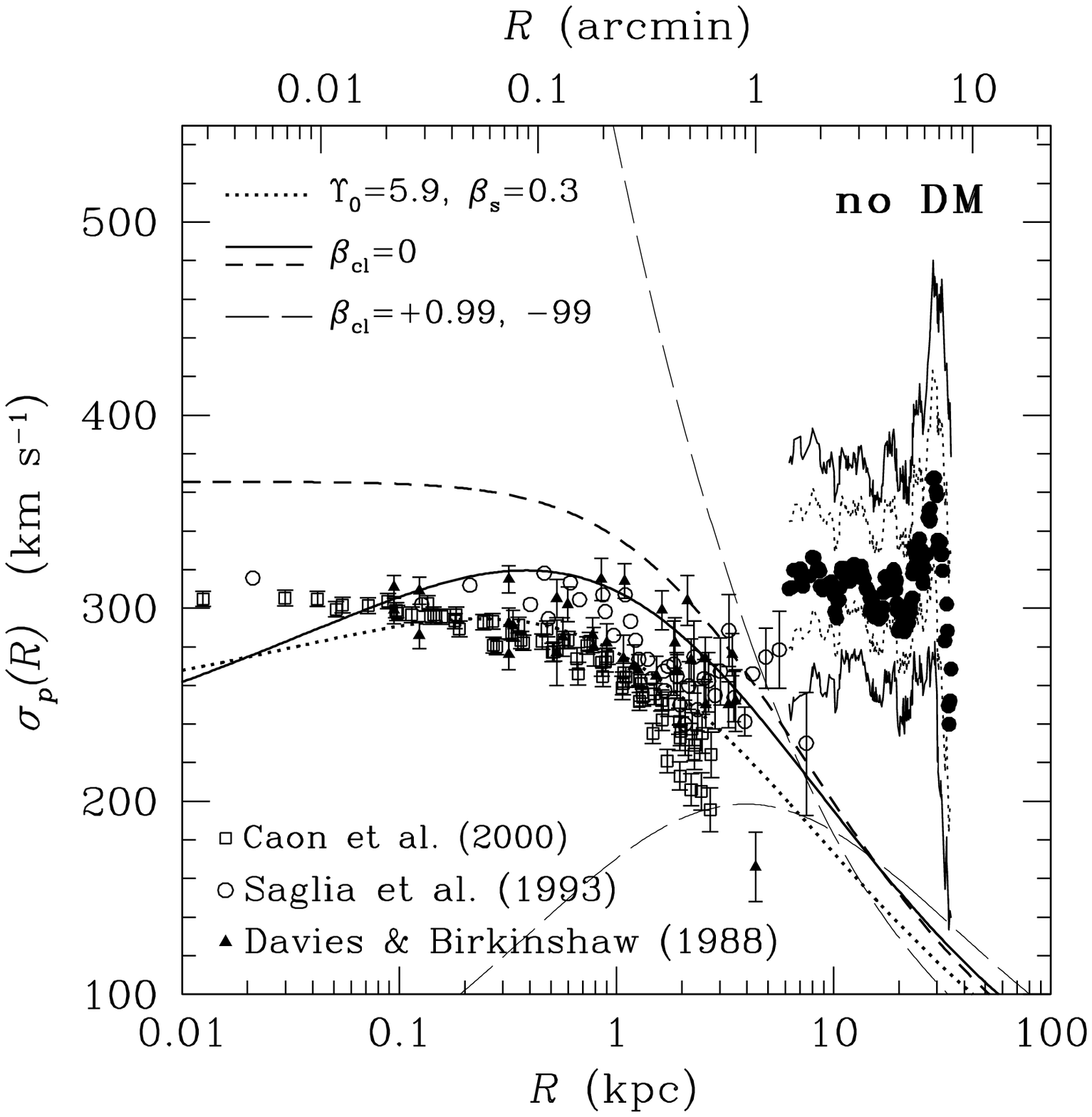}
 
\figcaption[fig16.eps]{\scriptsize Evidence from the motions of its globular clusters
that M49/Virgo B must contain a massive dark halo. Data points at small
galactocentric radii are stellar velocity-dispersion measurements from
Davies \& Birkinshaw (1988), Saglia \etal (1993), and Caon \etal (2000).
Filled circles show our velocity dispersion profile for the full sample of
globular clusters, taken from the upper panel of Figure~\ref{fig06}.
The heavy, dotted curve is the stellar velocity dispersion profile 
that would be produced in a self-consistent galaxy model: $i.e.,$ one with 
no dark-matter halo, a constant $R$-band mass-to-light ratio of 
$\Upsilon_0 = 5.9$ in solar units and a stellar velocity anisotropy $\beta_s=0.3$. 
The heavy solid line is the velocity dispersion profile predicted
for the full globular cluster system in such a model, if the density profile
$n_{\rm cl}(r)$ is given by equation (\ref{eq9}) and an isotropic velocity
dispersion is assumed. Lighter, long-dashed curves refer to models
incorporating strong radial and tangential velocity anisotropy in the 
globular cluster system. The heavy, dashed line is the prediction for an isotropic
globular cluster system if $n_{\rm cl}(r)$ is given instead by equation (\ref{eq10}).
\label{fig16}}
 
\clearpage
 
\plotone{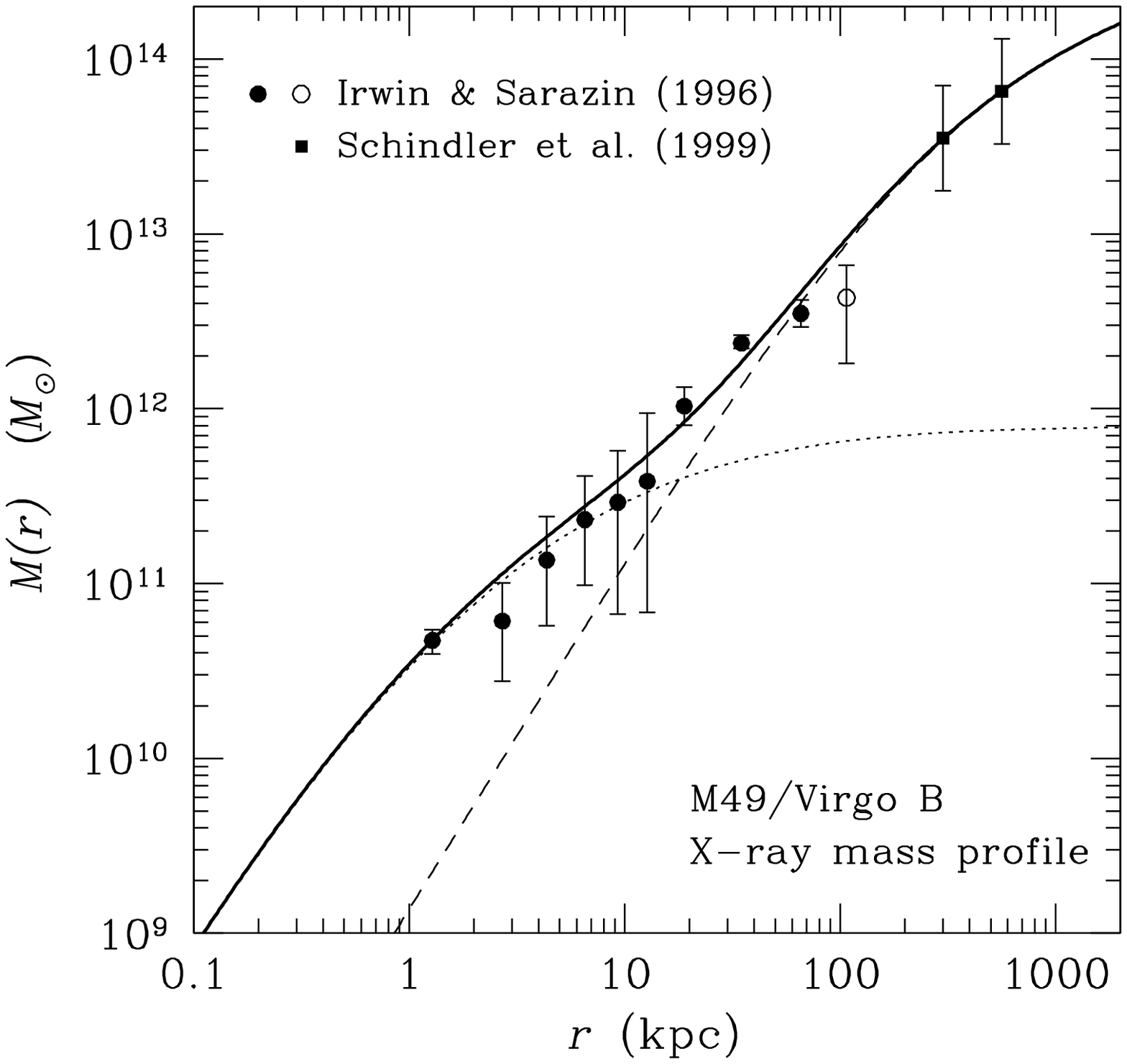}
 
\figcaption[fig17.eps]{\scriptsize 
Adopted mass model for M49 and the surrounding Virgo B subcluster, from
McLaughlin \& C\^ot\'e (2003). Dotted and dashed curves show the separate mass
profiles for the galaxy and the surrounding dark matter halo; their sum is
the total gravitating mass, $M_{\rm tot}(r)$, drawn as the bold, solid curve.
See equation (\ref{eq16}) of the text for details. Points represent the
X-ray mass measurements used to fit the model: filled circles are from Irwin
\& Sarazin (1996), with 2-$\sigma$ errorbars indicated (and one point
not used in the fit shown as an open circle), while squares are from Schindler
\etal (1999) with approximate, factor-of-two errorbars attached.
\label{fig17}}
 
\clearpage
 
\plotone{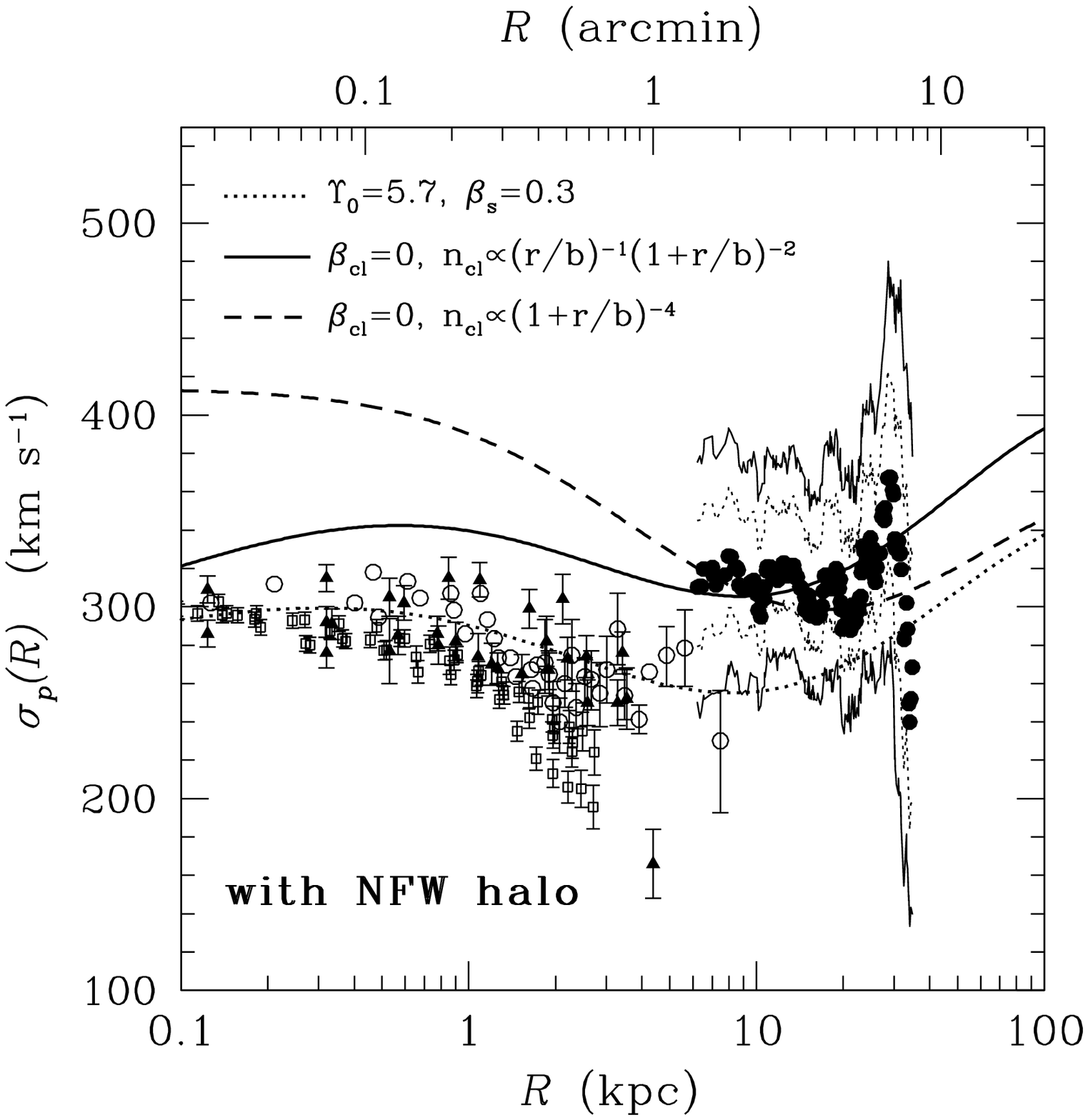}
 
\figcaption[fig18.eps]{\scriptsize Comparison of the observed stellar velocity dispersion profile 
(symbol types as in Figure~\ref{fig16}) with the predicted profile (heavy, dotted 
curve) given the mass model of Figure~\ref{fig17} and equation (\ref{eq16}), for a stellar velocity
anisotropy $\beta_s=0.3$. Velocity dispersion data for the full globular cluster system
are shown as filled circles, with 68\% and 95\% confidence bands indicated.
The bold solid and dashed curves are predicted $\sigma_p(R)$ profiles for the globular
cluster system assuming an isotropic velocity ellipsoid and adopting the
density-profile fits of equations (\ref{eq9}) and (\ref{eq10}).
\label{fig18}}
 
\clearpage
 
\plotone{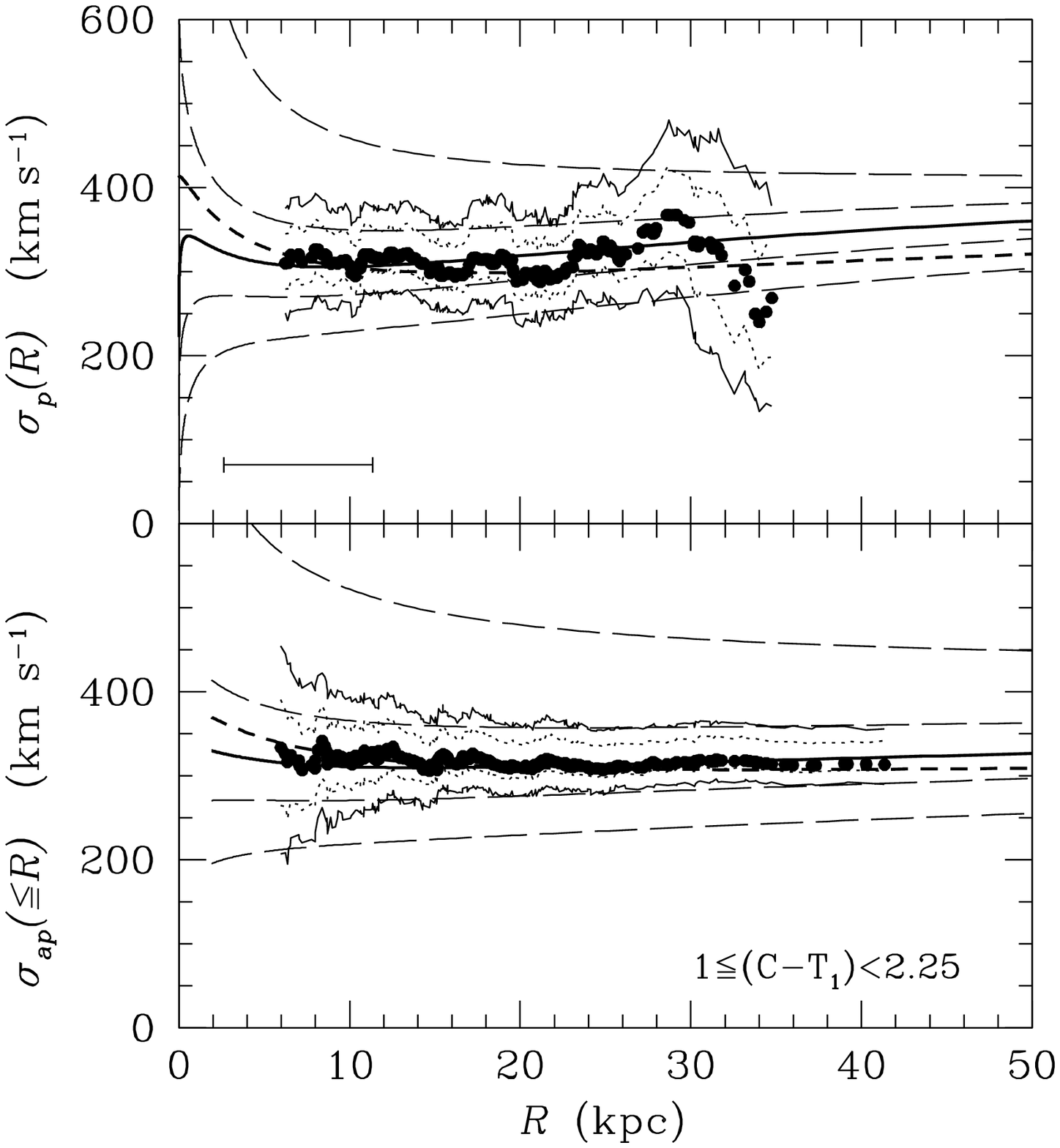}
 
\figcaption[fig19.eps]{\scriptsize {\it (Upper Panel)} Velocity dispersion profile,
${\sigma}_p(R)$, for the full sample of 263 globular clusters along with
bootstrap estimates of the 68\% and 95\% confidence intervals
(from Figure~\ref{fig06}).
The heavy solid curve is the predicted profile for clusters that are
embedded in the mass distribution given by equation (\ref{eq16}), follow the
density profile of equation (\ref{eq9}), and have an isotropic
velocity dispersion tensor. Lighter, long-dashed curves vary the
assumption on the velocity anisotropy: from top to bottom, $\beta_{\rm cl}(r)
\equiv 0.99$, 0.5, $-1$, and $-99$. The bold, short-dashed curve is a model
assuming isotropy but using the fit of equation (\ref{eq10}) to
$n_{\rm cl}(r)$. The horizontal errorbar in the lower left corner represents
the 2\arcmin\ smoothing width used to construct the empirical profile.
{\it (Lower Panel)} Aperture velocity dispersion profile for the complete
sample of globular clusters. Smooth curves are the models of the upper panel,
spatially averaged through equation (\ref{eq17}) in the text.
\label{fig19}}
 
\clearpage
 
\plotone{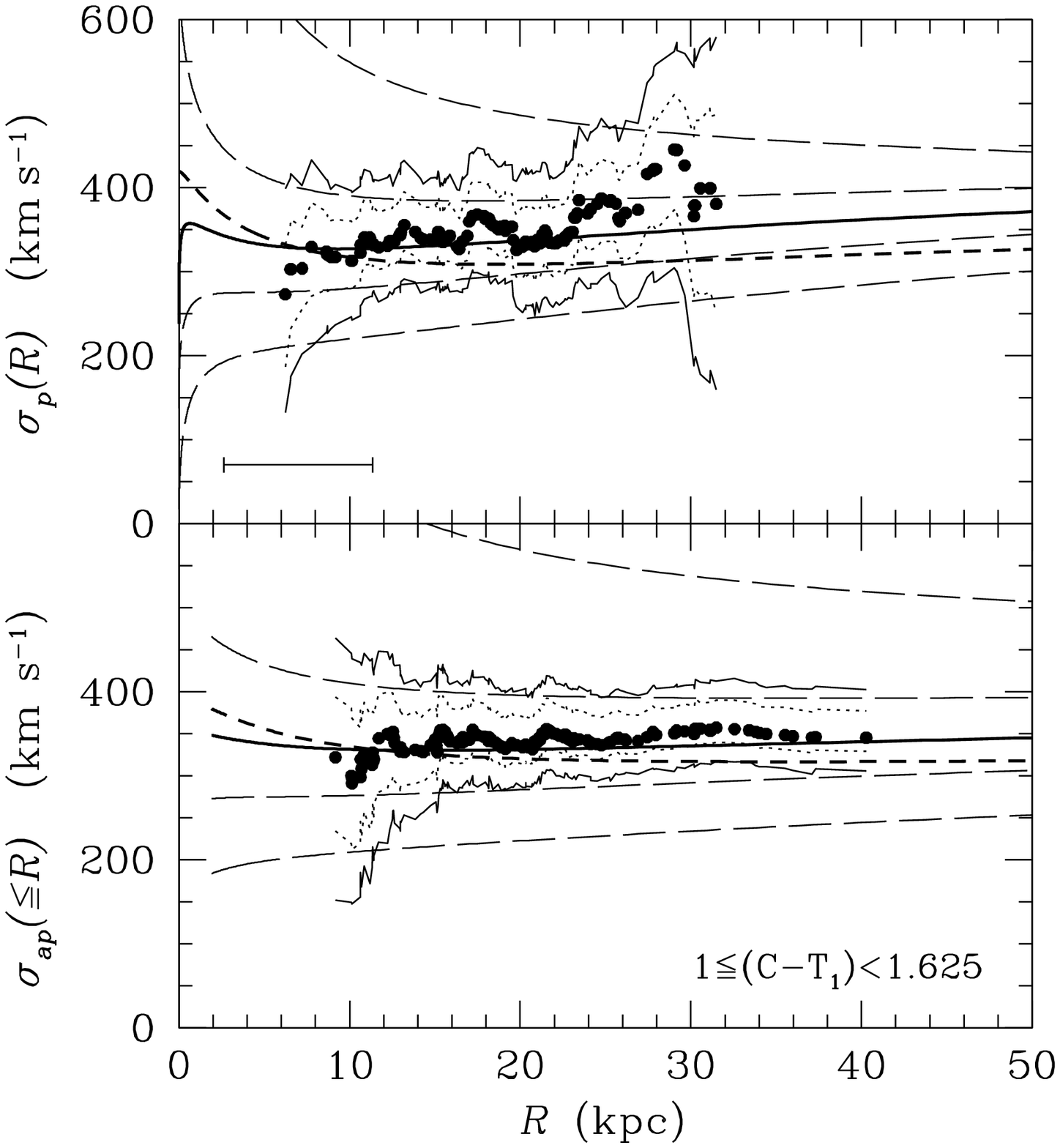}
 
\figcaption[fig20.eps]{\scriptsize Similar to Figure~\ref{fig19}, except for the sample
of 158 metal-poor globular clusters. Model curves now use the density-profile
fits for $n_{\rm cl}^{\rm MP}(r)$ in equations (\ref{eq11}) (bold solid and
light, long-dashed curves) and (\ref{eq12}) (bold, short-dashed curves).
\label{fig20}}
 
\clearpage
 
\plotone{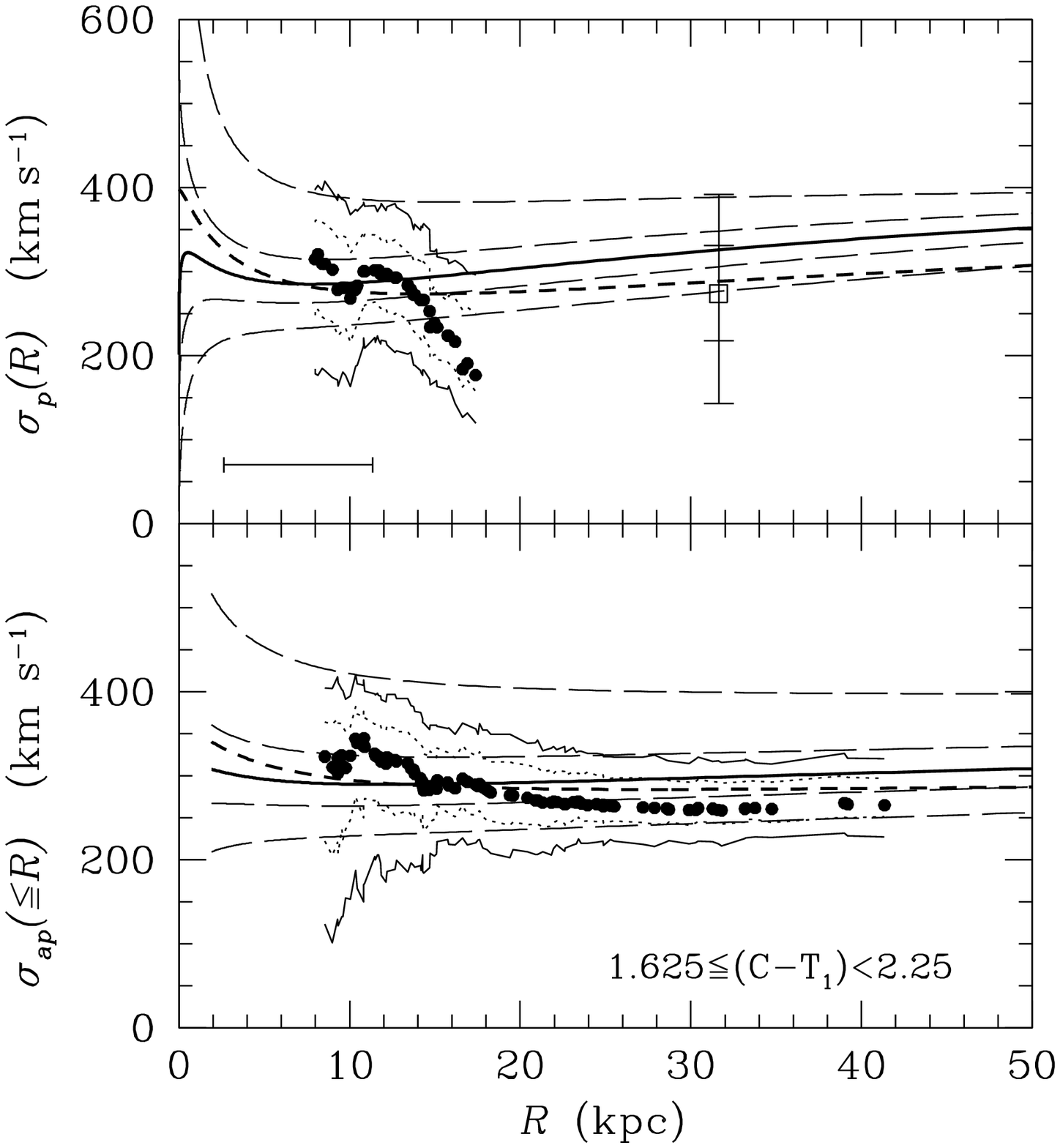}
 
\figcaption[fig21.eps]{\scriptsize Similar to Figures~\ref{fig19} and \ref{fig20}, but for
a sample of metal-rich globular clusters. Objects with $300\arcsec< R <
350\arcsec$ ($21.8\ {\rm kpc}< R < 25.5$ kpc) are excluded from the data
in the upper panel, though not from the spatially averaged, aperture
dispersion profile in the lower panel. The large, open square in the upper panel
represents the velocity dispersion (with 1- and 2-$\sigma$ errorbars) of
all metal-rich globular clusters beyond $R = 350\arcsec=25.5$ kpc. Model curves are
analogous to those in the previous Figures but employ the density-profile
fits for $n_{\rm cl}^{\rm MR}(r)$ in equations (\ref{eq11}) (bold solid and
light, long-dashed curves) and (\ref{eq12}) (bold, short-dashed curves).
\label{fig21}}
 
\end{document}